\documentclass[%
 reprint,
 amsmath,amssymb,
 aps,
 showkeys,
]{revtex4-2}

\usepackage{graphicx}
\usepackage{dcolumn}
\usepackage{bm}

\usepackage{xcolor}

\begin{document}

\preprint{APS/123-QED}

\title{Clustering of the extreme: A theoretical description of weak lensing critical points power spectra in the mildly nonlinear regime}

\author{Zhengyangguang Gong}
\email{lgong@usm.lmu.de}
\affiliation{
 Universit\"{a}ts-Sternwarte, Fakult\"{a}t f\"{u}r Physik, Ludwig-Maximilians-Universit\"{a}t M\"{u}nchen, Scheinerstra{\ss}e 1, 81679 M\"{u}nchen, Germany
 }
 \affiliation{
 Max Planck Institute for Extraterrestrial Physics, Giessenbachstra{\ss}e 1, 85748 Garching, Germany
 }
\author{Alexandre Barthelemy}
\affiliation{
 Universit\"{a}ts-Sternwarte, Fakult\"{a}t f\"{u}r Physik, Ludwig-Maximilians-Universit\"{a}t M\"{u}nchen, Scheinerstra{\ss}e 1, 81679 M\"{u}nchen, Germany
 }
 \affiliation{
 Universit\'e Paris-Saclay, Universit\'e Paris Cit\'e, CEA, CNRS, Astrophysique, Instrumentation et Mod\'elisation Paris-Saclay, 91191 Gif-sur-Yvette, France
}
\author{Sandrine Codis}
\affiliation{
 Universit\'e Paris-Saclay, Universit\'e Paris Cit\'e, CEA, CNRS, Astrophysique, Instrumentation et Mod\'elisation Paris-Saclay, 91191 Gif-sur-Yvette, France
}

\date{\today}

\begin{abstract}
In cosmic web analysis, complementary to traditional cosmological probes, the extrema (e.g. peaks and voids) two-point correlation functions (2PCFs) are of particular interest for the study of both astrophysical phenomena and cosmological structure formation. However most previous studies constructed those statistics via N-body simulations without a robust theoretical derivation from first principles. A strong motivation exists for analytically describing the 2PCFs of these local extrema, taking into account the nonlinear gravitational evolution in the late Universe. In this paper, we derive analytical formulae for the power spectra and 2PCFs of 2D critical points, including peaks (maxima), voids (minima) and saddle points, in mildly non-Gaussian weak gravitational lensing fields. We apply a perturbative bias expansion to model the clustering of 2D critical points. A generalized Gram-Charlier A series expansion is used to describe the probability density functional of the cosmic density field. We successfully derive the power spectrum of weak lensing critical points up to the next-to-next-to-leading order (NNLO) in gravitational perturbation theory, where trispectrum configurations of the weak lensing field have to be included. We numerically evaluate those power spectra up to the next-to-leading order (NLO), which correspond to the inclusion of bispectrum configurations, and transform them to the corresponding 2PCFs. An exact Monte Carlo (MC) integration is performed assuming a Gaussian distributed density field to validate our theoretical predictions. Overall, we find similar properties in 2D compared to the clustering of 3D critical points previously measured from N-body simulations. Contrary to standard lensing power spectra analysis, we find distinct BAO features in the lensing peak 2PCFs due to the gradient and curvature constraints, and we quantify that non-Gaussianity makes for $\sim 10\%$ of the signal at quasi-linear scales which could be important for current stage-IV surveys.
\end{abstract}

\maketitle

\section{Introduction}
\label{sec:intro}
The statistics of critical points, both in 3D and 2D, have attracted significant interests due to their applications to cosmology. In 3D, peaks in the initial Lagrangian density field are key sites for the nonlinear formation of dark matter halos (Ref.~\cite{galaxybias2018} and references therein) and their statistics, such as abundance and correlation functions, in Gaussian random fields have been extensively investigated in the past literature \cite{BBKS,Lumsden1989,peacock1990,Regos1995,matsubara2019,2021PhRvD.103h3530B}. Voids, which evolve in the quasi-linear regime, can be effectively modeled using relatively simple linear theory. As such, void statistics can serve as an effective cosmic laboratory for testing modified gravity and exploring dark energy phenomena \cite{Bos2012, Barreira2015, Perico2019,contarini2021, correa2023}. Additionally, they provide independent and complementary probes for constraining cosmological parameters \cite{Hamaus2020, contarini2023}. Cosmic filaments and walls, while being relatively less studied, provide valuable insights into phenomena such as matter transportation, cosmic web formation and galaxy evolution. Furthermore, their cross-correlations with peaks and voids offer a geometric characterization of the large-scale structure of the Universe \cite{Sousbie2011, Cautun2014, Sarron2019, Shim2021, Feldbrugge2024, Galrraga2024}.

In 2D, the statistics of both peaks (maxima) and voids (minima) in weak gravitational lensing have been widely studied and applied to infer cosmological parameters using data from Stage-III surveys, including the Dark Energy Survey (DES), the Kilo-Degree Survey (KiDS), and the Hyper Suprime-Cam SSP Survey (HSC) \citep{Davies2021, Harnois2021, Zurcher2022, Harnois2024,Marques2024}. These studies demonstrate that peaks and voids capture non-Gaussian information in the cosmological field. When combined with the weak lensing angular power spectrum, they significantly enhance parameter constraints compared to using the angular power spectrum alone. In particular for the parameter $S_8=\sigma_8\sqrt{\Omega_{\rm m}/0.3}$, several studies reported improvements in constraints of approximately $35\%\sim 40\%$ \cite{Zurcher2022,Marques2024}.

However, all these studies in weak lensing relied on a simulation-based inference approach, where the peak and void statistics were emulated using a grid of N-body simulations spanning various cosmologies, often with machine learning tools such as deep neural networks or Gaussian processes. This approach carries the risk of propagating numerical systematics inherent in the simulations into the emulated statistics, potentially biasing the resulting cosmological inferences \citep{Cannon2022,Horowitz2022}. Furthermore, from an analytical modeling perspective, this method lacks a robust theoretical foundation derived from first principles. On the other hand, most previous analytical modeling of peak and void statistics generally assumes a Gaussian cosmological density field. However, the weak lensing fields on our scales of interest are not Gaussian distributed as a consequence of the nonlinear gravitational evolution of the density field in the late Universe. 

There is thus a strong motivation to study the statistics of peaks and voids in non-Gaussian weak lensing fields in cosmology (as done for their number densities e.g. in Ref.~\cite{Gay2012}). Along with saddle points, these features are collectively referred to as critical points. In this work, we focus on the analytical modeling of the power spectrum which leads to the 2-point correlation functions (2PCFs) of critical points in non-Gaussian weak lensing fields, aiming to bridge the gap between current numerical approaches and a theoretical understanding of these statistics. Our analytical method builds on the general formalism proposed in Ref.~\cite{matsubara2020}, applied here to 2D weak lensing fields. The approach involves a direct perturbative bias expansion in the Eulerian density field. We derive analytical formulae with perturbative approximations up to the next-to-next-to-leading order (NNLO), which incorporates the trispectrum of the density field. For numerical computation, however, we limit our analysis to the next-to-leading order (NLO), which includes the bispectrum of the density field and represents the lowest-order non-Gaussian correction. All bias coefficients in the derived formulae can be computed analytically with an order-by-order correspondence to operators in the perturbative expansion. These coefficients can be interpreted as response functions of the critical point number densities to variations in the long-wavelength modes of the underlying density field.

Our paper is organized as follows: In Sec.~\ref{sec:readers_digest}, we first summarize our results and describe qualitatively the physical behaviour of the extrema 2PCFs before diving into the rigorous derivation of the plots and formulae we schematically present in this section. The following Sec.~\ref{sec:2d_critical_points} provides the formal definitions of general 2D critical points and their number density functions. It also presents the probability density function of critical points under the assumption of a Gaussian distributed density field. In Sec.~\ref{sec:2d_extrema_2pcf}, we derive the 2PCFs of 2D critical points, including both auto- and cross-correlations, up to the NNLO, and discuss the derivation of the bias coefficients. Next in Sec.~\ref{sec:2d_weak_lensing}, we incorporate weak lensing formalism into our analytical predictions and present numerical results for the 2PCFs of 2D weak lensing critical points up to the NLO. We then validate our perturbative bias expansion predictions by comparing them with results from computationally intensive high-dimensional numerical integrations in Sec.~\ref{sec:MC_integration}.  Finally,
conclusions are given in Sec.~\ref{sec:conclusion}.

\section{Summary and intuitive discussion of our results}
\label{sec:readers_digest}

Here we briefly discuss our main results, providing some qualitative arguments for readers to better understand the detailed analytical derivations in subsequent sections.

\subsection{2PCF of all critical points in mildly non-Gaussian weak-lensing fields}

\begin{figure}
    \centering
    \includegraphics[width=\linewidth]{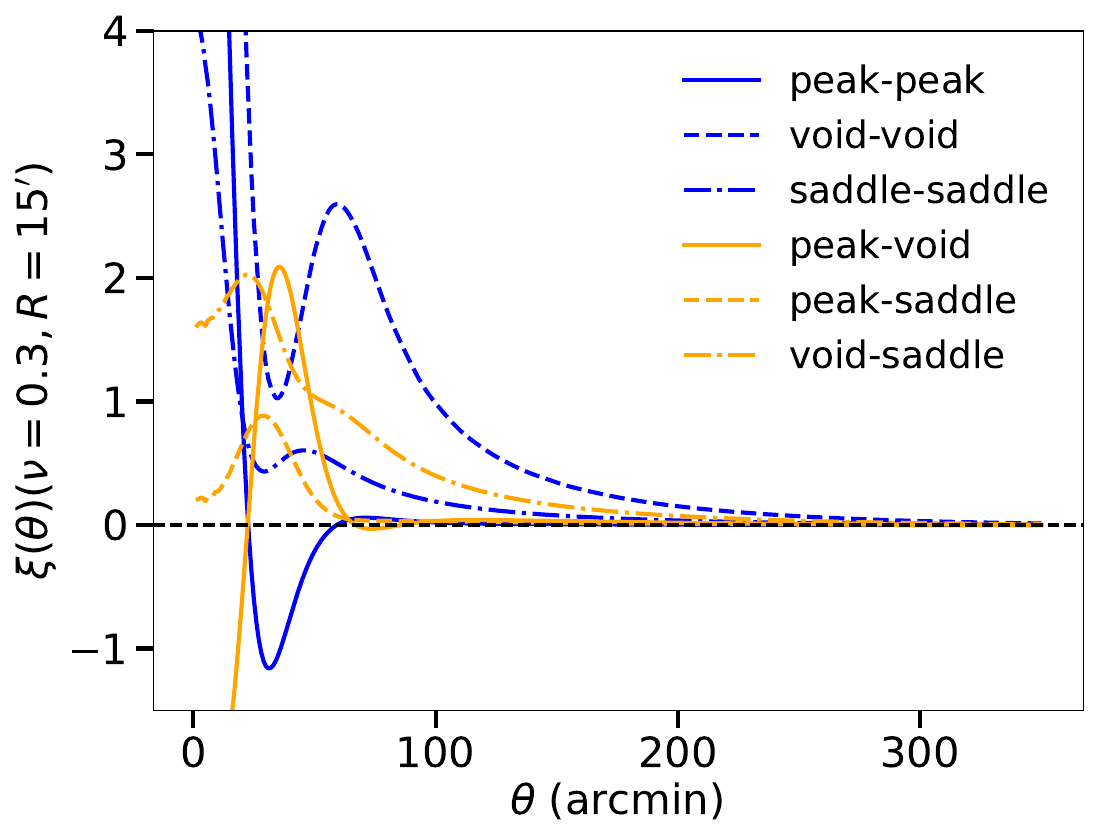}
    \caption{A summary of 2PCFs of all pairs of critical points in weak lensing convergence fields above a threshold of $\nu=0.3$. The convergence is smoothed with a Gaussian kernel at scale $R=15^{\prime}$. Blue curves represent auto 2PCFs while orange curves display cross 2PCFs (between different types of critical points). Within each color, different curve configurations represent different types of critical points in a 2PCF. Note that the chosen convergence field is for sources located at $z_s = 1.5$.}
    \label{fig:2pcf_summary}
\end{figure}

Our major result is a from-first-principles analytical expression for the power spectrum of all pairs of critical points in 2D mildly non-Gaussian fields, which we apply to the projected weak-lensing convergence. The full expression will be rigorously derived in Eq.~(\ref{eq:F_power_spectrum_nnlo}) but we display straight away the resulting 2PCFs among all pairs of critical points in weak lensing convergence field in Fig.~\ref{fig:2pcf_summary}. The $\nu$ variable in the figure, and the remainder of the text, is a parameter that allows us to characterize the ``rarity" of the considered extrema by only focusing on extrema of amplitude above the chosen $\nu \equiv \delta/\sigma_0$ where $\sigma_0$ can be expressed using Eq.~(\ref{eq:spectral_moment}). As such, fixing for instance $\nu = 0.3$ can be read as ``extrema whose amplitudes are larger than 0.3$\sigma_0$ above the mean density of the considered field", here the weak lensing convergence. This choice is arbitrary, and is usually more suited to the study of peaks of the field. Our derivations are generic enough so that another choice, e.g.~looking at extrema below a threshold, at a specific amplitude or within an interval, could straightforwardly be derived. We discuss below some symmetries in the formalism that enable to straightforwardly write down some of these cases without any additional derivations. On a side note, $\nu = 0.3$ corresponds to roughly $80\%$ of the total peaks and $6\%$ of voids in a Gaussian random field with standard cosmology and lensing parameters. Though subject to change, these order of magnitudes will be preserved for the mildly non-Gaussian fields that we will consider in this paper.

Schematically, our result in Eq.~(\ref{eq:F_power_spectrum_nnlo}) is the combination of two side-by-side perturbative expansions: One is in the development of the gravitational instability where orders are, in the underlying matter density field, expanded in powers of the traditional matter power spectrum (hence amplitude of fluctuations). The second expansion is a bias expansion for critical points that allows us to explicitly bypass the null gradient constraints of extrema by equivocating it to a series of responses of the extrema one-point distribution to changes in the underlying field at different (coupled) Fourier modes. At leading order (LO), the power spectrum between two critical points $i$ and $j$ is 
\begin{equation}
    P_{\rm extr}^{ij}(k)=g_1^i(\mathbf{k})g_1^j(\mathbf{k})P(k)
    \label{eq:simple1}
\end{equation}
where the bias function $g_1^i$ formally depends on the type of critical points, the chosen threshold $\nu$, and characteristics of the field. $P(k)$ is the usual power spectrum of the amplitude of the considered field, here the weak-lensing convergence.

\subsection{Interpretation of LO bias terms}

One specificity of our approach is that, without any loss of generality, our bias functions are that of a pure Gaussian field so that the non-Gaussian corrections needed in our formalism appear outside of the bias terms. This allows for a better interpretability of the behaviour of those bias terms. For example, the abundances of peaks and voids in Gaussian fields are symmetric with respect to the mean density, so that the computed biases for peaks above a threshold $\nu$, are exactly those of voids below a threshold $-\nu$. In higher dimensions than 2D, similar statements could be made for different saddle points with symmetric curvature signatures.

\begin{figure}
    \centering
    \includegraphics[width=\linewidth]{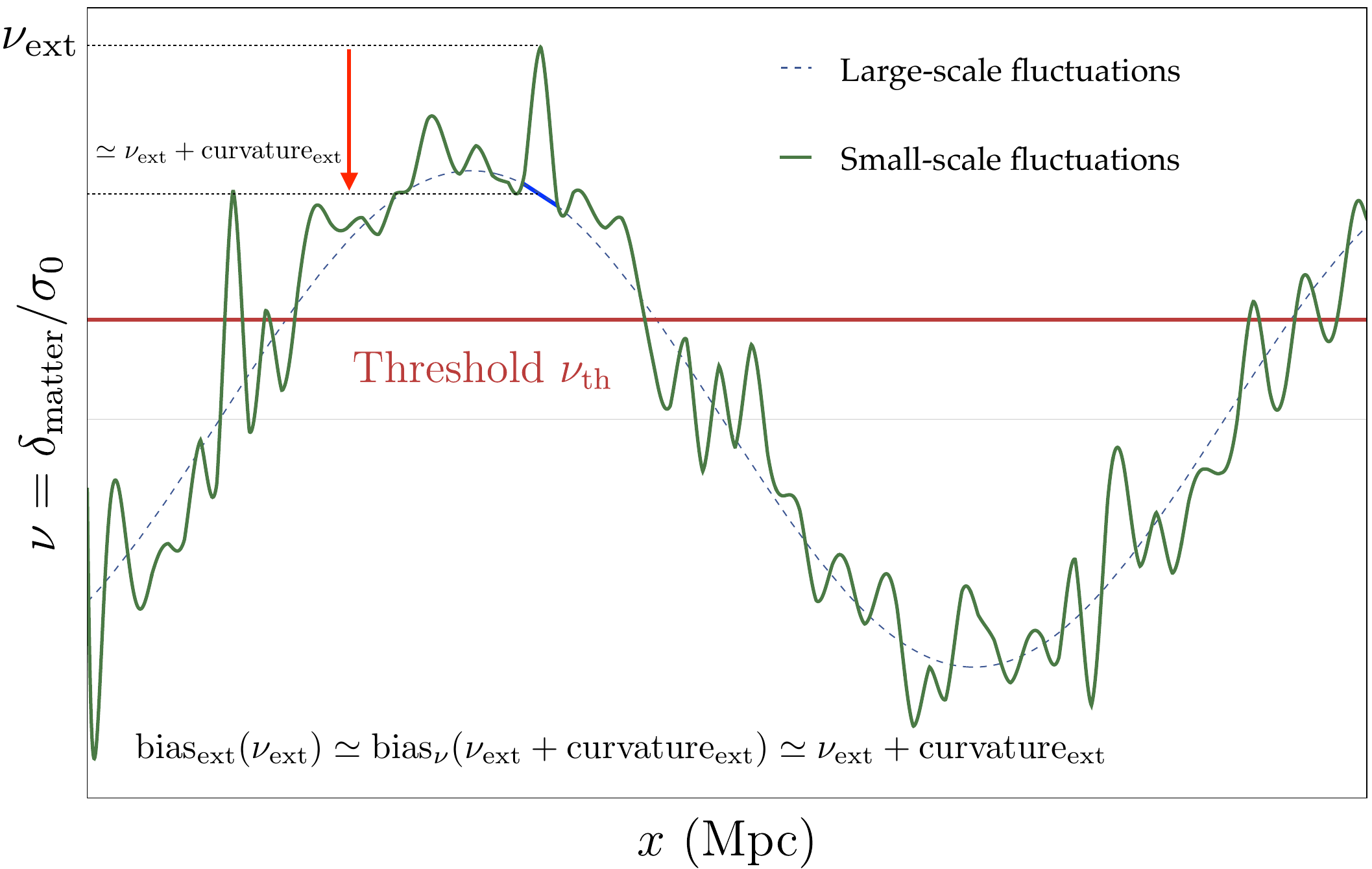}
    \caption{Illustration of the behaviour of the LO bias functions for extrema.
    We schematically decompose the field into its large and small-scales fluctuations. The critical point bias can at large $\nu$ be approximated by the value of the critical point amplitude plus its curvature.}
    \label{fig:bias_summary}
\end{figure}

Following Eq.~(\ref{eq:g1}), the LO bias term is decomposed into 
\begin{equation}
    g_1^i(\mathbf{k}) = \alpha_i(\nu) + \beta_i(\nu) k^2,
    \label{eq:simple2}
\end{equation}
where $\alpha_i$ and $\beta_i$ will b explicitly given in Eqs~(\ref{eq:g_ijklm_peaks}) to (\ref{eq:f_00_10_saddle}). Let us note several properties of this LO bias and their consequences for the extrema power spectra:
\begin{itemize}
    \item Both $\alpha_i$ and $\beta_i$ are integrals on the specific extrema constraints of the joint distribution of the field amplitude, gradient and second derivatives. Our expression seems to hide this fact through a change of variable that simplifies our calculations, but is effectively a combination of the responses (linear biases) of the amplitude, gradient and second derivatives to a mode fluctuation. 
    \item Following Eq.~(\ref{eq:simple1}), the small fluctuations in the usual convergence power spectrum are enhanced by $k^2$ and $k^4$ terms in the extrema power spectra. Performing the steps described in the above point, it formally allows to determine from which aspects of the field -- its amplitude, gradient, second derivatives, or a combination of those -- the amplifications come from. In particular, Ref.~\cite{Desjacques2008} demonstrated in their Eq.~(41) how the response to each successive spatial derivatives of a 3D field impacts the amplification of the power spectrum amplitude in the peak power spectrum calculation. It is apparent that similar expressions hold in our case, though considering all extrema \textit{above} our quoted threshold instead of at its value like in Ref.~\cite{Desjacques2008} add extra integrals that prevent us from having similar analytical expressions.
    \item The schematic behaviour of the bias terms can nevertheless be shown through qualitative arguments that we illustrate in Fig.~\ref{fig:bias_summary}. Indeed, at first order, an extrema of amplitude $\nu_{\rm ext}$ can be thought as tracing the overall matter density fluctuation of amplitude $\nu_{\rm ext}$ plus the curvature of the extrema. As such, the critical point bias will roughly behave as the overall bias of the field amplitude itself at the corresponding value of $\nu$, that is its response to a mode fluctuation of the density. Following Kaiser's formula \citep{1984ApJ...284L...9K}, we thus expect that for extrema of high amplitudes, which we control by a large chosen value of the threshold $\nu_{\rm th}$, the extrema bias will tend towards $\nu_{\rm ext} + {\rm curvature}_{\rm ext}$. The curvature of peaks being negative, and that of voids positive, this qualitatively explains why voids tend to be more biased than peaks for large values of the threshold. This is indeed what we observe in the rigorous result we will plot in Fig.~\ref{fig:7_g_ijklm} below, in its top left panel for large values of $\nu$.
\end{itemize}

\subsection{BAO features in weak-lensing critical points 2PCF}

Combining Eqs~(\ref{eq:simple1}) and (\ref{eq:simple2}), the LO critical point auto-spectrum can be written as
\begin{equation}
    P_{\rm extr}^{ii}(k)= \alpha_i^2(\nu) \left(1+ 2\frac{\beta_i(\nu)}{\alpha_i(\nu)}k^2 + \left(\frac{\beta_i(\nu)}{\alpha_i(\nu)}\right)^2k^4 \right) P(k).
    \label{eq:simple3}
\end{equation}
We now decompose the field power spectrum into a dark-matter component, schematically a power law with the index matched to the scales of interest, and a baryonic component, schematically a localized oscillation pattern coming from the Baryonic Acoustic Oscillations (BAO). For the sake of a qualitative argument, we will roughly consider this oscillation to be modeled by a \textit{Sinc} function but the important ingredient in our argument is the damping of the oscillations at large $k$. We thus have
\begin{equation}
    P(k) \approx k^n \left(1+\frac{\sin(ks)}{ks}\right),
\end{equation}
where $s$ is the typical scale of the BAO feature in the field. Remember that our application is the weak-lensing convergence field so that $s$ depends on the projection and more precisely on the redshift of the sources and the cosmology. This leads to
\begin{multline}
    P_{\rm extr}^{ii}(k) = \alpha_i(\nu)^2k^n \Biggl[1+ 2\frac{\beta_i(\nu)}{\alpha_i(\nu)}k^2 + \Biggl(\frac{\beta_i(\nu)}{\alpha_i(\nu)}\Biggr)^2k^4 \\ + 2\left|\frac{\beta_i(\nu)}{\alpha_i(\nu)}\right|\frac{k}{s}\sin(ks)\Biggl(\frac{|\alpha_i(\nu)|}{2|\beta_i(\nu)|k^2} \\ + {\rm sign}\left(\frac{\beta_i(\nu)}{\alpha_i(\nu)}\right) +\left|\frac{\beta_i(\nu)}{\alpha_i(\nu)}\right|\frac{k^2}{2} \Biggr) \Biggr],
    \label{eq:simple4}
\end{multline}
where ``$\rm sign$" in the above equation is the \textit{Sign} function. Eq.~(\ref{eq:simple4}) shows how the oscillatory behaviour in the field power spectrum can be enhanced in critical point power spectra. This in particular leads to the wiggles in the peak 2PCF that we observe in Fig.~\ref{fig:xi_pp_bao}. However, those wiggles are not observed in the void 2PCF shown in Fig.~\ref{fig:xi_critical_points_wl} in our particular setting of a Gaussian smoothing of the field at $15^{\prime}$ and $\nu = 0.3$. This can also be explained in our simplified model. Indeed, the oscillations would be dampened to the same level as in the field power spectrum if the terms in the last line of Eq.~(\ref{eq:simple4}) tend to be 0. This would happen if both $\beta_{\rm void}(\nu = 0.3)/\alpha_{\rm void}(\nu = 0.3)$ were negative and $|\beta_{\rm void}(\nu = 0.3)/\alpha_{\rm void}(\nu = 0.3)| \times k^2/2$ at the $k$ of interest were to be of order unity. For the weak-lensing convergence field in the hereby considered case, we indeed have $\beta_{\rm void}(\nu = 0.3)/\alpha_{\rm void}(\nu = 0.3) \sim - 3\times 10^{-5}$ and $k \sim 10^2 - 10^3$ for the wiggles, which qualitatively explains the absence of BAO features in our plots of the void 2PCF, while highlighting the fact that this is specific to the chosen configuration, threshold and smoothing, and does not hold in general in our formalism.

\subsection{Amplitude of the non-Gaussian corrections, impact of our results}

One of the other main results of this work is the inclusion of mild (gravitational) non-Gaussianities in the studied fields. In order to illustrate the relative importance of this effect, we plot in Fig.~\ref{fig:2pcf_smoothing} the ratio of our computed non-Gaussian peak 2PCF and its Gaussian counterpart computed at the same order in the critical point bias expansion, as a function of the applied Gaussian smoothing to the convergence field. We perform this comparison at a fixed separation where our formalism typically applies, $\theta = 150^{\prime}$ given the range of smoothing, the threshold and the redshift of the sources. As expected, the importance of the non-Gaussian terms diminishes as the smoothing increases, but it is worth noting that a smoothing of $\sim 10^{\prime}$ still shows a difference of 10\% at a separation which could be considered important, and thus close to the Gaussian regime, in the cosmological context. A careful analysis including the expected error bars of the critical point 2PCFs in a current stage-IV survey is beyond the scope of this paper and left to future studies.

\section{2D critical points}
\label{sec:2d_critical_points}
To define a critical point in a 2D random field $f$, one must consider the field amplitude $f$ itself, its derivatives $\partial_i f$ and $\partial_i\partial_j f$ up to the second order, as all local extrema require the gradient to vanish and the curvatures to comply to certain conditions. To be more explicit, the Hessian matrix at a peak location should be negative definite, it is positive definite at a void location and has both positive and negative eigenvalues at a saddle point. Based on the above discussion, we adopt the following three corresponding random variables
\begin{equation}
    \label{eq:random_variables}
    \alpha = \frac{f}{\sigma_0} \ , \hspace{3mm} \eta_i=\frac{\partial_i f}{\sigma_1} \ , \hspace{3mm} \zeta_{ij}=\frac{\partial_i\partial_j f}{\sigma_2} \ ,
\end{equation}
where $\sigma_n$ acts as a normalization constant and is defined as the spectral moment of the field
\begin{equation}
    \label{eq:spectral_moment}
    \sigma_n^2=\int\frac{{\rm d}^2k}{(2\pi)^2}k^{2n}P(k) \ ,
\end{equation}
in which $P(k)$ is the power spectrum of the 2D random field and is only a function of the magnitude of the wave vector $k=|\mathbf{k}|$ due to the supposedly statistical isotropy of the Universe. These normalization factors are chosen because we have $\langle f^2\rangle=\sigma_0^2$, $\langle(\nabla f)^2\rangle=\sigma_1^2$ and $\langle(\Delta f)^2\rangle=\sigma_2^2$ where $\Delta$ represents a Laplacian operator. The power spectrum $P(k)$ is expressed as
\begin{equation}
    \label{eq:power_spectrum_def}
    \langle\tilde{f}(\mathbf{k})\tilde{f}(\mathbf{k}^{\prime})\rangle = (2\pi)^2\delta_{\rm D}(\mathbf{k}+\mathbf{k}^{\prime})P(k) \ ,
\end{equation}
where we use $\tilde{f}$ to represent the Fourier counterpart of the random field and the Dirac delta function is a result of the statistical homogeneity.

With the above random variables defined, the number density function of critical points above a given threshold $\nu$ (meaning $f\geq \nu\sigma_0$), for example peaks $n_p(\nu)$, can be explicitly expressed as \citep{Kac1943, Rice1945, Longuet1957, BBKS}: 
\begin{equation}
    \label{eq:n_peak}
    n_p(\nu)=\left(\frac{\sigma_2}{\sigma_1}\right)^2\Theta(\alpha-\nu)\delta_{\rm D}(\boldsymbol{\eta})\Theta(\lambda_2)|\mathrm{det}\,\zeta| \ ,
\end{equation}
where $\Theta$ is the Heaviside step function and $\lambda_2$ is the smallest eigenvalue of the $2\times2$ matrix $(-\zeta)$ (without loss of generality, we assume $\lambda_1>\lambda_2$ in the discussion below). For the other two types of critical points, voids and saddle points, their respective number density functions can be derived by modifying the constraint on the eigenvalues of the Hessian matrix $\zeta$ as specified in the above equation. For voids, $\lambda_1$ must be negative, whereas for saddle points, $\lambda_1$ must be positive and $\lambda_2$ negative. Eventually, these different constraints will enter the calculation of bias coefficients as will be shown in Sec.~\ref{sec:2d_extrema_2pcf}.

Let us denote the full set of random variables which describes the critical points in 2D as $\mathbf{X}=(\alpha, \eta_1, \eta_2, \zeta_{11},\zeta_{12},\zeta_{22})$.
The statistics of this multi-variate random vector assuming a Gaussian distributed density field are solely determined by their covariances and reads
\begin{eqnarray}
    \label{eq:covariance_matrix_gaussian_entries}
    \langle\alpha^2\rangle &=& 1 \ , \quad \langle\alpha\eta_i\rangle = 0 \ , \quad \langle\alpha\zeta_{ij}\rangle = -\frac{\gamma}{2}\delta_{ij} \ , \nonumber \\
    \langle\eta_i\eta_j\rangle &=& \frac{\delta_{ij}}{2} \ , \quad \langle\eta_i\zeta_{jk}\rangle = 0 \ , \nonumber \\
    \langle\zeta_{ij}\zeta_{kl}\rangle &=& \frac{1}{8}(\delta_{ij}\delta_{kl} + \delta_{ik}\delta_{jl} + \delta_{il}\delta_{jk}) \ ,
\end{eqnarray}
where 
\begin{equation}
    \label{eq:gamma}
    \gamma=\frac{\sigma_1^2}{\sigma_0\sigma_2} \ ,
\end{equation}
and $\delta_{ij}$ is the Kronecker delta. 
The multivariate Gaussian distribution function for the random vector variable $\mathbf{X}$ is therefore:
\begin{equation}
    \label{eq:multivariate_gaussian}
    P_{\rm G}(\mathbf{X})=\frac{1}{(2\pi)^3\sqrt{\mathrm{det}\mathbf{M}}}\mathrm{exp}\left(\frac{1}{2}\mathbf{X}^{\rm T}\mathbf{M}^{-1}\mathbf{X}\right) \ ,
\end{equation}
where the covariance matrix $\mathbf{M}$ of the data vector $  \mathbf{X}$ reads
\begin{equation}
    \label{eq:covariance_matrix_gaussian}
    \mathbf{M} = \begin{pmatrix}
1 & 0 & 0 & -\frac{\gamma}{2} & 0 & -\frac{\gamma}{2}\\
0 & \frac{1}{2} & 0 & 0 & 0 & 0\\
0 & 0 & \frac{1}{2} & 0 & 0 & 0\\
-\frac{\gamma}{2} & 0 & 0 & \frac{3}{8} & 0 & \frac{1}{8} \\
0 & 0 & 0 & 0 & \frac{1}{8} & 0\\
-\frac{\gamma}{2} & 0 & 0 &\frac{1}{8} & 0 & \frac{3}{8}
\end{pmatrix} \ .
\end{equation}
It is very helpful to transform the above probability density function into another representation in terms of rotationally invariant random variables \citep{Pogosyan2009, Gay2012}:
\begin{equation}
    \label{eq:multivariate_gaussian_rotationally_invariant}
    P_{\rm G}(\mathbf{X})\propto\mathcal{N}(\alpha,J_1)\mathrm{exp}\left(-\eta^2-J_2\right) \ ,
\end{equation}
up to a normalization constant, where $\mathcal{N}(\alpha,J_1)$ is a Gaussian joint distribution of $\alpha$ and $J_1$
\begin{equation}
    \label{eq:N_function}
    \mathcal{N}(\alpha,J_1)=\frac{1}{2\pi\sqrt{1-\gamma^2}}\mathrm{exp}\left[-\frac{\alpha^2+J_1^2-2\gamma\alpha J_1}{2(1-\gamma^2)}\right] \ .
\end{equation}

In the above Eqs.~(\ref{eq:multivariate_gaussian_rotationally_invariant}) and (\ref{eq:N_function}), the new random variables $\eta$, $J_1$, $J_2$ are defined as:
\begin{eqnarray}
    \label{eq:rotationally_invariant_variable}
    \eta &&\equiv \boldsymbol{\eta} \cdot \boldsymbol{\eta} = \eta_1^2 + \eta_2^2 \ , \quad J_1 \equiv -\zeta_{ii} = \lambda_1 + \lambda_2 \ , \nonumber \\
    J_2 &&\equiv 2\tilde{\zeta}_{ij}\tilde{\zeta}_{ji} = \lambda_1^2 + \lambda_2^2 - 2\lambda_1\lambda_2 \ ,
\end{eqnarray}
where the repeated indices follow the Einstein summation convention. The random variable $\tilde{\zeta}_{ij}$ represents the traceless part of the Hessian matrix $\zeta$, $\tilde{\zeta}_{ij}=\zeta_{ij}+\delta_{ij}J_1/2$, and $J_1$ is the negative trace of the Hessian matrix.

\section{2PCFs of 2D critical points in mildly non-Gaussian fields}
\label{sec:2d_extrema_2pcf}
In this paper, our aim is to predict correlation functions of critical points. To do so, let us introduce a generic functional $\mathcal{F}$ of the density field $f$. Its power spectrum can be written down as \citep{Peebles1980}:
\begin{equation}
    \label{eq:F_power_spectrum}
    \frac{\langle\tilde{\mathcal{F}}(\mathbf{k})\tilde{\mathcal{F}}(\mathbf{k}^{\prime})\rangle_{\rm c}}{\langle\mathcal{F}\rangle^2}=(2\pi)^2\delta_{\rm D}(\mathbf{k}+\mathbf{k}^{\prime})P_{\mathcal{F}}(k) \ ,
\end{equation}
where $\tilde{\mathcal{F}}$ is the Fourier transform of the real space functional $\mathcal{F}$ and $\langle\ldots\rangle_{\rm c}$ denotes the connected part of the corresponding computed moment. The squared mean of the real space functional $\langle\mathcal{F}\rangle^2$ is a normalization factor to maintain the consistency between the definition of 2PCF and the inverse Fourier transform of the power spectrum. The connected part in Eq.~(\ref{eq:F_power_spectrum}) in Fourier space is $\langle\tilde{\mathcal{F}}(\mathbf{k})\tilde{\mathcal{F}}(\mathbf{k}^{\prime})\rangle_{\rm c}=\langle\tilde{\mathcal{F}}(\mathbf{k})\tilde{\mathcal{F}}(\mathbf{k}^{\prime})\rangle - \langle\tilde{\mathcal{F}}\rangle^2$ where 
\begin{equation}
    \label{eq:F_2nd_moment_Fourier}
    \langle\tilde{\mathcal{F}}(\mathbf{k})\tilde{\mathcal{F}}(\mathbf{k}^{\prime})\rangle=\int\mathcal{D}\tilde{f}\tilde{\mathcal{F}}(\mathbf{k})\tilde{\mathcal{F}}(\mathbf{k}^{\prime})\mathcal{P}
\end{equation}
is the 2nd-order moment of the functional $\mathcal{F}$ in Fourier space. In the above Eq.~(\ref{eq:F_2nd_moment_Fourier}), $\mathcal{D}\tilde{f}$ represents the volume element of the functional integral over $\tilde{f}$ with appropriate measures and $\mathcal{P}$ is the probability density functional of $\tilde{f}$. We apply the general formalism presented in Ref.~\cite{matsubara2020} to expand both $\mathcal{P}$ and $\tilde{\mathcal{F}}$ in Eq.~(\ref{eq:F_2nd_moment_Fourier}) with a basis composed of the Wiener-Hermite functionals defined as:
\begin{equation}
    \label{eq:Wiener-Hermite_functionals}
    \mathcal{H}_n(\mathbf{k}_1,\ldots,\mathbf{k}_n)\equiv\frac{(-1)^n}{\mathcal{P}_{\rm G}}\frac{\partial^n\mathcal{P}_{\rm G}}{\partial\tilde{f}(\mathbf{k}_1)\cdots\partial\tilde{f}(\mathbf{k}_n)} \ ,
\end{equation}
where $\mathcal{P}_{\rm G}$ here is the Gaussian probability density functional of $\tilde{f}$, different from those in Eq.~(\ref{eq:multivariate_gaussian}) and (\ref{eq:multivariate_gaussian_rotationally_invariant}). The Wiener-Hermite polynomials $\mathcal{H}_n$ contains the n-point response of the Gaussian probability density functional to the density field and $\mathcal{H}_0=1$ when $n=0$. With this mathematical tool, we can write the expansion of $\mathcal{P}$ and $\tilde{\mathcal{F}}$ as:
\begin{widetext}
\begin{eqnarray}
\label{eq:fourier_P_expansion}
    \mathcal{P}(\tilde{f})&&= \mathcal{H}_0\mathcal{P}_{\mathrm{G}} + \frac{1}{6}\int\mathrm{d}^2k_1\mathrm{d}^2k_2\mathrm{d}^2k_3\langle\tilde{f}(\mathbf{k}_1)\tilde{f}(\mathbf{k}_2)\tilde{f}(\mathbf{k}_3)\rangle_\mathrm{c}\mathcal{H}_3(\mathbf{k}_1, \mathbf{k}_2, \mathbf{k}_3)\mathcal{P}_{\mathrm{G}} \nonumber\\
    && + \frac{1}{24}\int\mathrm{d}^2k_1\ldots\mathrm{d}^2k_4\langle\tilde{f}(\mathbf{k}_1)\cdots\tilde{f}(\mathbf{k}_4)\rangle_\mathrm{c}\mathcal{H}_4(\mathbf{k}_1, \mathbf{k}_2, \mathbf{k}_3, \mathbf{k}_4)\mathcal{P}_\mathrm{G} + \frac{1}{120}\int\mathrm{d}^2k_1\ldots\mathrm{d}^2k_5\langle\tilde{f}(\mathbf{k}_1)\cdots\tilde{f}(\mathbf{k}_5)\rangle_\mathrm{c}\mathcal{H}_5(\mathbf{k}_1, \ldots, \mathbf{k}_5)\mathcal{P}_\mathrm{G} \nonumber\\
    && + \frac{1}{720}\int\mathrm{d}^2k_1\ldots\mathrm{d}^2k_6\langle\tilde{f}(\mathbf{k}_1)\cdots\tilde{f}(\mathbf{k}_6)\rangle_\mathrm{c}\mathcal{H}_6(\mathbf{k}_1, \ldots, \mathbf{k}_6)\mathcal{P}_\mathrm{G} + \cdots \nonumber\\
    && + \frac{1}{72}\int\mathrm{d}^2k_1\ldots\mathrm{d}^2k_6\langle\tilde{f}(\mathbf{k}_1)\tilde{f}(\mathbf{k}_2)\tilde{f}(\mathbf{k}_3)\rangle_\mathrm{c}\langle\tilde{f}(\mathbf{k}_4)\tilde{f}(\mathbf{k}_5)\tilde{f}(\mathbf{k}_6)\rangle_\mathrm{c}\mathcal{H}_6(\mathbf{k}_1, \ldots, \mathbf{k}_6)\mathcal{P}_\mathrm{G} + \cdots \ ,
\end{eqnarray}
\end{widetext}
and
\begin{eqnarray}
\label{eq:fourier_F_wiener_hermite_expansion}
    \tilde{\mathcal{F}}&&(\mathbf{k}) \nonumber\\
    &&= \sum_{n=0}^{\infty}\frac{1}{n!}\int\frac{\mathrm{d}^2k_1}{(2\pi)^2}\ldots\frac{\mathrm{d}^2k_n}{(2\pi)^2}(2\pi)^2\delta^2_\mathrm{D}(\mathbf{k}_1+\cdots+\mathbf{k}_n-\mathbf{k}) \nonumber\\
    && \times \mathcal{G}_n(\mathbf{k}_1,\ldots,\mathbf{k}_n)\mathcal{H}_n^\star(\mathbf{k}_1,\ldots,\mathbf{k}_n) \ .
\end{eqnarray}

The detailed derivation of Eq.~(\ref{eq:fourier_P_expansion}) can be found in  Appendix.~\ref{appendix:A} of this paper or in Appendix.~A of Ref.~\cite{matsubara2020}. The expansion in Eq.~(\ref{eq:fourier_P_expansion}) is a generalization of the Gram-Charlier A (GCA) series \citep{Blinnikov1998}. The coefficients $\langle\tilde{f}(\mathbf{k}_1)\ldots\tilde{f}(\mathbf{k}_n)\rangle_{\rm c}$ in the expansion are the corresponding n-th order cumulants of the Fourier density field. They are related to the definition of the higher-order spectrum of the density field, for example:
\begin{eqnarray}
    \label{eq:bispectrum}
    \langle\tilde{f}(\mathbf{k}_1)\tilde{f}(\mathbf{k}_2)\tilde{f}(\mathbf{k}_3)\rangle_\mathrm{c} 
    &=& (2\pi)^2 \delta_{\rm D}(\mathbf{k}_1+\mathbf{k}_2+\mathbf{k}_3) \nonumber \\ 
    && \times B(\mathbf{k}_1,\mathbf{k}_2,\mathbf{k}_3) \ ,
\end{eqnarray}
where $B(\mathbf{k}_1,\mathbf{k}_2,\mathbf{k}_3)$ is the bispectrum of the density field and
\begin{eqnarray}
    \label{eq:trispectrum}
    \langle\tilde{f}(\mathbf{k}_1)\cdots\tilde{f}(\mathbf{k}_4)\rangle_\mathrm{c} &=& (2\pi)^2\delta_{\rm D}(\mathbf{k}_1+\cdots+\mathbf{k}_4)\nonumber \\
    && \times T(\mathbf{k}_1,\ldots,\mathbf{k}_3) \ ,
\end{eqnarray}
where $T(\mathbf{k}_1,\ldots,\mathbf{k}_4)$ is the trispectrum. In Eq.~(\ref{eq:fourier_F_wiener_hermite_expansion}) we use the dual Wiener-Hermite functionals $\mathcal{H}_n^{\star}$ whose definition is:
\begin{eqnarray}
    \label{eq:dual_Wiener-Hermite_functionals}
    \mathcal{H}_n^{\star}(\mathbf{k}_1,\ldots,\mathbf{k}_n)&=&(2\pi)^{2n}P(k_1)\cdots P(k_n) \nonumber \\
    && \times \mathcal{H}_n(-\mathbf{k}_1,\ldots,-\mathbf{k}_n) \ ,
\end{eqnarray}
and it has a convenient property of being orthogonal to the Wiener-Hermite functionals with respect to the Gaussian probability density functional $\mathcal{P}_{\rm G}$ \citep{Matsubara1995}
\begin{eqnarray}
    \label{eq:Wiener-Hermite_orthogonality}
    \langle\mathcal{H}_n^{\star}(\mathbf{k}_1,\ldots,\mathbf{k}_n)&&\mathcal{H}_m(\mathbf{k}_1^{\prime},\ldots,\mathbf{k}_m^{\prime})\rangle_{\rm G}\nonumber\\
    &&=\delta_{nm}\bigg[\delta_{\rm D}(\mathbf{k}_1-\mathbf{k}_1^{\prime})\cdots\delta_{\rm D}(\mathbf{k}_n-\mathbf{k}_m^{\prime})\nonumber\\
    &&+ {\rm perm}(\mathbf{k}_1,\ldots,\mathbf{k}_n)\bigg] 
\end{eqnarray}
where ${\rm perm}(\mathbf{k}_1,\ldots,\mathbf{k}_n)$ stands for the $(n!-1)$ terms to symmetrize the previous term $\delta_{\rm D}(\mathbf{k}_1-\mathbf{k}_1^{\prime})\cdots\delta_{\rm D}(\mathbf{k}_n-\mathbf{k}_m^{\prime})$ with respect to the permutations of its arguments $\mathbf{k}_1,\ldots,\mathbf{k}_n$. The Dirac delta function appears in Eq.~(\ref{eq:fourier_F_wiener_hermite_expansion}) is due to the (statistical) translational invariance of space. The expansion coefficient functions $\mathcal{G}_n(\mathbf{k}_1,\ldots,\mathbf{k}_n)$ can be derived based on the orthogonality relation in Eq.~(\ref{eq:Wiener-Hermite_orthogonality}). If we multiply $\mathcal{H}_m(\mathbf{k}_1,\ldots,\mathbf{k}_m)$ on both sides of Eq.~(\ref{eq:fourier_F_wiener_hermite_expansion}) and take their expectation value with respect to the Gaussian probability density functional, we would have 
\begin{align}
    \label{eq:G_equation_1}
    &(2\pi)^2\delta^2_\mathrm{D}(\mathbf{k}_1+\cdots+\mathbf{k}_n-\mathbf{k})\mathcal{G}_n(\mathbf{k}_1,\ldots,\mathbf{k}_n) \nonumber\\
    &= (2\pi)^{2n}\langle\tilde{\mathcal{F}}(\mathbf{k})\mathcal{H}_n(\mathbf{k}_1,\ldots,\mathbf{k}_n)\rangle_{\rm G} \nonumber\\
    &=(2\pi)^{2n}\left\langle\frac{\partial^n\tilde{\mathcal{F}}(\mathbf{k})}{\partial\tilde{f}(\mathbf{k}_1)\cdots\partial\tilde{f}(\mathbf{k}_n)}\right\rangle_{\rm G} \ ,
\end{align}
where the first equation is based on the orthogonal relation in Eq.~(\ref{eq:Wiener-Hermite_orthogonality}) and the second equation makes use of the definition of the Wiener-Hermite functional in Eq.~(\ref{eq:Wiener-Hermite_functionals}) followed by an integration by parts. To derive the final expression for $\mathcal{G}_n(\mathbf{k}_1,\ldots,\mathbf{k}_n)$, we apply Fourier transform to the above equation with respect to $\mathbf{k}$:
\begin{align}
    \label{eq:G_equation_2}
    &\mathcal{G}_n(\mathbf{k}_1,\ldots,\mathbf{k}_n) \nonumber\\
    &=(2\pi)^{2n}e^{i(\mathbf{k}_1+\cdots+\mathbf{k}_n)\cdot\mathbf{x}}\langle\mathcal{F}(\mathbf{x})\mathcal{H}_n(\mathbf{k}_1,\ldots,\mathbf{k}_n)\rangle_{\rm G} \nonumber\\
    &=(2\pi)^{2n}e^{i(\mathbf{k}_1+\cdots+\mathbf{k}_n)\cdot\mathbf{x}}\left\langle\frac{\partial^n\mathcal{F}(\mathbf{x})}{\partial\tilde{f}(\mathbf{k}_1)\cdots\partial\tilde{f}(\mathbf{k}_n)}\right\rangle_{\rm G} \ ,
\end{align}
where we can further set $\mathbf{x}=0$ due to the translational invariance of $\mathcal{G}_n(\mathbf{k}_1,\ldots,\mathbf{k}_n)$ and conveniently evaluate it to be
\begin{equation}
    \label{eq:G_equation_3}
    \mathcal{G}_n(\mathbf{k}_1,\ldots,\mathbf{k}_n)=(2\pi)^{2n}\left\langle\frac{\partial^n\mathcal{F}(\mathbf{x})}{\partial\tilde{f}(\mathbf{k}_1)\cdots\partial\tilde{f}(\mathbf{k}_n)}\right\rangle_{\rm G} \ .
\end{equation}
Following the above equation, the expansion coefficient functions can be interpreted as the Gaussian n-point response of the 2D functional $\mathcal{F}$ to the underlying density field. Conceptually this is analogous to the large-scale galaxy bias and therefore can be thought of in the same way for 2D critical points in this work.

With a proper understanding of Eq.~(\ref{eq:fourier_P_expansion}) and (\ref{eq:fourier_F_wiener_hermite_expansion}), we can substitute them with the corresponding terms in Eq.~(\ref{eq:F_2nd_moment_Fourier}) and expand the whole equation. During the process, one recurrent term is $\langle\mathcal{H}^{\star}_n(\mathbf{k}_1,\ldots,\mathbf{k}_n)\mathcal{H}^{\star}_m(\mathbf{k}_1^{\prime},\ldots,\mathbf{k}_m^{\prime})\mathcal{H}_l(\mathbf{k}_1^{''},\ldots,\mathbf{k}_l^{''})\rangle_{\rm G}$. One can compute these terms by solving both $\mathcal{H}_n$ and $\mathcal{H}_n^{\star}$ at each order explicitly using Eq.~(\ref{eq:Wiener-Hermite_functionals}
)and (\ref{eq:dual_Wiener-Hermite_functionals}). Here we give examples of the  first few $\mathcal{H}_n^{\star}$ expressions:
\begin{widetext}
\begin{eqnarray}
    \label{eq:H_star_example}
     \mathcal{H}_0^{\star}&&=1 \ , \nonumber\\
     \mathcal{H}_1^{\star}(\mathbf{k})&&=\tilde{f}(\mathbf{k}) \ ,\nonumber\\
    \mathcal{H}_2^{\star}(\mathbf{k}_1,\mathbf{k}_2)&&=\tilde{f}(\mathbf{k}_1)\tilde{f}(\mathbf{k}_2)-(2\pi)^2\delta_{\rm D}(\mathbf{k}_1+\mathbf{k}_2)P(k_1) \ , \nonumber\\
    \mathcal{H}_3^{\star}(\mathbf{k}_1,\mathbf{k}_2, \mathbf{k}_3)&&=\tilde{f}(\mathbf{k}_1)\tilde{f}(\mathbf{k}_2)\tilde{f}(\mathbf{k}_3)-\left[(2\pi)^2\delta_{\rm D}(\mathbf{k}_1+\mathbf{k}_2)P(k_1)\tilde{f}(\mathbf{k}_3) + {\rm sym}\right] \ , \nonumber\\
    \mathcal{H}_4^{\star}(\mathbf{k}_1,\ldots,\mathbf{k}_4)&&=\tilde{f}(\mathbf{k}_1)\tilde{f}(\mathbf{k}_2)\tilde{f}(\mathbf{k}_3)\tilde{f}(\mathbf{k}_4)+\left[(2\pi)^4\delta(\mathbf{k}_1+\mathbf{k}_2)\delta(\mathbf{k}_3+\mathbf{k}_4)P(k_1)P(k_3)+\mathrm{sym}\right] \nonumber\\
    && -\left[(2\pi)^2\delta(\mathbf{k}_1+\mathbf{k}_2)P(k_1)\tilde{f}(\mathbf{k}_3)\tilde{f}(\mathbf{k}_4) + \mathrm{sym}\right] \ ,
\end{eqnarray}
\end{widetext}
where ``sym" stands for all non-repeating symmetric expressions of the previous term with respect to the $\mathbf{k}_1,\ldots,\mathbf{k}_n$ arguments. Such terms can be inserted into $\langle\mathcal{H}^{\star}_n(\mathbf{k}_1,\ldots,\mathbf{k}_n)\mathcal{H}^{\star}_m(\mathbf{k}_1^{\prime},\ldots,\mathbf{k}_m^{\prime})\mathcal{H}_l(\mathbf{k}_1^{''},\ldots,\mathbf{k}_l^{''})\rangle_{\rm G}$ and the whole expression can be evaluated by applying Wick's theorem for Gaussian statistics. As a result, $\langle\mathcal{H}^{\star}_n\mathcal{H}^{\star}_m\mathcal{H}_l\rangle_{\rm G}$ has nonzero value only when $n+m+l$ is an even number. Though straightforward, this computation becomes tedious very quickly. For example, in the case of NNLO where $n+m+l=8$, Wick's theorem predicts 105 terms from the contraction of eight density field $\tilde{f}$ alone. It is more convenient to evaluate such expressions using the diagrammatic method. We show the formalism in Appendix.~\ref{appendix:B} where we also derive all existing $\langle\mathcal{H}^{\star}_n\mathcal{H}^{\star}_m\mathcal{H}_l\rangle_{\rm G}$ factors up to NNLO. Readers can also refer to Appendix.~A in Ref.~\cite{Matsubara1995} where the same formalism is presented but in real space.

With the above discussion, we can now show the resulting equation of $\langle\tilde{\mathcal{F}}(\mathbf{k})\tilde{\mathcal{F}}(\mathbf{k}^{\prime})\rangle$ in Eq.~(\ref{eq:F_2nd_moment_Fourier}) up to NNLO
\begin{widetext}
\begin{eqnarray}
    \label{eq:F_2nd_moment_Fourier_nnlo}
    && \langle\tilde{\mathcal{F}}^i(\mathbf{k})\tilde{\mathcal{F}}^j(\mathbf{k}^{\prime})\rangle = (2\pi)^4\mathcal{G}^i_0\mathcal{G}^j_0 + (2\pi)^2\delta_\mathrm{D}^2(\mathbf{k} + \mathbf{k}^{\prime})\mathcal{G}^i_1(\mathbf{k})\mathcal{G}^j_1(\mathbf{k})P(k)\nonumber\\
    && + \frac{(2\pi)^2}{2}\delta_\mathrm{D}^2(\mathbf{k} + \mathbf{k}^{\prime})\int\frac{\mathrm{d}^2k_1}{(2\pi)^2}\mathcal{G}^i_2(\mathbf{k}_1, \mathbf{k}-\mathbf{k}_1)\mathcal{G}^j_2(\mathbf{k}_1, \mathbf{k}-\mathbf{k}_1)P(k_1)P(|\mathbf{k}-\mathbf{k}_1|) \nonumber\\
    && + \frac{1}{6}\left[(2\pi)^2\mathcal{G}^i_0\int\frac{\mathrm{d}^2k_1}{(2\pi)^2}\cdots\int\frac{\mathrm{d}^2k_3}{(2\pi)^2}(2\pi)^2\delta_\mathrm{D}^2(\mathbf{k}_1+\mathbf{k}_2+\mathbf{k}_3-\mathbf{k})\mathcal{G}^j_3(\mathbf{k}_1, \mathbf{k}_2, \mathbf{k}_3)\langle\tilde{f}(\mathbf{k}_1)\tilde{f}(\mathbf{k}_2)\tilde{f}(\mathbf{k}_3)\rangle_\mathrm{c} + (i\leftrightarrow j) \right]\nonumber\\
    && + \frac{(2\pi)^2}{2}\delta_\mathrm{D}^2(\mathbf{k} + \mathbf{k}^{\prime})\left[\mathcal{G}^i_1(\mathbf{k})\int\frac{\mathrm{d}^2k_1}{(2\pi)^2}\mathcal{G}^j_2(\mathbf{k}_1, \mathbf{k}-\mathbf{k}_1)B(-\mathbf{k},\mathbf{k}_1,\mathbf{k}-\mathbf{k}_1) + (i\leftrightarrow j)\right] \nonumber\\
    && + \frac{(2\pi)^2}{6}\delta_\mathrm{D}^2(\mathbf{k} + \mathbf{k}^{\prime})\int\frac{\mathrm{d}^2k_1}{(2\pi)^2}\int\frac{\mathrm{d}^2k_2}{(2\pi)^2}\mathcal{G}_3^i(\mathbf{k}_1, \mathbf{k}_2, \mathbf{k}-\mathbf{k}_1-\mathbf{k}_2)\mathcal{G}_3^j(\mathbf{k}_1, \mathbf{k}_2, \mathbf{k}-\mathbf{k}_1-\mathbf{k}_2)P(k_1)P(k_2)P(|\mathbf{k}-\mathbf{k}_1-\mathbf{k}_2|) \nonumber\\
    && + \frac{(2\pi)^2}{6}\delta_\mathrm{D}^2(\mathbf{k} + \mathbf{k}^{\prime})\left[\mathcal{G}^i_1(\mathbf{k})P(k)\int\frac{\mathrm{d}^2k_1}{(2\pi)^2}\int\frac{\mathrm{d}^2k_2}{(2\pi)^2}\mathcal{G}^j_4(\mathbf{k},-\mathbf{k}_1, -\mathbf{k}_2, \mathbf{k}_1+\mathbf{k}_2)B(-\mathbf{k}_1, -\mathbf{k}_2, \mathbf{k}_1+\mathbf{k}_2)+(i\leftrightarrow j)\right] \nonumber\\
    && + \frac{(2\pi)^2}{2}\delta_\mathrm{D}^2(\mathbf{k} + \mathbf{k}^{\prime})\left[\int\frac{\mathrm{d}^2k_1}{(2\pi)^2}\int\frac{\mathrm{d}^2k_2}{(2\pi)^2}\mathcal{G}^i_2(\mathbf{k}_1,\mathbf{k}-\mathbf{k}_1)\mathcal{G}^j_3(\mathbf{k}-\mathbf{k}_1,-\mathbf{k}_2,\mathbf{k}_1+\mathbf{k}_2)P(|\mathbf{k}-\mathbf{k}_1|)B(\mathbf{k}_1,\mathbf{k}_2,-\mathbf{k}_1-\mathbf{k}_2)+(i\leftrightarrow j)\right] \nonumber\\
    &&+\frac{(2\pi)^4}{24}\left[\mathcal{G}^i_0\int\frac{\mathrm{d}^2k_1}{(2\pi)^2}\cdots\int\frac{\mathrm{d}^2k_4}{(2\pi)^2}\delta_\mathrm{D}^2(\mathbf{k}_1+\mathbf{k}_2+\mathbf{k}_3+\mathbf{k}_4-\mathbf{k})\mathcal{G}^j_4(\mathbf{k}_1,\mathbf{k}_2,\mathbf{k}_3,\mathbf{k}_4)\langle\tilde{f}(\mathbf{k}_1)\tilde{f}(\mathbf{k}_2)\tilde{f}(\mathbf{k}_3)\tilde{f}(\mathbf{k}_4)\rangle_\mathrm{c}+(i\leftrightarrow j)\right] \nonumber\\
    &&+\frac{(2\pi)^2}{6}\delta_\mathrm{D}^2(\mathbf{k} + \mathbf{k}^{\prime})\left[\mathcal{G}^i_1(\mathbf{k})\int\frac{\mathrm{d}^2k_1}{(2\pi)^2}\int\frac{\mathrm{d}^2k_2}{(2\pi)^2}\mathcal{G}^j_3(-\mathbf{k}_1,-\mathbf{k}_2,\mathbf{k}+\mathbf{k}_1+\mathbf{k}_2)T(-\mathbf{k}, -\mathbf{k}_1,-\mathbf{k}_2,\mathbf{k}+\mathbf{k}_1+\mathbf{k}_2)+(i\leftrightarrow j)\right] \nonumber\\
    &&+\frac{(2\pi)^2}{4}\delta_\mathrm{D}^2(\mathbf{k} + \mathbf{k}^{\prime})\left[\int\frac{\mathrm{d}^2k_1}{(2\pi)^2}\int\frac{\mathrm{d}^2k_2}{(2\pi)^2}\mathcal{G}^i_2(\mathbf{k}-\mathbf{k}_1,\mathbf{k}_1)\mathcal{G}^j_2(\mathbf{k}-\mathbf{k}_2, \mathbf{k}_2)T(\mathbf{k}-\mathbf{k}_1, \mathbf{k}_1,\mathbf{k}_2,-\mathbf{k}-\mathbf{k}_2)+(i\leftrightarrow j)\right] \ ,
\end{eqnarray}
\end{widetext}
where we exploited the following parity symmetries: 
\begin{align}
    \label{eq:parity_symmetry}
    \mathcal{G}_n(-\mathbf{k}_1,\ldots,-\mathbf{k}_n)&=\mathcal{G}_n(\mathbf{k}_1,\ldots,\mathbf{k}_n) \nonumber\\
    B(-\mathbf{k}_1,-\mathbf{k}_2,-\mathbf{k}_3)&=B(\mathbf{k}_1,\mathbf{k}_2,\mathbf{k}_3) \nonumber\\
    T(-\mathbf{k}_1,\ldots,-\mathbf{k}_4)&=T(\mathbf{k}_1,\ldots,\mathbf{k}_4) \ .
\end{align}
There are two terms in the above equation which contain $\langle\tilde{f}(\mathbf{k}_1)\tilde{f}(\mathbf{k}_2)\tilde{f}(\mathbf{k}_3)\rangle_\mathrm{c}$ and $\langle\tilde{f}(\mathbf{k}_1)\tilde{f}(\mathbf{k}_2)\tilde{f}(\mathbf{k}_3)\tilde{f}(\mathbf{k}_4)\rangle_\mathrm{c}$ respectively. We use such notations to distinguish them from the rest as these two terms represent the unconnected parts in the 2nd-order moment of $\tilde{\mathcal{F}}$ and will be subtracted off as we will show below. Additionally, in Eq.~(\ref{eq:F_2nd_moment_Fourier_nnlo}) we characterize the functional $\tilde{\mathcal{F}}$ with indices $i$, $j$ which denote different types of critical points (i.e., $i$,$j=$ peaks, voids, saddle points) through which we can construct both auto- and cross-2PCFs. The symbol $(i\leftrightarrow j)$ denotes the addition of a term possessing the same form but exchanging the correspondence of the  expansion coefficient function $\mathcal{G}_n$ to the other functional type. From Eq.~(\ref{eq:F_2nd_moment_Fourier_nnlo}) we observe that the leading-order (LO) result is composed of the power spectrum $P(k)$ of the underlying density field. The NLO result is proportional to $P(k)^2$ including the bispectrum contribution (at tree-level) which is also the lowest-order non-Gaussian correction. Finally the NNLO result is proportional to $P(k)^3$ which includes contribution from $P\times B$ and $T$ terms where both bispectrum and trispectrum are at tree-level. 

To compute the connected part of the 2nd-order moment, we need $\langle\tilde{\mathcal{F}}\rangle$ which can be calculated based on Eqs.~(\ref{eq:fourier_P_expansion}), (\ref{eq:fourier_F_wiener_hermite_expansion}) and (\ref{eq:Wiener-Hermite_orthogonality})
\begin{widetext}
\begin{eqnarray}
\label{eq:fourier_F_expectation}
    &&\langle\tilde{\mathcal{F}}\rangle = (2\pi)^2\mathcal{G}_0 + \frac{1}{6}\int\frac{\mathrm{d}^2k_1}{(2\pi)^2}\frac{\mathrm{d}^2k_2}{(2\pi)^2}\frac{\mathrm{d}^2k_3}{(2\pi)^2}(2\pi)^2\delta^2_{\mathrm{D}}(\mathbf{k}_1+\mathbf{k}_2+\mathbf{k}_3-\mathbf{k})\langle\tilde{f}(\mathbf{k}_1)\tilde{f}(\mathbf{k}_2)\tilde{f}(\mathbf{k}_3)\rangle_\mathrm{c}\mathcal{G}_3(\mathbf{k}_1,\mathbf{k}_2,\mathbf{k}_3) \nonumber\\
    &&+ \frac{1}{24}\int\frac{\mathrm{d}^2k_1}{(2\pi)^2}\cdots\frac{\mathrm{d}^2k_4}{(2\pi)^2}(2\pi)^2\delta^2_{\mathrm{D}}(\mathbf{k}_1+\cdots+\mathbf{k}_4-\mathbf{k})\langle\tilde{f}(\mathbf{k}_1)\cdots\tilde{f}(\mathbf{k}_4)\rangle_\mathrm{c}\mathcal{G}_4(\mathbf{k}_1, \mathbf{k}_2, \mathbf{k}_3, \mathbf{k}_4) \nonumber\\
    && + \frac{1}{72}\int\frac{\mathrm{d}^2k_1}{(2\pi)^2}\cdots\frac{\mathrm{d}^2k_6}{(2\pi)^2}(2\pi)^2\delta^2_{\mathrm{D}}(\mathbf{k}_1+\cdots+\mathbf{k}_6-\mathbf{k})\langle\tilde{f}(\mathbf{k}_1)\tilde{f}(\mathbf{k}_2)\tilde{f}(\mathbf{k}_3)\rangle_\mathrm{c}\langle\tilde{f}(\mathbf{k}_4)\tilde{f}(\mathbf{k}_5)\tilde{f}(\mathbf{k}_6)\rangle_\mathrm{c}\mathcal{G}_6(\mathbf{k}_1, \ldots, \mathbf{k}_6) \nonumber\\
    &&+ \cdots \ ,
\end{eqnarray}
\end{widetext}
where we can easily observe that the first three lowest-order terms in the expansion of $\langle\tilde{\mathcal{F}}^i\rangle\langle\tilde{\mathcal{F}}^j\rangle$ are the constant $(2\pi)^4\mathcal{G}^i_0\mathcal{G}^j_0$ and the two terms we mentioned above in Eq.~(\ref{eq:F_2nd_moment_Fourier_nnlo}). By subtracting them from Eq.~(\ref{eq:F_2nd_moment_Fourier_nnlo}), We can summarize the connected 2nd-order moment of the functional $\tilde{\mathcal{F}}$ to be
\begin{widetext}
\begin{eqnarray}
    \label{eq:F_connected_2nd_moment_Fourier_nnlo}
    && \langle\tilde{\mathcal{F}}^i(\mathbf{k})\tilde{\mathcal{F}}^j(\mathbf{k}^{\prime})\rangle_{\rm c} = (2\pi)^2\delta_\mathrm{D}^2(\mathbf{k} + \mathbf{k}^{\prime})\mathcal{G}^i_1(\mathbf{k})\mathcal{G}^j_1(\mathbf{k})P(k)\nonumber\\
    && + \frac{(2\pi)^2}{2}\delta_\mathrm{D}^2(\mathbf{k} + \mathbf{k}^{\prime})\int\frac{\mathrm{d}^2k_1}{(2\pi)^2}\mathcal{G}^i_2(\mathbf{k}_1, \mathbf{k}-\mathbf{k}_1)\mathcal{G}^j_2(\mathbf{k}_1, \mathbf{k}-\mathbf{k}_1)P(k_1)P(|\mathbf{k}-\mathbf{k}_1|) \nonumber\\
    && + \frac{(2\pi)^2}{2}\delta_\mathrm{D}^2(\mathbf{k} + \mathbf{k}^{\prime})\left[\mathcal{G}^i_1(\mathbf{k})\int\frac{\mathrm{d}^2k_1}{(2\pi)^2}\mathcal{G}^j_2(\mathbf{k}_1, \mathbf{k}-\mathbf{k}_1)B(-\mathbf{k},\mathbf{k}_1,\mathbf{k}-\mathbf{k}_1) + (i\leftrightarrow j)\right] \nonumber\\
    && + \frac{(2\pi)^2}{6}\delta_\mathrm{D}^2(\mathbf{k} + \mathbf{k}^{\prime})\int\frac{\mathrm{d}^2k_1}{(2\pi)^2}\int\frac{\mathrm{d}^2k_2}{(2\pi)^2}\mathcal{G}_3^i(\mathbf{k}_1, \mathbf{k}_2, \mathbf{k}-\mathbf{k}_1-\mathbf{k}_2)\mathcal{G}_3^j(\mathbf{k}_1, \mathbf{k}_2, \mathbf{k}-\mathbf{k}_1-\mathbf{k}_2)P(k_1)P(k_2)P(|\mathbf{k}-\mathbf{k}_1-\mathbf{k}_2|) \nonumber\\
    && + \frac{(2\pi)^2}{6}\delta_\mathrm{D}^2(\mathbf{k} + \mathbf{k}^{\prime})\left[\mathcal{G}^i_1(\mathbf{k})P(k)\int\frac{\mathrm{d}^2k_1}{(2\pi)^2}\int\frac{\mathrm{d}^2k_2}{(2\pi)^2}\mathcal{G}^j_4(\mathbf{k},-\mathbf{k}_1, -\mathbf{k}_2, \mathbf{k}_1+\mathbf{k}_2)B(-\mathbf{k}_1, -\mathbf{k}_2, \mathbf{k}_1+\mathbf{k}_2)+(i\leftrightarrow j)\right] \nonumber\\
    && + \frac{(2\pi)^2}{2}\delta_\mathrm{D}^2(\mathbf{k} + \mathbf{k}^{\prime})\left[\int\frac{\mathrm{d}^2k_1}{(2\pi)^2}\int\frac{\mathrm{d}^2k_2}{(2\pi)^2}\mathcal{G}^i_2(\mathbf{k}_1,\mathbf{k}-\mathbf{k}_1)\mathcal{G}^j_3(\mathbf{k}-\mathbf{k}_1,-\mathbf{k}_2,\mathbf{k}_1+\mathbf{k}_2)P(|\mathbf{k}-\mathbf{k}_1|)B(\mathbf{k}_1,\mathbf{k}_2,-\mathbf{k}_1-\mathbf{k}_2)+(i\leftrightarrow j)\right] \nonumber\\
    &&+\frac{(2\pi)^2}{6}\delta_\mathrm{D}^2(\mathbf{k} + \mathbf{k}^{\prime})\left[\mathcal{G}^i_1(\mathbf{k})\int\frac{\mathrm{d}^2k_1}{(2\pi)^2}\int\frac{\mathrm{d}^2k_2}{(2\pi)^2}\mathcal{G}^j_3(-\mathbf{k}_1,-\mathbf{k}_2,\mathbf{k}+\mathbf{k}_1+\mathbf{k}_2)T(-\mathbf{k}, -\mathbf{k}_1,-\mathbf{k}_2,\mathbf{k}+\mathbf{k}_1+\mathbf{k}_2)+(i\leftrightarrow j)\right] \nonumber\\
    &&+\frac{(2\pi)^2}{4}\delta_\mathrm{D}^2(\mathbf{k} + \mathbf{k}^{\prime})\left[\int\frac{\mathrm{d}^2k_1}{(2\pi)^2}\int\frac{\mathrm{d}^2k_2}{(2\pi)^2}\mathcal{G}^i_2(\mathbf{k}-\mathbf{k}_1,\mathbf{k}_1)\mathcal{G}^j_2(\mathbf{k}-\mathbf{k}_2, \mathbf{k}_2)T(\mathbf{k}-\mathbf{k}_1, \mathbf{k}_1,\mathbf{k}_2,-\mathbf{k}-\mathbf{k}_2)+(i\leftrightarrow j)\right] \ .
\end{eqnarray}
\end{widetext}
The last piece of element needed to complete the derivation of $P_{\mathcal{F}}(k)$ is $\langle\mathcal{F}\rangle$ which can be calculated based on Eqs.~(\ref{eq:fourier_P_expansion}) and (\ref{eq:G_equation_2})
\begin{widetext}
\begin{eqnarray}
\label{eq:real_F_expectation}
    \langle\mathcal{F}\rangle =&& \mathcal{G}_0 + \frac{1}{6}\int\frac{\mathrm{d}^2k_1}{(2\pi)^2}\frac{\mathrm{d}^2k_2}{(2\pi)^2}\frac{\mathrm{d}^2k_3}{(2\pi)^2}\langle\tilde{f}(\mathbf{k}_1)\tilde{f}(\mathbf{k}_2)\tilde{f}(\mathbf{k}_3)\rangle_\mathrm{c}\mathcal{G}_3(\mathbf{k}_1,\mathbf{k}_2,\mathbf{k}_3) \nonumber\\
    &&+ \frac{1}{24}\int\frac{\mathrm{d}^2k_1}{(2\pi)^2}\ldots\frac{\mathrm{d}^2k_4}{(2\pi)^2}\langle\tilde{f}(\mathbf{k}_1)\ldots\tilde{f}(\mathbf{k}_4)\rangle_\mathrm{c}\mathcal{G}_4(\mathbf{k}_1, \mathbf{k}_2, \mathbf{k}_3, \mathbf{k}_4) \nonumber\\
    && + \frac{1}{72}\int\frac{\mathrm{d}^2k_1}{(2\pi)^2}\ldots\frac{\mathrm{d}^2k_6}{(2\pi)^2}\langle\tilde{f}(\mathbf{k}_1)\tilde{f}(\mathbf{k}_2)\tilde{f}(\mathbf{k}_3)\rangle_\mathrm{c}\langle\tilde{f}(\mathbf{k}_4)\tilde{f}(\mathbf{k}_5)\tilde{f}(\mathbf{k}_6)\rangle_\mathrm{c}\mathcal{G}_6(\mathbf{k}_1, \ldots, \mathbf{k}_6) + \ldots \ .
\end{eqnarray}
\end{widetext}
By combining Eqs.~(\ref{eq:F_connected_2nd_moment_Fourier_nnlo}) and (\ref{eq:real_F_expectation}), we can compute $P_{\mathcal{F}}(k)$ in Eq.~(\ref{eq:F_power_spectrum}). One notice is that the non-Gaussian corrections of Eq.~(\ref{eq:real_F_expectation}) in the denominator of Eq.~(\ref{eq:F_power_spectrum}) would not contribute to the NLO but they contribute to higher-order results in general. With this in mind, we can derive the expression for the power spectrum up to NNLO as
\begin{widetext}
\begin{eqnarray}
    \label{eq:F_power_spectrum_nnlo}
    P_{\mathcal{F}}^{ij}(k)&&=g_1^i(\mathbf{k})g_1^j(\mathbf{k})P(k) \nonumber\\
    && +\frac{1}{2}\int\frac{\mathrm{d}^2k_1}{(2\pi)^2}g^i_2(\mathbf{k}_1, \mathbf{k}-\mathbf{k}_1)g^j_2(\mathbf{k}_1, \mathbf{k}-\mathbf{k}_1)P(k_1)P(|\mathbf{k}-\mathbf{k}_1|) \nonumber\\
    && + \frac{1}{2}\left[g^i_1(\mathbf{k})\int\frac{\mathrm{d}^2k_1}{(2\pi)^2}g^j_2(\mathbf{k}_1, \mathbf{k}-\mathbf{k}_1)B(-\mathbf{k},\mathbf{k}_1,\mathbf{k}-\mathbf{k}_1) + (i\leftrightarrow j)\right] \nonumber\\
    && + \frac{1}{6}\int\frac{\mathrm{d}^2k_1}{(2\pi)^2}\int\frac{\mathrm{d}^2k_2}{(2\pi)^2}g_3^i(\mathbf{k}_1, \mathbf{k}_2, \mathbf{k}-\mathbf{k}_1-\mathbf{k}_2)g_3^j(\mathbf{k}_1, \mathbf{k}_2, \mathbf{k}-\mathbf{k}_1-\mathbf{k}_2)P(k_1)P(k_2)P(|\mathbf{k}-\mathbf{k}_1-\mathbf{k}_2|) \nonumber\\
    && + \frac{1}{6}\left[g^i_1(\mathbf{k})P(k)\int\frac{\mathrm{d}^2k_1}{(2\pi)^2}\int\frac{\mathrm{d}^2k_2}{(2\pi)^2}g^j_4(\mathbf{k},-\mathbf{k}_1, -\mathbf{k}_2, \mathbf{k}_1+\mathbf{k}_2)B(-\mathbf{k}_1, -\mathbf{k}_2, \mathbf{k}_1+\mathbf{k}_2)+(i\leftrightarrow j)\right] \nonumber\\
    && + \frac{1}{2}\left[\int\frac{\mathrm{d}^2k_1}{(2\pi)^2}\int\frac{\mathrm{d}^2k_2}{(2\pi)^2}g^i_2(\mathbf{k}_1,\mathbf{k}-\mathbf{k}_1)g^j_3(\mathbf{k}-\mathbf{k}_1,-\mathbf{k}_2,\mathbf{k}_1+\mathbf{k}_2)P(|\mathbf{k}-\mathbf{k}_1|)B(\mathbf{k}_1,\mathbf{k}_2,-\mathbf{k}_1-\mathbf{k}_2)+(i\leftrightarrow j)\right] \nonumber\\
    &&+\frac{1}{6}\left[g^i_1(\mathbf{k})\int\frac{\mathrm{d}^2k_1}{(2\pi)^2}\int\frac{\mathrm{d}^2k_2}{(2\pi)^2}g^j_3(-\mathbf{k}_1,-\mathbf{k}_2,\mathbf{k}+\mathbf{k}_1+\mathbf{k}_2)T(-\mathbf{k}, -\mathbf{k}_1,-\mathbf{k}_2,\mathbf{k}+\mathbf{k}_1+\mathbf{k}_2)+(i\leftrightarrow j)\right] \nonumber\\
    &&+\frac{1}{4}\left[\int\frac{\mathrm{d}^2k_1}{(2\pi)^2}\int\frac{\mathrm{d}^2k_2}{(2\pi)^2}g^i_2(\mathbf{k}-\mathbf{k}_1,\mathbf{k}_1)g^j_2(\mathbf{k}-\mathbf{k}_2, \mathbf{k}_2)T(\mathbf{k}-\mathbf{k}_1, \mathbf{k}_1,\mathbf{k}_2,-\mathbf{k}-\mathbf{k}_2)+(i\leftrightarrow j)\right] \nonumber\\
    &&-\frac{1}{6}\left[g_1^i(\mathbf{k})g_1^j(\mathbf{k})P(k)\int\frac{\mathrm{d}^2k_1}{(2\pi)^2}\int\frac{\mathrm{d}^2k_2}{(2\pi)^2}g^j_3(\mathbf{k}_1,\mathbf{k}_2,-\mathbf{k}_1-\mathbf{k}_2)B(\mathbf{k}_1,\mathbf{k}_2,-\mathbf{k}_1-\mathbf{k}_2)+(i\leftrightarrow j)\right] \ ,
\end{eqnarray}
\end{widetext}
where 
\begin{equation}
    \label{eq:g_function}
    g_n(\mathbf{k}_1,\ldots,\mathbf{k}_n)\equiv\frac{\mathcal{G}_n(\mathbf{k}_1,\ldots,\mathbf{k}_n)}{\mathcal{G}_0} \ ,
\end{equation}
and the last line is the contribution from non-Gaussian corrections in the denominator of Eq.~(\ref{eq:F_power_spectrum}). One can then apply a Hankel transformation to Eq.~(\ref{eq:F_power_spectrum_nnlo}) to evaluate the correlation function
\begin{equation}
    \label{eq:correlation_function}
    \xi_{\mathcal{F}}^{ij}(r)=\int\frac{k\mathrm{d}k}{2\pi}J_0(kr)P_{\mathcal{F}}^{ij}(k) \ ,
\end{equation}
where $J_n(x)$ is the Bessel function. Note that in the above Eq.~(\ref{eq:F_power_spectrum_nnlo}), if the underlying density field is Gaussian distributed, all terms containing bispectrum or trispectrum would vanish and we will recover the perturbative bias expansion in the context of Gaussian approximation.

So far, all the above results are general for any 2D functionals $\mathcal{F}$ of a density field $f$. We still need to answer one question before we can actually compute the 2PCFs for 2D critical points in a mildly non-Gaussian regime using Eqs.~(\ref{eq:F_power_spectrum_nnlo}) and (\ref{eq:correlation_function}), that is what are the expressions for $g_n(\mathbf{k}_1,\ldots,\mathbf{k}_n)$  for 2D critical points. From the previous Eq.~(\ref{eq:G_equation_3}), we know that they can be derived directly by substituting $\mathcal{F}$ in the equation with the corresponding number density function of a critical point type, such as the one for peaks in Eq.~(\ref{eq:n_peak}). The required functional derivatives contain very technical calculation and for our purpose of numerically evaluating Eq.~(\ref{eq:F_power_spectrum_nnlo}) to the NLO, we directly present the results below. Readers who are interested in the derivation details can refer to the method presented in Appendix.~B of Ref.~\cite{matsubara2020}.
\begin{equation}
    \label{eq:g1}
    g_1(\mathbf{k})=g_{10000}+g_{01000}k^2 \ ,
\end{equation}
\begin{align}
    \label{eq:g2}
    g_2(\mathbf{k}_1,\mathbf{k}_2)=&g_{20000}+g_{11000}(k_1^2+k_2^2)+g_{02000}k_1^2k_2^2\nonumber\\
    &-2g_{00100}\mathbf{k}_1\cdot\mathbf{k}_2\nonumber\\
    &+4g_{00010}\left[(\mathbf{k}_1\cdot\mathbf{k}_2)^2-\frac{1}{2}k_1^2k_2^2\right] \ ,
\end{align}
where the coefficients $g_{ijklm}$ include the constraints imposed on the density field by the critical point functional and can be expressed as
\begin{equation}
    \label{eq:g_ijklm_1}
    g_{ijklm} = \frac{G_{ijklm}}{\sigma_0^i\sigma_1^{2k}\sigma_2^{j+2l+3m}G_{00000}} \ ,
\end{equation}
and the numerator factor $G_{ijklm}$ is
\begin{align}
    \label{eq:G_ijklm}
    G_{ijklm}(\nu)=\frac{1}{2\pi}\left(\frac{\sigma_2}{\sqrt{2}\sigma_1}\right)^2X_k&\int{\rm d}x\,H_{i-1,j}(\nu, x)\nonumber\\
    &\times\mathcal{N}(\nu,x)f_{lm}(x) \ ,
\end{align}
which is a function of the threshold $\nu$ as discussed in Eq.~(\ref{eq:n_peak}). In the integration above, $x$ is exactly the trace of the negative Hessian matrix $(-\zeta)$ and equivalent to $J_1$ defined in Eq.~(\ref{eq:rotationally_invariant_variable}). $X_k$ is a constant and from Eqs.~(\ref{eq:g1}) and (\ref{eq:g2}) we only need $X_0$ and $X_1$ which are $1$ and $-1$ respectively. The function $H_{ij}(\alpha, J_1)$ is the multivariate Hermite polynomials defined as
\begin{equation}
    \label{eq:multivariate_hermite}
    H_{ij}(\alpha, J_1)=\frac{1}{\mathcal{N}(\alpha,J_1)}\left(-\frac{\partial}{\partial\alpha}\right)^i\left(-\frac{\partial}{\partial J_1}\right)^j\mathcal{N}(\alpha, J_1) \ ,
\end{equation}
where the $\mathcal{N}(\alpha,J_1)$ function is defined previously in Eq.~(\ref{eq:N_function}). In the case of $g_{0jklm}$, we need to calculate $H_{-1,j}$ which is
\begin{equation}
    \label{eq:multivariate_hermite_minus1}
    H_{-1,j}(\alpha, J_1)=\frac{1}{\mathcal{N}(\alpha,J_1)}\int_{\alpha}^{\infty}{\rm d}\beta H_{0j}(\beta, J_1)\mathcal{N}(\beta,J_1) \ .
\end{equation}
Again from Eqs.~(\ref{eq:g1}) and (\ref{eq:g2}) we observe that we only need functions $f_{l0}(x)$ which are (for the general definition equation of $f_{lm}$, please refer to Ref.~\cite{matsubara2020})
\begin{equation}
    \label{eq:f_lm}
    f_{l0}=8\int{\rm d}yye^{-4y^2}(x^2-4y^2)(-1)^l L_l(4y^2) \ ,
\end{equation}
where $y$ is defined as $y\equiv (\lambda_1-\lambda_2)/2$ and $L_l$ is the generalized Laguerre polynomial $L_l^{(n)}$ with the index $n=1$
\begin{equation}
    \label{eq:generalized_Laguerre_polynomial}
    L_l(x) = \frac{e^x}{xl!}\frac{{\rm d}^l}{{\rm d}x^l}(x^{l+1}e^{-x}) \ .
\end{equation}
By replacing $G_{ijklm}$ factors in Eq.~(\ref{eq:g_ijklm_1}) with Eq.~(\ref{eq:G_ijklm}), we can have the expression
\begin{equation}
    \label{eq:g_ijklm_2}
    g_{ijklm}(\nu)=\frac{X_k\int{\rm d}xH_{i-1,j}(\nu, x)\mathcal{N}(\nu,x)f_{lm}(x)}{\sigma_0^i\sigma_1^{2k}\sigma_2^{j+2l+3m}\int{\rm d}xH_{-1,0}(\nu, x)\mathcal{N}(\nu,x)f_{00}(x)} \ ,
\end{equation}
which can be evaluated by Eq.~(\ref{eq:N_function}) and Eqs.~(\ref{eq:multivariate_hermite}) to (\ref{eq:generalized_Laguerre_polynomial}).

As discussed in Sec.~\ref{sec:2d_critical_points}, different critical points are characterized by their eigenvalues of the Hessian matrix. This characterization is reflected in the integration limits of Eq.~(\ref{eq:g_ijklm_2}) through which we can then compute the $g_{ijklm}$ factors for peaks, voids and saddle points separately \citep{Gay2012}. As discussed above, the integration variable $x=J_1=\lambda_1+\lambda_2$, and ranges from 0 to $\infty$ for peaks, $-\infty$ to 0 for voids and $-\infty$ to $\infty$ for saddle points. Another integration limit needs considering is in Eq.~(\ref{eq:f_lm}), where $y$ is strictly positive as we have already assumed $\lambda_1 > \lambda_2$. Furthermore we have $y^2=\left[(\lambda_1+\lambda_2)^2-4\lambda_1\lambda_2\right]/4=(x^2-4\lambda_1\lambda_2)/4$ where both peaks and voids have $\lambda_1\lambda_2>0$, therefore we have $y<x/2$ $(x>0)$ for peaks and $y<-x/2$ $(x<0)$ for voids. On the other hand, saddle points always have $\lambda_1\lambda_2<0$, thus the integration limit in Eq.~(\ref{eq:f_lm}) would become $y>|x|/2$. To summarize, we have the following equations for $g_{ijklm}$ factors for different types of critical points
\begin{equation}
    \label{eq:g_ijklm_peaks}
    g^{peak}_{ijklm} = \frac{X_k\int_0^{\infty}\mathrm{d}xH_{i-1,j}(\nu, x)\mathcal{N}(\nu, x)f_{lm}(x)}{\sigma_0^i\sigma_1^{2k}\sigma_2^{j+2l+3m}\int_0^{\infty}\mathrm{d}xH_{-1,j}(\nu, x)\mathcal{N}(\nu, x)f_{00}(x)} \ ,
\end{equation}
\begin{equation}
    \label{eq:f_peaks}
    f^{peak}_{lm}=f_{l0}^{peak}=8\int_0^{\frac{x}{2}}\mathrm{d}y ye^{-4y^2}(x^2-4y^2)(-1)^lL_l(4y^2) \ ,
\end{equation}
\begin{equation}
    \label{eq:g_factors_voids}
    g^{void}_{ijklm} = \frac{X_k\int_{-\infty}^0\mathrm{d}xH_{i-1,j}(\nu, x)\mathcal{N}(\nu, x)f_{lm}(x)}{\sigma_0^i\sigma_1^{2k}\sigma_2^{j+2l+3m}\int_{-\infty}^0\mathrm{d}xH_{-1,j}(\nu, x)\mathcal{N}(\nu, x)f_{00}(x)} \ ,
\end{equation}
\begin{equation}
    \label{eq:f_voids}
    f^{void}_{lm}=f_{l0}^{void}=8\int_0^{-\frac{x}{2}}\mathrm{d}y ye^{-4y^2}(x^2-4y^2)(-1)^lL_l(4y^2) \ ,
\end{equation}
and the analytical integration results of the $f_{lm}$ function are the same for both peaks and voids
\begin{eqnarray}
    \label{eq:f_00_10_peak_void}
    &&f_{00}^{peak/void}(x)=e^{-x^2}+x^2-1 \nonumber\\
    &&f_{10}^{peak/void}(x)=(1+x^2)e^{-x^2}-1 \ ,
\end{eqnarray}
whereas
\begin{equation}
    \label{eq:g_factors_saddle}
    g^{saddle}_{ijklm} = \frac{X_k\int_{-\infty}^{\infty}\mathrm{d}xH_{i-1,j}(\nu, x)\mathcal{N}(\nu, x)f_{lm}(x)}{\sigma_0^i\sigma_1^{2k}\sigma_2^{j+2l+3m}\int_{-\infty}^{\infty}\mathrm{d}xH_{-1,j}(\nu, x)\mathcal{N}(\nu, x)f_{00}(x)} \ ,
\end{equation}
\begin{equation}
    \label{eq:f_saddle}
    f^{saddle}_{lm}=f_{l0}^{saddle}=8\int_{\frac{|x|}{2}}^{\infty}\mathrm{d}y ye^{-4y^2}(x^2-4y^2)(-1)^lL_l(4y^2) \ ,
\end{equation}
and
\begin{eqnarray}
    \label{eq:f_00_10_saddle}
    &&f^{saddle}_{00} = -e^{-x^2} \nonumber\\
    &&f^{saddle}_{10} = -e^{-x^2}(1+x^2) \ .
\end{eqnarray}
In Appendix~\ref{appendix:C}, we display plots of the seven $g_{ijklm}$ factors in Eqs.~(\ref{eq:g1}) and (\ref{eq:g2}) as functions of the threshold $\nu$ for different types of critical points.

\section{Results for the 2PCFs of 2D weak lensing critical points}
\label{sec:2d_weak_lensing}
Building on the formalism for calculating the 2PCFs of 2D critical points in a mildly non-Gaussian regime introduced in the previous section, we apply it to the weak lensing convergence field $\kappa$. This particular 2D field can be understood as the weighted line-of-sight projection of the 3D cosmic matter density contrast field \citep{Bartelmann2001, Schneider2006}
\begin{equation}
    \label{eq:kappa_field}
    \kappa=\int{\rm d}\chi q(\chi)\delta(\chi) \ ,
\end{equation}
where $\delta(\chi)$ is the 3D matter density contrast at a comoving radial distance $\chi$ and $q(\chi)$ is the weight function of the convergence field along the line-of-sight \citep{Kilbinger2015}
\begin{equation}
    \label{eq:lensing_kernel}
    q(\chi)\equiv \frac{3H_0^2\Omega_{m,0}}{2a(\chi)c^2}\frac{\chi(\chi_s-\chi)}{\chi_s} \ ,
\end{equation}
where $H_0$, $\Omega_{m,0}$ and $c$ are the Hubble constant, matter density parameter at the present and the speed of light respectively. The $a(\chi)$ function is the scale factor of the Universe and $\chi_s$ is the comoving distance to the source galaxies. Here we only consider the case where all source galaxies are located at a Dirac delta like source redshift distribution, but it is straightforward to extend the description of $q(\chi)$ to a general distribution of source galaxies \citep{Schneider2006}.

Adopting the flat-sky and Limber approximations \citep{Limber1953}, one can derive the power spectrum and bispectrum of the convergence field \citep{Kaiser1998}
\begin{equation}
    \label{eq:kappa_power_spectrum}
    P_{\kappa}(k)=\int{\rm d}\chi\frac{q^2(\chi)}{\chi^2}P_{m}\left(\frac{k}{\chi},\chi\right) \ ,
\end{equation}
\begin{equation}
    \label{eq:kappa_bispectrum}
    B_{\kappa}(k_1,k_2,k_3)=\int{\rm d}\chi\frac{q^3(\chi)}{\chi^4}B_m\left(\frac{k_1}{\chi},\frac{k_2}{\chi},\frac{k_3}{\chi},\chi\right) \ ,
\end{equation}
where $P_m$ and $B_m$ are the 3D matter power spectrum and bispectrum.

We smooth the convergence field with a smoothing kernel $W(kR)$, where $R$ is the smoothing angle, before we apply Eqs.~(\ref{eq:kappa_power_spectrum}) and (\ref{eq:kappa_bispectrum}) to Eq.~(\ref{eq:F_power_spectrum_nnlo}). The resulting smoothed convergence power spectrum and bispectrum then read $P(k)=W(kR)^2P_{\kappa}(k)$ and $B(k_1,k_2,k_3)=W(k_1R)W(k_2R)W(k_3R)B_{\kappa}(k_1,k_2,k_3)$. In practice, we use a Gaussian smoothing kernel, $W(kR)=e^{-k^2R^2/2}$. From there, the power spectrum of weak lensing critical points up to NLO reads
\begin{widetext}
\begin{eqnarray}
    \label{eq:wl_power_spectrum_nlo}
    P^{ij}(k)&&=g_1^i(\mathbf{k})g_1^j(\mathbf{k})W(kR)^2P_{\kappa}(k) \nonumber\\
    && +\frac{1}{2}\int\frac{\mathrm{d}^2k_1}{(2\pi)^2}g^i_2(\mathbf{k}_1, \mathbf{k}-\mathbf{k}_1)g^j_2(\mathbf{k}_1, \mathbf{k}-\mathbf{k}_1)W(k_1R)^2W(|\mathbf{k}-\mathbf{k}_1|R)^2P_{\kappa}(k_1)P_{\kappa}(|\mathbf{k}-\mathbf{k}_1|) \nonumber\\
    && + \frac{1}{2}\left[g^i_1(\mathbf{k})\int\frac{\mathrm{d}^2k_1}{(2\pi)^2}g^j_2(\mathbf{k}_1, \mathbf{k}-\mathbf{k}_1)W(kR)W(k_1R)W(|\mathbf{k}-\mathbf{k}_1|R)B_{\kappa}(-\mathbf{k},\mathbf{k}_1,\mathbf{k}-\mathbf{k}_1) + (i\leftrightarrow j)\right] \ .
\end{eqnarray}
\end{widetext}
In the above equation, the spectral moments from Eq.~(\ref{eq:spectral_moment}) exploited in $g_n$ functions has the following form
\begin{equation}
    \label{eq:spectral_moment_wl}
    \sigma_n^2=\int{\rm d}\chi\frac{q^2(\chi)}{\chi^2}\int\frac{k{\rm d}k}{2\pi}k^{2n}W(kR)^2P_{m}\left(\frac{k}{\chi},\chi\right) \ ,
\end{equation}
and we are going to use the tree-level 3D matter bispectrum in Eq.~(\ref{eq:kappa_bispectrum}) \citep{Scoccimaro1997}
\begin{equation}
    \label{eq:tree-level_bispectrum}
    B_m(k_1,k_2,k_3)=2F_2(\mathbf{k}_1,\mathbf{k}_2)P_m(k_1)P_m(k_2) + {\rm  perm} \ ,
\end{equation}
where $F_2(\mathbf{k}_1,\mathbf{k}_2)$ is the symmetric coupling kernel 
\begin{equation}
    \label{eq:F2_kernel}
    F_2(\mathbf{k}_1,\mathbf{k}_2)=\frac{5}{7}+\frac{2}{7}\frac{(\mathbf{k}_1\cdot\mathbf{k}_2)^2}{k_1^2k_2^2}+\frac{1}{2}\mathbf{k}_1\cdot\mathbf{k}_2\left(\frac{k_1}{k_2}+\frac{k_2}{k_1}\right) \ ,
\end{equation}
and ``perm" in Eq.~(\ref{eq:tree-level_bispectrum}) represents the same term but with cyclic permutations on arguments $\mathbf{k}_1$, $\mathbf{k}_2$ and $\mathbf{k}_3$ (such that the bispectrum ends up being a sum of three such terms). Another notice is that we use the nonlinear 3D matter power spectrum in Eqs.~(\ref{eq:kappa_power_spectrum}) and (\ref{eq:tree-level_bispectrum}). For this purpose we apply the fitting function \textsc{HaloFit} \citep{Smith2003, Takahashi2012} adopted in the Boltzmann solver package \textsc{Class} \citep{Blas2011}.

In the following computation, we adopt the flat $\Lambda$CDM model with Planck 2018 cosmological parameters \citep{Planck2018}: $\Omega_{\rm cdm}h^2=0.1201$, $\Omega_{b}h^2=0.02238$, $h=0.6732$, $n_{\rm s}=0.9660$ and $\sigma_8=0.8120$. We assume that all source galaxies are located at $z=1.5$. The smoothing scale is $R=15^{\prime}$
for the Gaussian kernel which corresponds to approximately 20 Mpc at the source redshift with the above background cosmology. We applied a fast and accurate numerical evaluation of Eq.~(\ref{eq:wl_power_spectrum_nlo}) using the method of separation of integration variables. After the separation we perform angular integrations first and the resulting expression can be computed by multiple one-dimensional Fourier transform. Readers can refer to Appendix.~\ref{appendix:Senpai_derivation} or Sec.~IIIB in Ref.~\cite{matsubara2020} for more technical details.

\begin{figure*}[t!]
\begin{minipage}{.5\linewidth}
\centering
{\label{main:Pk_pp}\includegraphics[scale=.45]{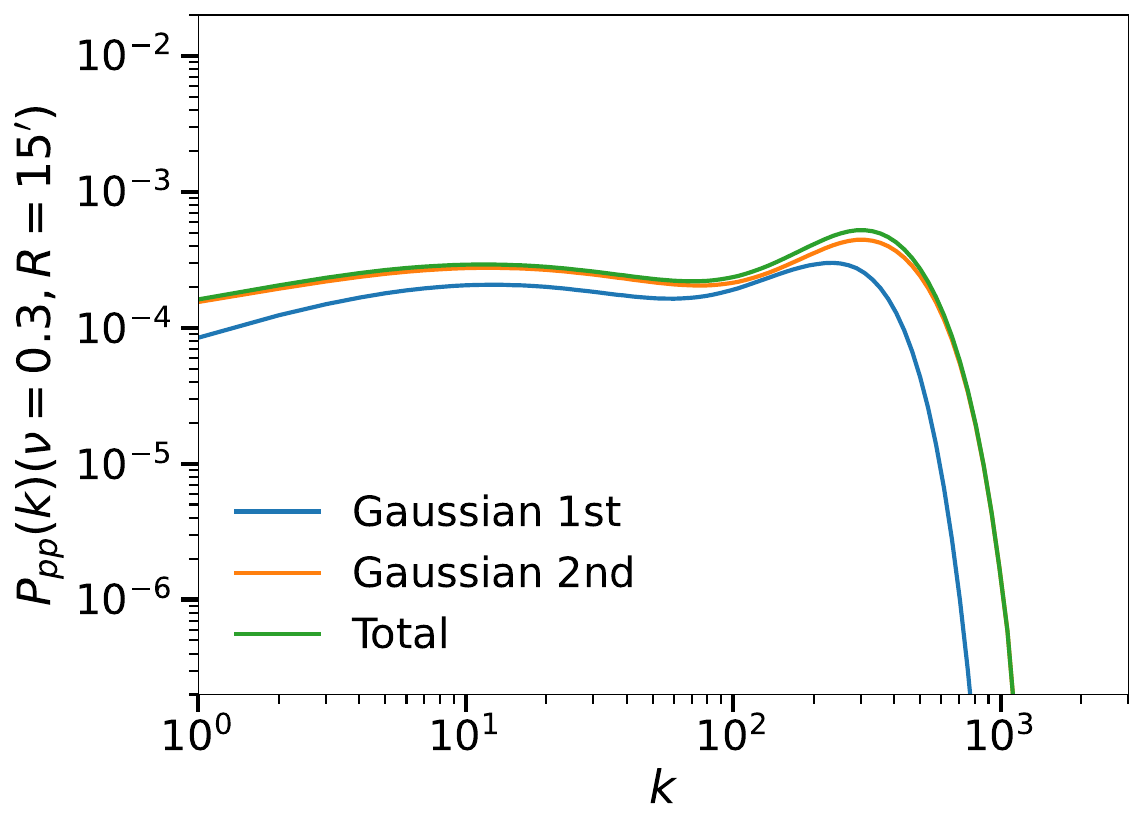}}
\end{minipage}%
\begin{minipage}{.5\linewidth}
\centering
{\label{main:Pk_pv}\includegraphics[scale=.45]{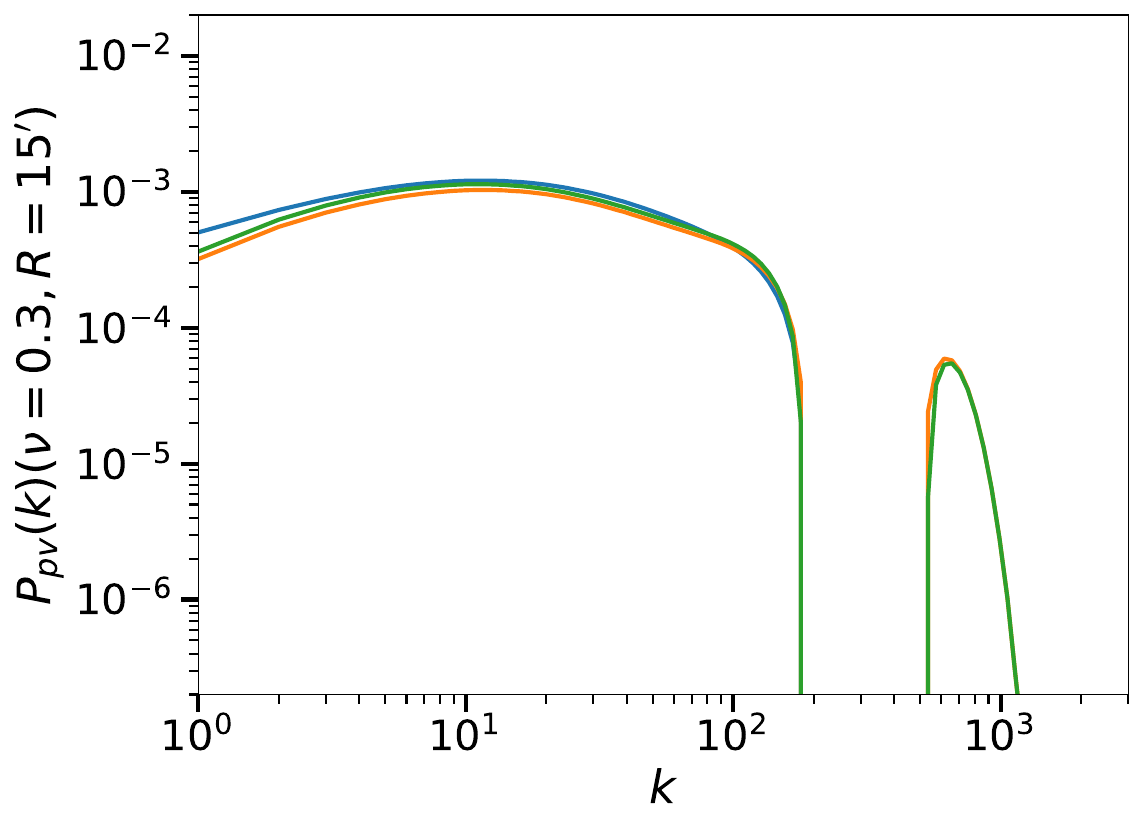}}
\end{minipage} \par\medskip
\begin{minipage}{.5\linewidth}
\centering
{\label{main:Pk_vv}\includegraphics[scale=.45]{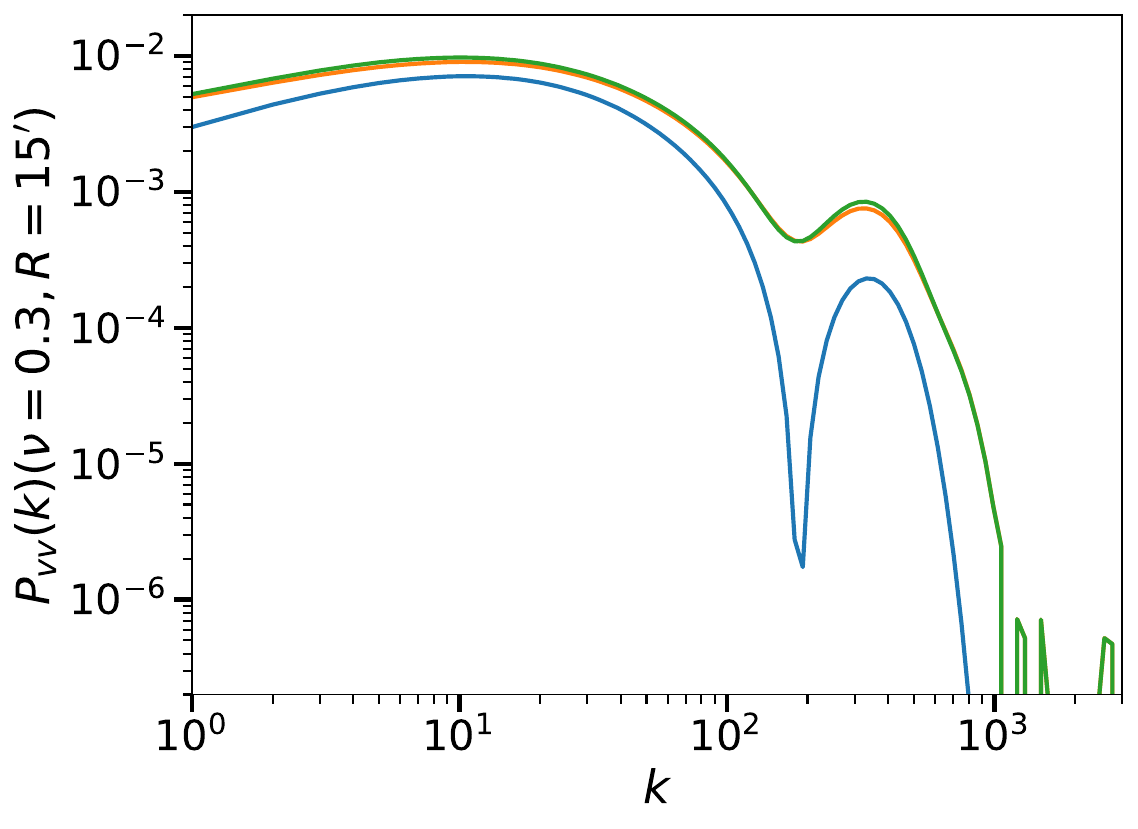}}
\end{minipage}%
\begin{minipage}{.5\linewidth}
\centering
{\label{main:Pk_ps}\includegraphics[scale=.45]{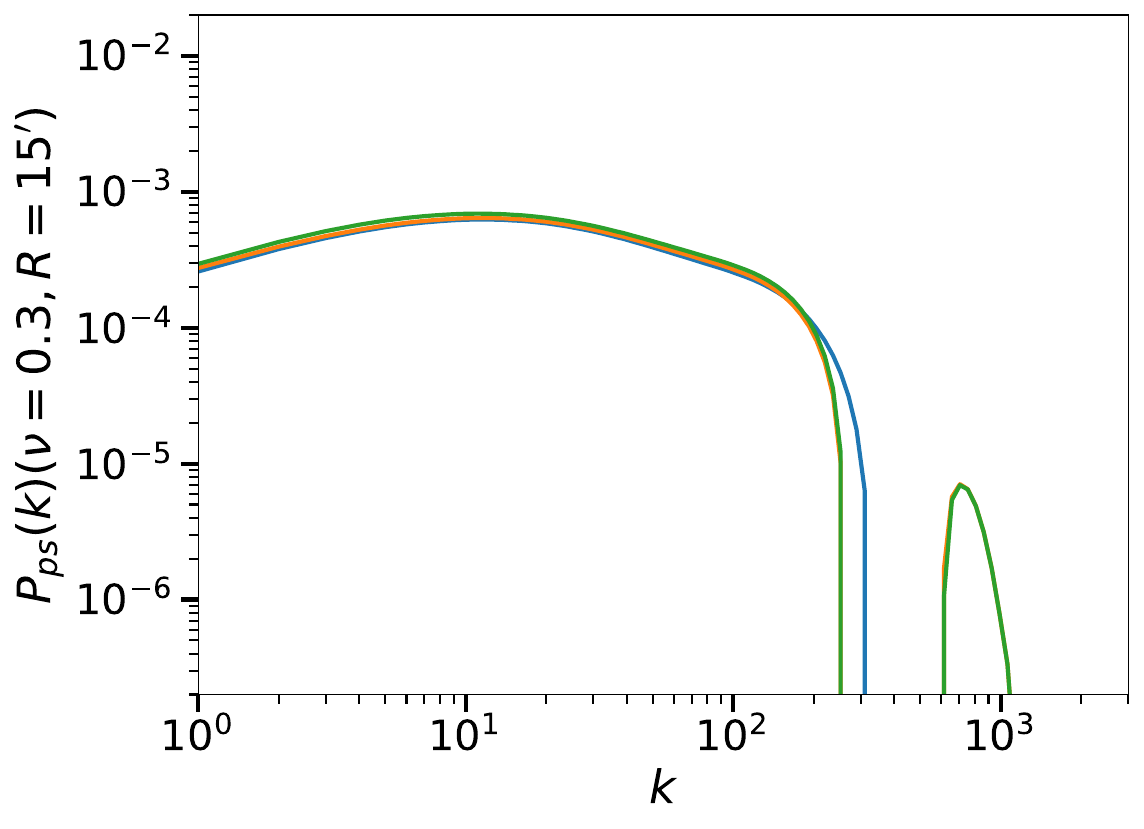}}
\end{minipage} \par\medskip
\begin{minipage}{.5\linewidth}
\centering
{\label{main:Pk_ss}\includegraphics[scale=.45]{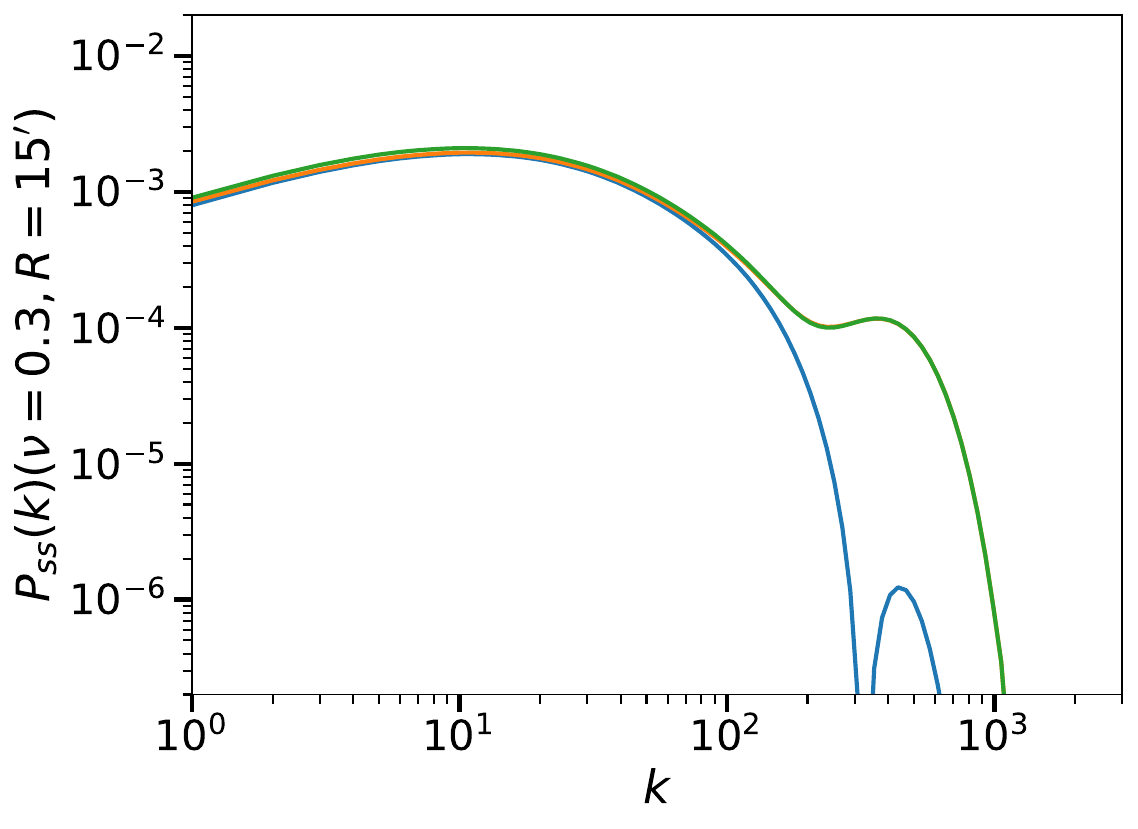}}
\end{minipage}%
\begin{minipage}{.5\linewidth}
\centering
{\label{main:Pk_vs}\includegraphics[scale=.45]{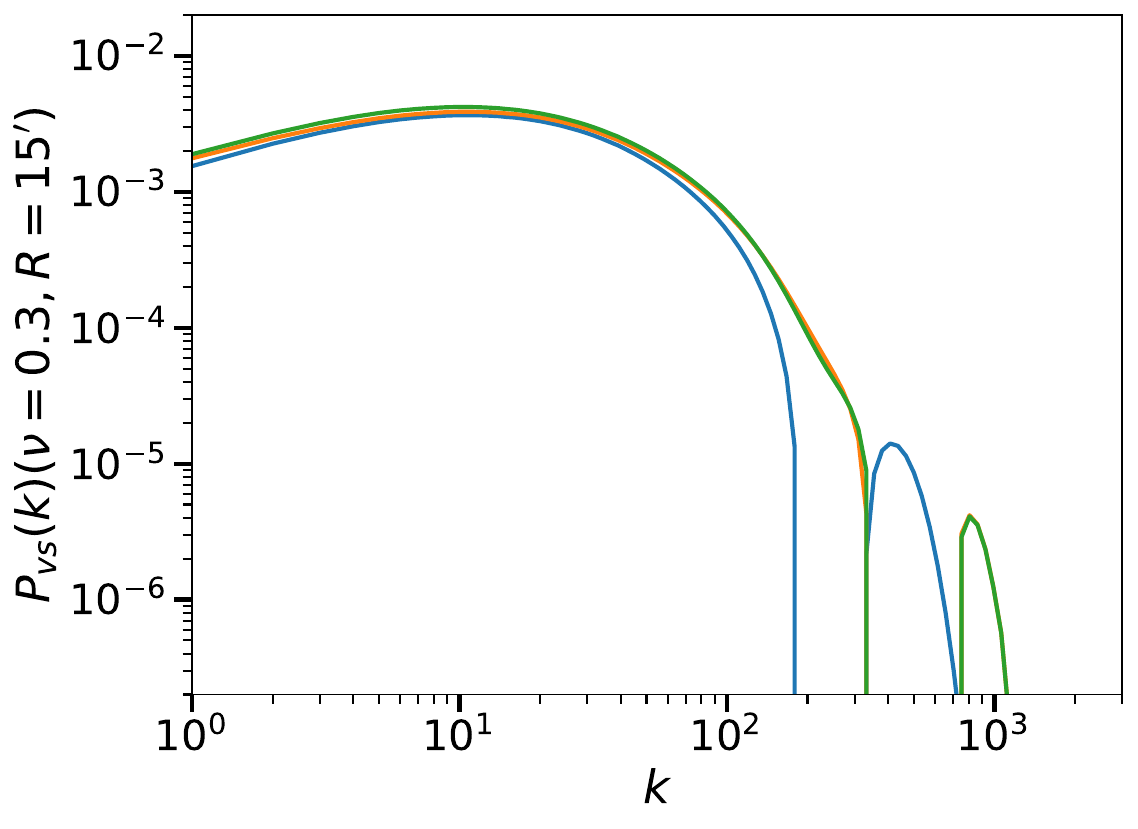}}
\end{minipage}
\caption{Auto and cross power spectrum of different critical points in 2D weak lensing fields above a given threshold $\nu=0.3$ and a smoothing scale $R=15'$. The subscript ``p" represents peaks while ``v" and ``s" stand for voids and saddle points respectively. Blue curve represents the LO in the power spectrum, corresponding to the first line on the right hand side of Eq.~(\ref{eq:wl_power_spectrum_nlo}). Orange curve is the sum of the LO and the 2nd-order Gaussian approximation ($\propto P(k)^2$ term) in the NLO, which is the second line term in Eq.~(\ref{eq:wl_power_spectrum_nlo}). Green curve is the full NLO prediction including the bispectrum correction expressed by the third line term in Eq.~(\ref{eq:wl_power_spectrum_nlo}). The color curves in all the other sub-panels have
the same representation as that denoted in the top left subplot. The fluctuations on large $k$ scales are the residuals of the unphysical components from the perturbative bias expansion after smoothing. They will not impact the 2PCFs on intermediate and large angular separations after the Hankel transform as we will show in Sec.~\ref{sec:MC_integration} with the peak 2PCF as an example.}
\label{fig:P_critical_points_wl}
\end{figure*}

\begin{figure*}[t!]
\begin{minipage}{.5\linewidth}
\centering
{\label{main:xi_pp}\includegraphics[scale=.45]{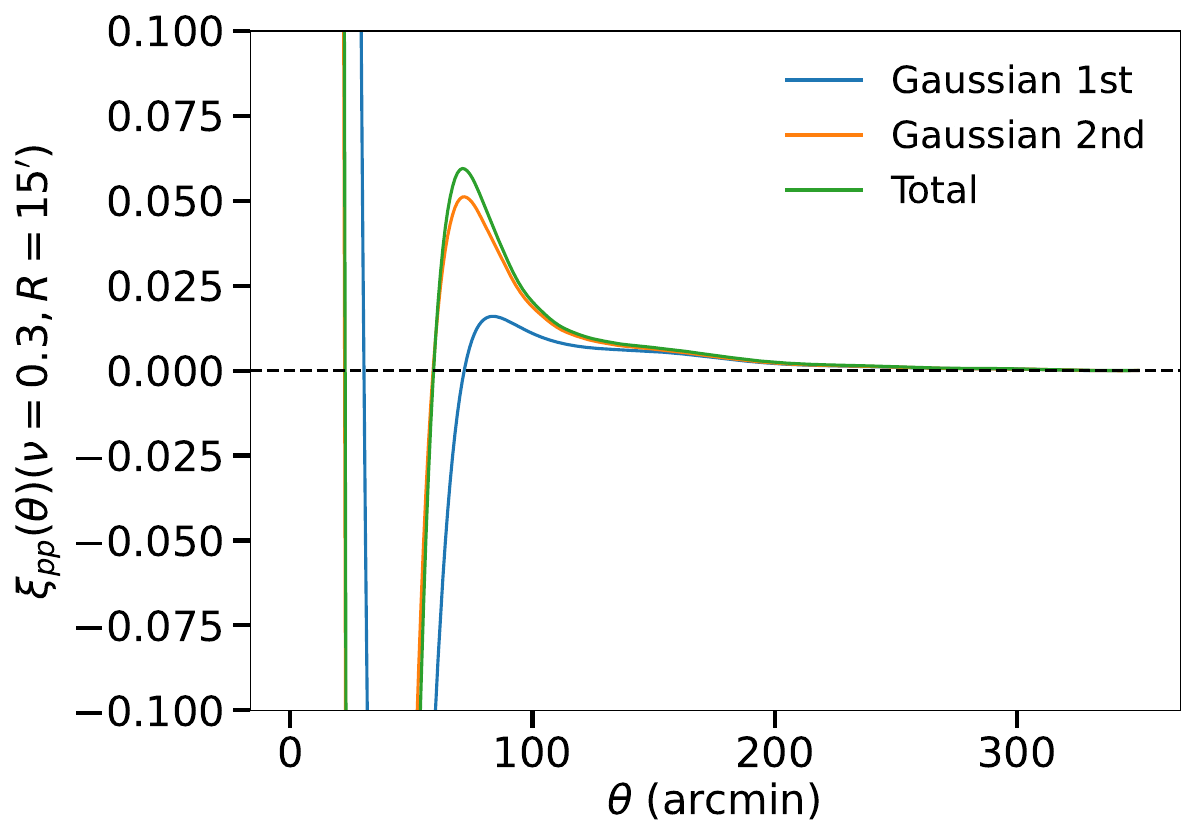}}
\end{minipage}%
\begin{minipage}{.5\linewidth}
\centering
{\label{main:xi_pv}\includegraphics[scale=.45]{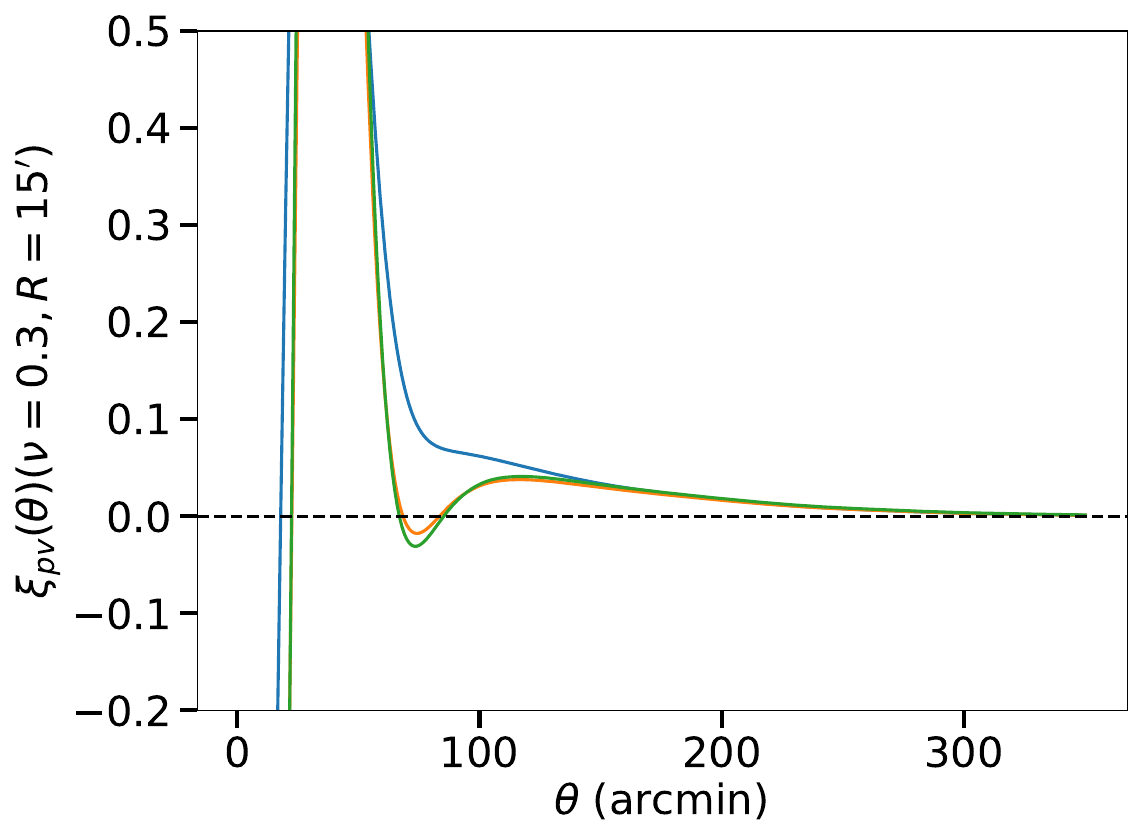}}
\end{minipage} \par\medskip
\begin{minipage}{.5\linewidth}
\centering
{\label{main:xi_vv}\includegraphics[scale=.45]{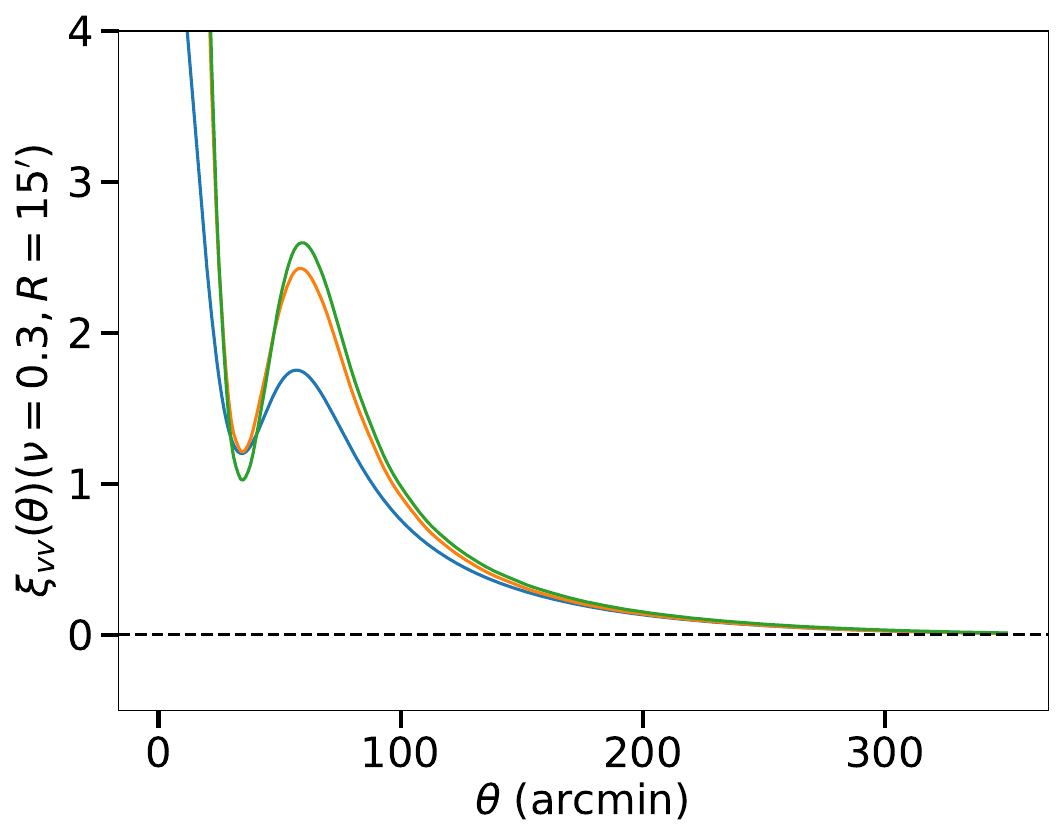}}
\end{minipage}%
\begin{minipage}{.5\linewidth}
\centering
{\label{main:xi_ps}\includegraphics[scale=.45]{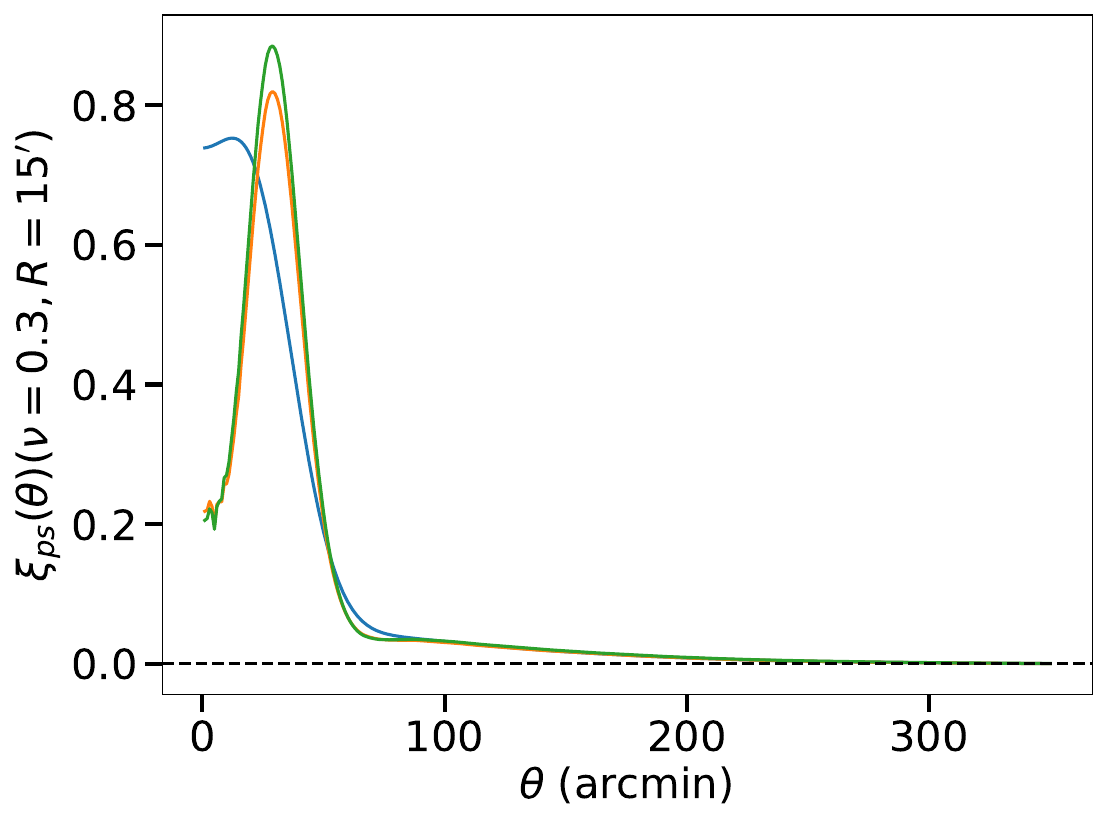}}
\end{minipage} \par\medskip
\begin{minipage}{.5\linewidth}
\centering
{\label{main:xi_ss}\includegraphics[scale=.45]{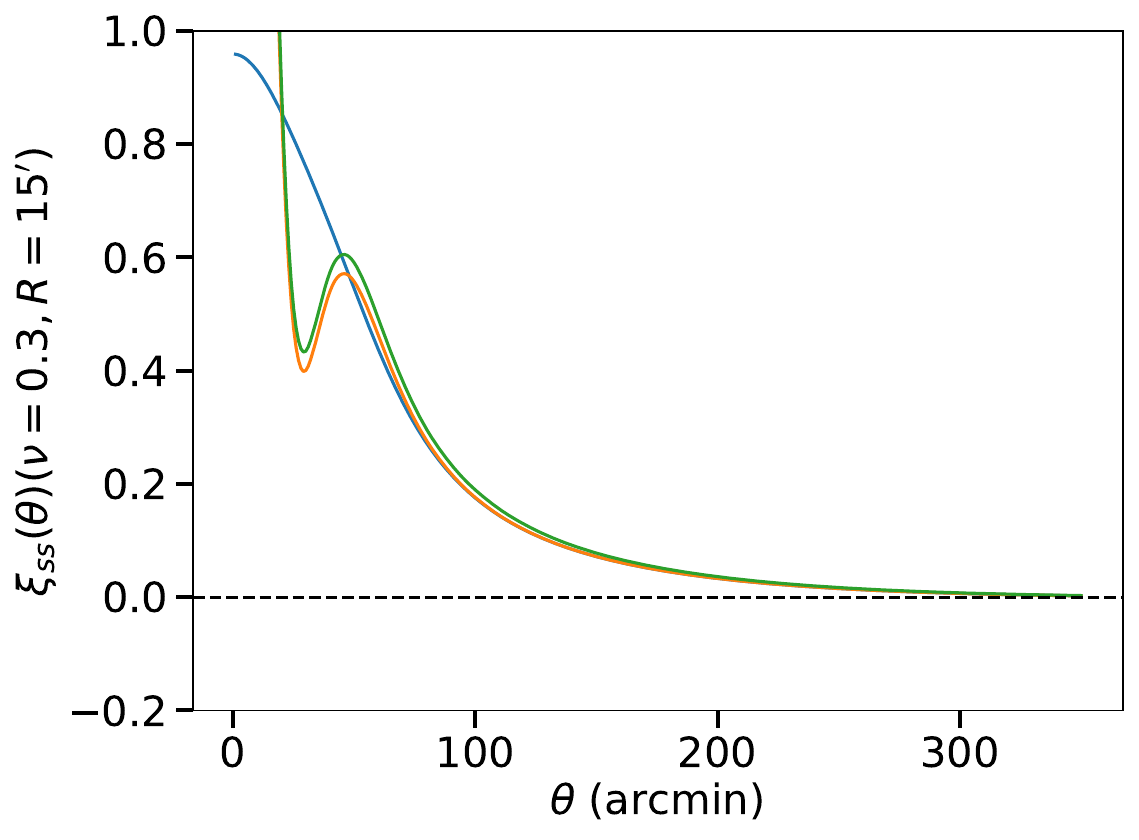}}
\end{minipage}%
\begin{minipage}{.5\linewidth}
\centering
{\label{main:xi_vs}\includegraphics[scale=.45]{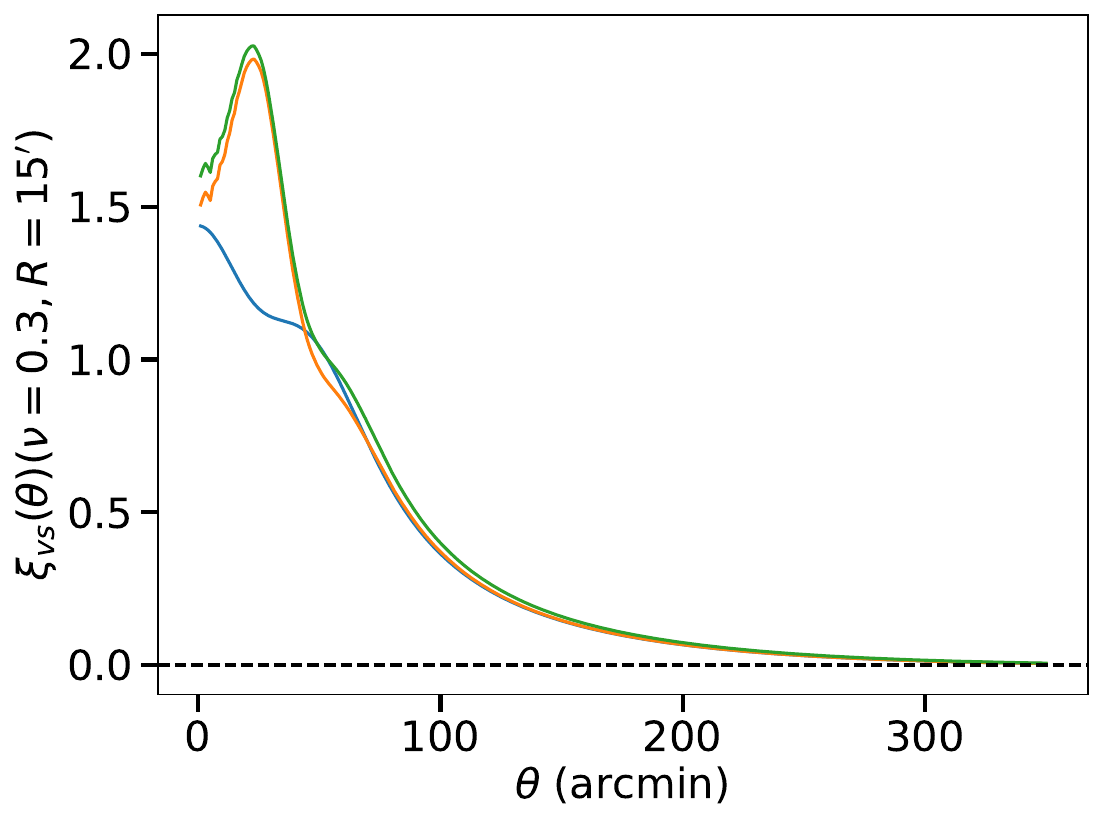}}
\end{minipage}
\caption{Auto and cross 2PCF of different critical points in 2D weak lensing fields above a given threshold $\nu=0.3$ and a smoothing scale $R=15'$. The subscript ``p" represents peaks while ``v" and ``s" stand for voids and saddle points respectively. Blue curve represents the LO in the 2PCF, which is Hankel transformed from the first line on the right hand side of Eq.~(\ref{eq:wl_power_spectrum_nlo}). The orange curve is the sum of the LO and the 2nd-order Gaussian approximation (whose Fourier counterpart is the $\propto P(k)^2$ term) in the NLO, which is the sum of the Hankel transform of the first two line terms in Eq.~(\ref{eq:wl_power_spectrum_nlo}). The green curve is the full NLO prediction including the bispectrum correction, i.e.~the Hankel transform of the complete expression of Eq.~(\ref{eq:wl_power_spectrum_nlo}). The color curves in all the other sub-panels have
the same representation as that denoted in the top left subplot.}
\label{fig:xi_critical_points_wl}
\end{figure*}

We show in Fig.~\ref{fig:P_critical_points_wl} the predicted auto power spectrum of peak-peak, void-void, and saddle-saddle, as well as the cross power spectrum of peak-void, peak-saddle and void-saddle, at a threshold $\nu=0.3$ where the value is taken with respect to the smoothed $\sigma_0$ shown in Eq.~(\ref{eq:spectral_moment_wl}). Above this given threshold, together with the above cosmology and lensing parameters, we compute the abundances of peaks and voids to be approximately $80\%$ and $6\%$ of the total number using a Monte Carlo (MC) integration method, assuming a underlying Gaussian random field.
We first notice that there is a discrepancy on $k\rightarrow 0$ scale between the LO and 2nd-order Gaussian approximation for peaks and voids. It has been suggested in previous works that the exclusion zone in 2PCFs for peaks and halos would non-trivially impact the power spectrum on large scales \citep{Baldauf2013,Baldauf2016,Codis2018,matsubara2019}. It has also been shown in Ref.~\cite{matsubara2020} for 3D peaks that this nonzero value in the limit of $k\rightarrow 0$, corresponding to unphysical component in the perturbative expansion, only exists in the 2nd-order Gaussian approximation term in the NLO, but not in other components. In Fig.~\ref{fig:P_critical_points_wl}, we observe this effect not only in 2D weak lensing peak power spectrum $P_{pp}(k)$, but also in voids $P_{vv}(k)$ which is caused by the exclusion zone between two voids. Meanwhile, such effect exists but not significant for the saddle point power spectrum $P_{ss}(k)$ and its cross power spectrum with peaks and voids $P_{ps}(k)$, $P_{vs}(k)$. This suggests that there is no strong exclusion effect between saddle points and other types of critical points (at the same threshold) since the matter flows through filaments (a type of saddle point) that are closely connected to either peaks or voids. This is because when the thresholds are the same, curvature and gradient constraints can be smoothly mapped from one to the other, contrary to peak-void for instance where the gradient constraint and the sign of the curvature impose two configurations that are incompatible in the zero separation limit. When transformed to real space for 2PCFs calculation, the above mentioned zero-lag value would turn into a Dirac delta-like function on small angular separations and thus not impact the convergence among different orders of perturbative bias expansion on large angular scales. This is confirmed in Fig.~\ref{fig:xi_critical_points_wl} where we show the corresponding 2PCFs.

For all 2PCFs, results from different orders of perturbative bias expansion converge with respect to each other on large angular separations.
There are amplitude increments for $\xi_{pp}$, $\xi_{vv}$, $\xi_{ss}$ and $\xi_{pv}$ towards small angular scales starting between $30'$ and $70'$, those are caused by the unphysical component in the perturbative expansion prediction which cannot correctly capture the non perturbative exclusion zones. Compared to the 2nd-order Gaussian approximation, the shape of the non-Gaussian correction is almost proportional to the Gaussian contribution on most of the scales. Thereby, the total shape of the extreme 2PCF does not change much by the effect of non-Gaussianity, but the amplitude does change, especially around their maxima beyond the unphysical angular scales. Among the auto 2PCFs, $\xi_{pp}$ reaches its maximum around $70'$ which is larger than that of $\xi_{vv}$ at $60'$ and in turn larger than $45'$ for $\xi_{ss}$. This relationship is consistent with what was found in 3D \citep{Shim2021} where peak 2PCF has the largest maximum separation, followed by that of voids and then saddle points. This might be due to the different rarity of critical points above the same threshold, thus causes different characteristic separations in their clustering.

If we examine $\xi_{ps}$ and $\xi_{vs}$ instead, we would find that their 2PCF maxima are on much smaller angular separations compared to the other four 2PCFs. This again indicates that saddle points have a weak exclusion effect with respect to peaks and voids above the same threshold, reflecting a shell-like structure in the distribution of saddle points around a given extremum, reminiscent of a cubic crystal lattice as pointed out by Refs.~\cite{Shim2021,Codis2018}. It implies that cross 2PCFs between saddle points and other extreme may better explore small-scale physics. 

\begin{figure*}[t!]
\begin{minipage}{.5\linewidth}
\centering
{\label{main:xi_pp_theta2}\includegraphics[scale=.45]{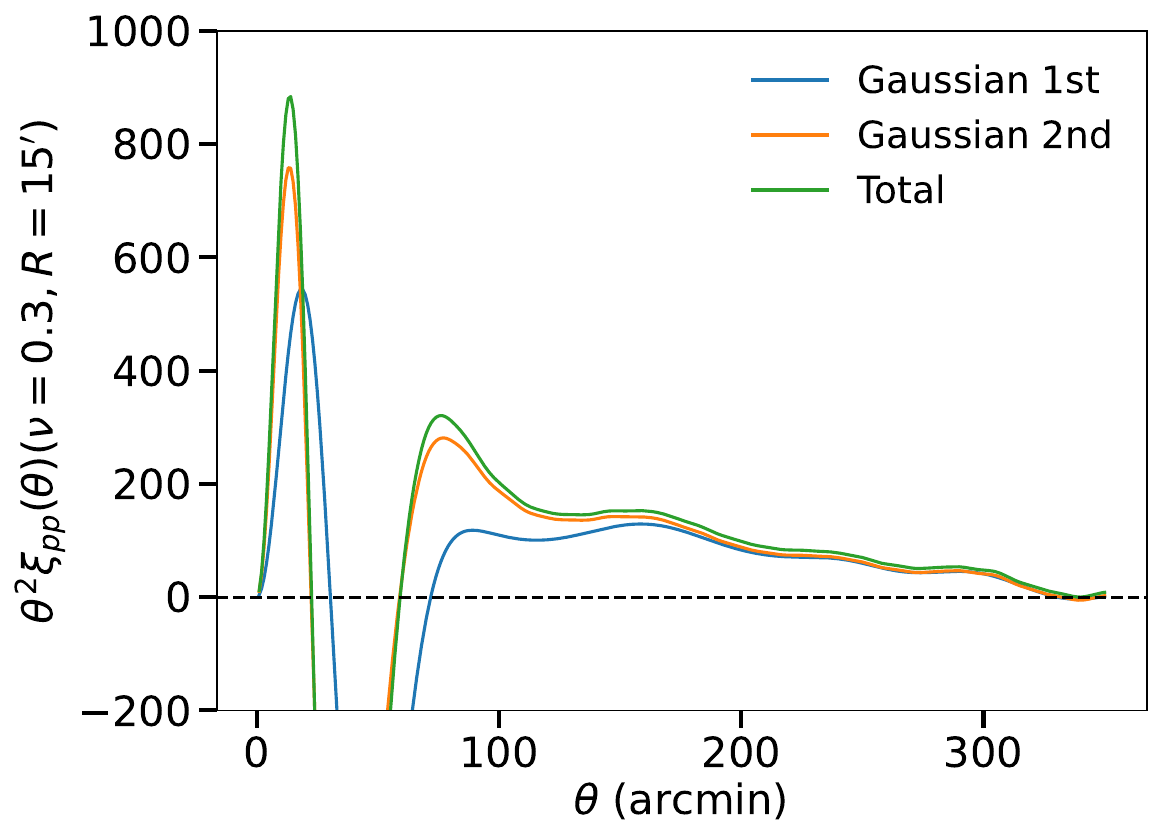}}
\end{minipage}%
\begin{minipage}{.5\linewidth}
\centering
{\label{main:xi_pp_nobaryon}\includegraphics[scale=.45]{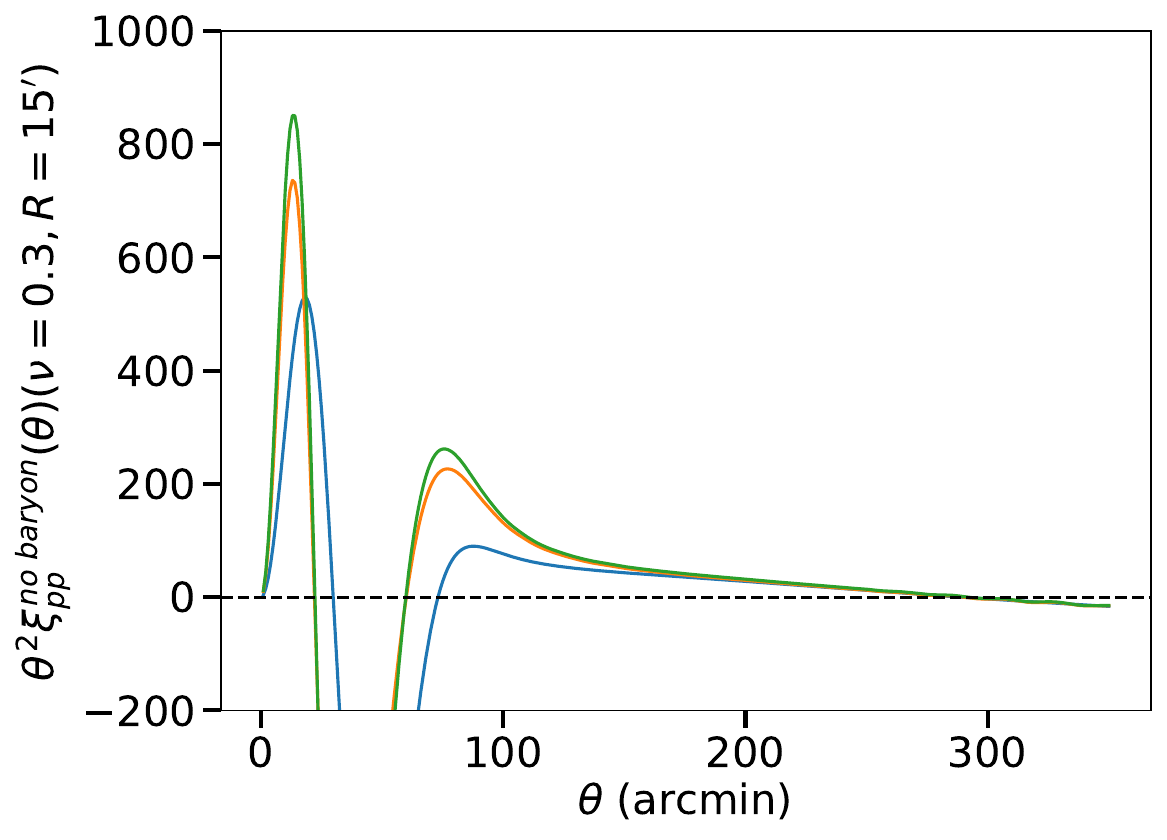}}
\end{minipage}
\caption{\textit{Left}: The same peak 2PCF as in Fig.~\ref{fig:xi_critical_points_wl} but multiplied by $\theta^2$. \textit{Right}: Same as the left panel, but the underlying matter power spectrum is calculated without baryons while kept at the same total matter density parameter.}
\label{fig:xi_pp_bao}
\end{figure*}

Another interesting feature to observe is the existence of oscillations in $\xi_{pp}$ on scales beyond $\theta=100'$ once we multiply the signal by $\theta^2$. The corresponding 2PCF is plotted in Fig.~\ref{fig:xi_pp_bao}.
At least two additional oscillation peaks at $\theta\approx 150'$ and $300'$ can be observed. This wiggly feature is related to the effect of BAOs in the underlying matter power spectrum. If we remove the baryon component from the underlying matter power spectrum and recalculate the peak 2PCF, we would obtain the result shown in the right panel of Fig.~\ref{fig:xi_pp_bao}. The peaks at $150'$ and $300'$ both vanish and overall amplitude of the correlation function is reduced. It is well-known that baryonic features are highly suppressed in weak lensing power spectrum due to the line-of-sight projection shown in Eq.~(\ref{eq:kappa_power_spectrum}). However, these features can be enhanced in 3D peak correlation functions for both Gaussian and mildly non-Gaussian density field as pointed out in Refs.~\cite{Desjacques2008, Desjacques2010, matsubara2020}. In this paper we confirm this property in 2D mildly non-Gaussian weak lensing fields. We do not observe such significant BAO related features in other types of correlation functions. We believe this is because peaks have larger curvatures in very overdense regions compared to other critical points, e.g.~voids, and therefore can better amplify the strongly suppressed BAO features in the weak lensing convergence power spectrum. Since our $g_{n}(\mathbf{k}_1,\ldots,\mathbf{k}_n)$ functions are computed with respect to the Gaussian random field (Eqs.~(\ref{eq:G_equation_3}) and (\ref{eq:g_function})), based on the symmetry argument, the 2PCF of voids as a critical point should be able to exhibit BAO features if we modify our modeling and probe voids below an underdense threshold. However, we leave this aspect to future investigations.

\begin{figure}[h!]
    \centering
    \includegraphics[width=\linewidth]{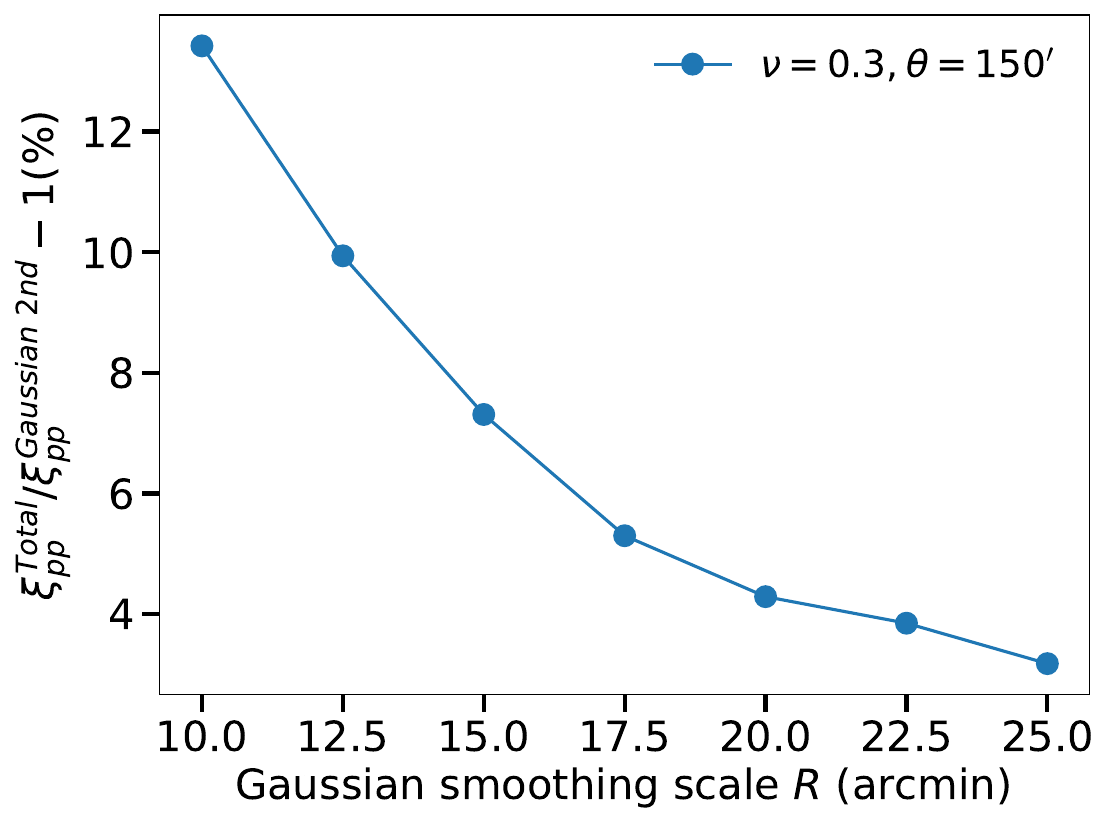}
    \caption{The fractional difference in percentage between the full NLO computation of peak 2PCF and its corresponding 2nd-order Gaussian approximation at different smoothing scales, where the Gaussian smoothing scale $R$ varies from $10^{\prime}$ to $25^{\prime}$. All fractional differences are calculated at a single angular separation $\theta=150^{\prime}$ and above the same threshold $\nu=0.3$. Note that the spectral moments that enter 
    for different smoothing scales follow that in Eq.~(\ref{eq:spectral_moment_wl}).}
    \label{fig:2pcf_smoothing}
\end{figure}

We explore the influence of the Gaussian smoothing kernel scale, $R$, on the computed 2PCFs, focusing on the non-Gaussian effects arising from the inclusion of the bispectrum correction compared to the second-order Gaussian approximation. In Fig.~\ref{fig:2pcf_smoothing}, we vary $R$ in the computation of the peak power spectrum (Eq.~(\ref{eq:wl_power_spectrum_nlo})) from $10^{\prime}$ to $25^{\prime}$. Using an angular separation of $150^{\prime}$, which is significantly larger than the range of smoothing scales, we calculate the difference between the full NLO computation of the peak 2PCF and its second-order Gaussian approximation. The results show that the fractional difference decreases as the smoothing scale increases, indicating that stronger smoothing of the underlying cosmic density field reduces the non-Gaussian effects introduced by the bispectrum correction.

With the above discussion, we establish the analytical 2PCFs, including the mildly non-Gaussian correction, among all pairs of critical points in 2D weak lensing convergence field on large angular separations. Previous literature such as those cited in Sec.~\ref{sec:intro} showed that peak 2PCF is sensitive to cosmological parameters such as $\Omega_m$ and $\sigma_8$, and it can add constraining power to the inference of these parameters complementary to peak number count. However, those simulation-based models for the peak 2PCF do not extend to large angular separations where our model is valid and fast in its predictions. Therefore our model can contribute effectively to the cosmological inference. Additionally, the BAO features  features on large scales for peak 2PCFs can serve as an independent standard ruler \citep{Desjacques2010} without galaxy bias by just looking at the amplitude of the fields in weak lensing maps, to study the evolution of LSS and test different cosmological models, an area that is currently not sufficiently explored.

\section{Numerical predictions compared to Monte Carlo integrations}
\label{sec:MC_integration}
In this section, we aim to validate our perturbative bias expansion approach. In order to do this, we compare one of our predictions for 2PCFs of critical points in the previous section, the peak 2PCF $\xi_{pp}(\theta)$, to a full numerical integration of the peak 2PCF obtained by a MC integration method in \textsc{Mathematica} (for comparison of other critical point 2PCFs, please refer to Appendix.~\ref{appendix:MC_comparison}). In the MC integration, we assume a Gaussian probability density distribution for the underlying density field. This assumption guarantees that our full numerical integration result is exact and can be used to validate our theoretical prediction from perturbative bias expansion approach on large angular separations.
\begin{figure}[h!]
    \centering
    \includegraphics[width=\linewidth]{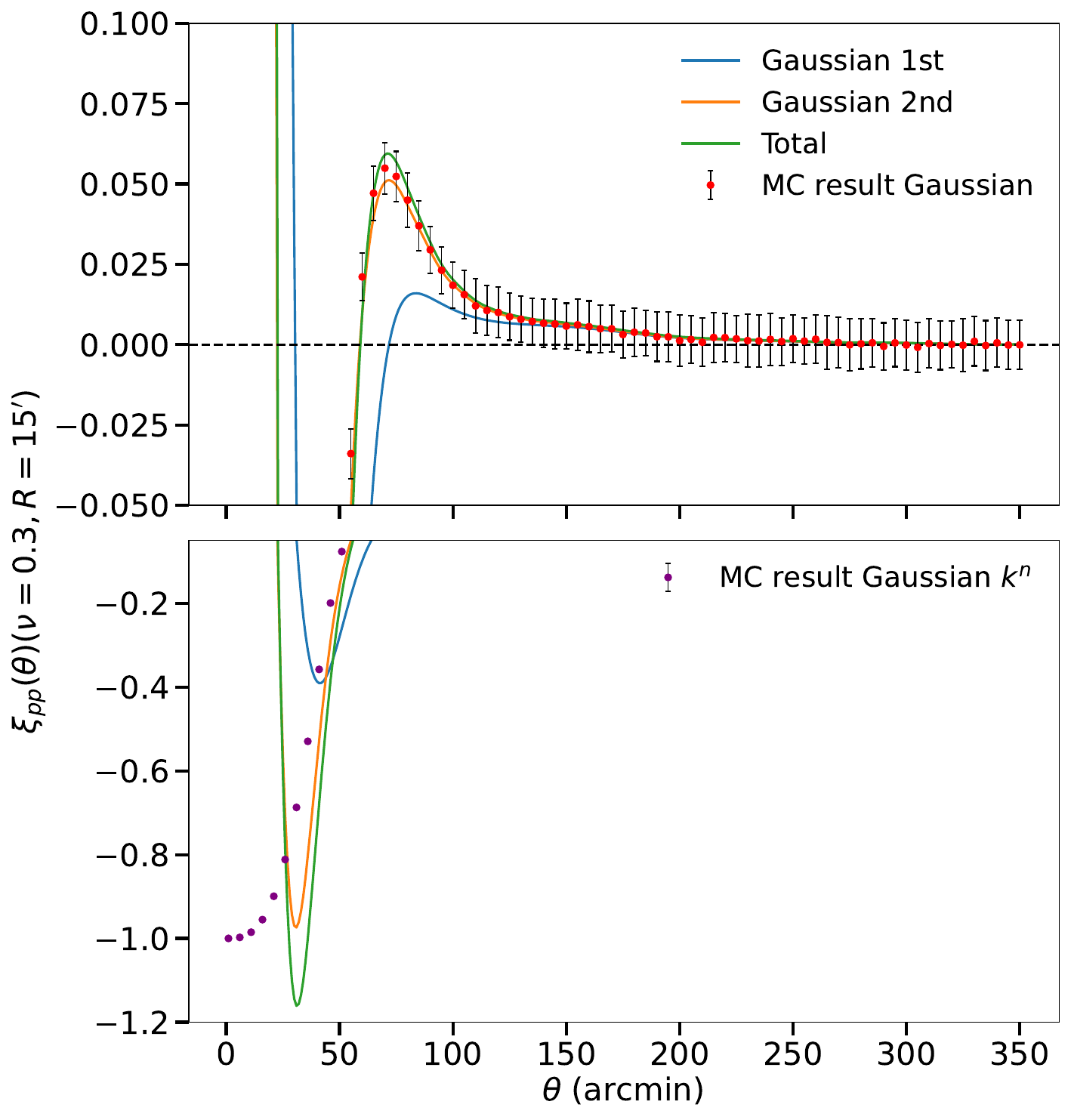}
    \caption{2PCF for peaks above the threshold $\nu=0.3$ with a flat $\Lambda$CDM model under Planck 2018 cosmological parameters. The underlying density field is smoothed by a Gaussian kernel with a smoothing scale $R=15'$. The color curves exploit the same convention to those in Fig.~\ref{fig:xi_critical_points_wl}. Red dots and the corresponding error bars in the top panel are the mean and standard deviation from 60 estimations of the MC integration. Purple dots and the corresponding error bars in the bottom panel are the mean and standard deviation from 60 estimations of the MC integration but with a power law approximation for the weak lensing convergence power spectrum. The full 2PCF is divided into two panels, each covering a different range of linear scales, allowing for a clearer examination of the small amplitude at larger separations.}
    \label{fig:MC_result}
\end{figure}

In practice, we use the same MC integration method as that presented in Sec.~IV of Ref.~\cite{matsubara2019} but in 2D. We draw random numbers of dimension 8 from the joint Gaussian conditional probability of $(\alpha, \zeta_{ij})$ at position $\mathbf{x}_1$ and $(\alpha, \zeta_{ij})$ at position $\mathbf{x}_{2}$ which satisfy $\eta_i=0$. We only keep the drawn sample if $\alpha$ is above the amplitude threshold $\nu$ and eigenvalues of $\zeta$ are negative. With ${\rm det}|\zeta_{ij}^{(k)}(\mathbf{x}_1)|$ and ${\rm det}|\zeta_{ij}^{(k)}(\mathbf{x}_2)|$ computed, we have
\begin{align}
    \label{eq:MC_numerator}
    &\langle n_{p}(\nu,\mathbf{x}_1)n_{p}(\nu,\mathbf{x}_2)\rangle\nonumber\\
    &\approx \frac{\mathcal{P}_G\left[\boldsymbol{\eta}(\mathbf{x}_1)=\boldsymbol{\eta}(\mathbf{x}_2)=\mathbf{0}\right]}{N}\sum_{k\in \mathcal{S}}{\rm det}|\zeta_{ij}^{(k)}(\mathbf{x}_1)|{\rm det}|\zeta_{ij}^{(k)}(\mathbf{x}_2)| \ ,
\end{align}
where $N$ is the total number of drawn sample, $\mathcal{S}$ is the subset of the drawn indices that correspond to the sample satisfying the conditions on the eigenvalues and amplitude. We can use the same procedure to evaluate the expectation value of the peak number density $\langle n_p(\nu,\mathbf{x})\rangle$. The peak 2PCF $\xi_{pp}(\theta)$ would therefore be
\begin{equation}
    \label{eq:MC_2PCF}
    \xi_{pp}(\theta) = \frac{\langle n_{p}(\nu,\mathbf{x}_1)n_{p}(\nu,\mathbf{x}_2)\rangle}{\langle n_p(\nu,\mathbf{x})\rangle^2}-1 \ ,
\end{equation}
where the angular $\theta$ dependence is from the covariance matrix of the joint Gaussian conditional probability in $\langle n_{p}(\nu,\mathbf{x}_1)n_{p}(\nu,\mathbf{x}_2)\rangle$. The high dimensionality of the above integration makes the computation expensive, however, we parallelized the algorithm on a local cluster such that the calculation is completed within a reasonable period of time. Another subtlety in our MC integration method is that on small angular separations, $\theta \lesssim 40'$, the covariance matrix between two points at $\mathbf{x}_1$ and $\mathbf{x}_2$ would become non-invertible due to the numerical instability in the integrand of some entries. In order to qualitatively show the exclusion effect of the critical point clustering, which our perturbative bias expansion is incapable of fully capturing, we approximate the weak lensing power spectrum by a power law $P_{\kappa}(k) \propto k^n$ when performing MC integration on these scales. The power index $n$ is determined by solving $\gamma = \sqrt{(n+2)/(n+4)}$, where $\gamma$ is from Eq.~(\ref{eq:gamma}) computed with the weak lensing power spectrum from Boltzmann solver, and the right side comes from expressing $\gamma$ in terms of $P_{\kappa}(k) \propto k^n$.

In Fig.~\ref{fig:MC_result}, for each angular separation, we perform 60 estimations of the 2PCF to obtain the mean value and the associated estimated standard deviation. For each MC integration, we draw 20 million times 8 random numbers for which evaluation is parallelized on 16 cores. On the local cluster, one such estimation for all angular separations took averagely half an hour (with some variability).

We observe from Fig.~\ref{fig:MC_result} that on angular scales $\theta\geq 100'$, the theoretical predictions are almost identical to the MC integration result. This proves that the convergence behavior of our theoretical prediction among different orders of approximation is correct. On angular scales $\theta\approx 55'$, the 2nd-order Gaussian approximation already reproduces quite accurately the 2D weak lensing peak exclusion effect \citep{Baldauf2016, Codis2018} as demonstrated by the MC result on the same scale. The non-Gaussian correction from bispectrum does not add significant changes to the predicted exclusion scale if one compares the orange to the green curve on $\theta\approx 55'$. On further smaller scales, there is an increase of the predicted peak 2PCF and the convergence of predictions from different orders of approximation is very poor. This is a well-known feature that the perturbative bias expansion on small scales cannot capture the non-perturbative exclusion zone, as demonstrated by the power law approximation MC result in the bottom panel of Fig.~\ref{fig:MC_result}.

On angular scales between $50'$ and $100'$, the theoretical prediction from 2nd-order Gaussian approximation is much closer to the MC integration result compared to the LO prediction as expected. Note that in principle one could extend the perturbative expansion to higher orders in the context of Gaussian approximation (e.g, the 3rd-order Gaussian approximation would include the $P(k)^3$ term in NNLO in Eq.~(\ref{eq:F_power_spectrum_nnlo})) but the convergence of such high-order bias expansion is known to be slow because of the non-perturbative nature of the small-scale exclusion zone (as also shown in Ref.~\cite{matsubara2019} for 3D peaks in Gaussian distributed density field). Adding a bispectrum correction on top of it leads to an excess of 2PCF amplitude which gradually deviates from the exact Gaussian MC result on $\theta\approx 100'$ and reaches its maximum around $70'$, within the context of our chosen smoothing scale and threshold. Around $\theta=70^{\prime}$, the amplitude of the 2PCF including the bispectrum (non-Gaussian) correction is about $17\%$ larger than that from the 2nd-order Gaussian approxiamtion. This discrepancy reduces to about $8\%$ on $\theta\approx 100^{\prime}$. When comparing the theoretical NLO prediction to the exact MC integration result under Gaussian assumption, the discrepancy due to non-Gaussianity at tree-level bispectrum is within the MC sample standard deviation. However, note that with higher-order non-Gaussian corrections or more accurate numerical and survey experiments, we will be able to statistically distinguish them. A further investigation of how well the bispectrum correction characterizes the non-Gaussian part in the 2PCFs of critical points requires a detailed comparison to the N-body simulations. This is beyond the scope of this paper and we leave it to future works.

\section{Conclusion}
\label{sec:conclusion}
In this paper, we extended the general formalism presented in Ref.~\cite{matsubara2020} for peak statistics in mildly non-Gaussian density field to 2PCFs of 2D critical points, including peaks, voids and saddle points. We applied this formalism to the case of mildly non-Gaussian weak lensing convergence field. Analytically we derived the perturbative bias expansion up to the NNLO, taking the linear terms of trispectrum induced by nonlinear evolution of gravitational instability into account, in Eq.~(\ref{eq:F_power_spectrum_nnlo}). For the numerical calculation, we only consider the lowest-order non-Gaussian correction as in Eq.~(\ref{eq:wl_power_spectrum_nlo}), which is composed of linear terms of the bispectrum. In order to evaluate correlation functions for different types of critical points in 2D weak lensing fields, one needs to compute the $g_{ijklm}$ terms in Eq.~(\ref{eq:g_ijklm_2}) and adjust the integration limits within there to a specific critical point type accordingly (Eqs.~(\ref{eq:g_ijklm_peaks}), (\ref{eq:g_factors_voids}) and (\ref{eq:g_factors_saddle})), where in Appendix.~\ref{appendix:C} we show plots of them as functions of the density field threshold $\nu$.

As a demonstration, we calculated six power spectra and their corresponding 2PCFs from all possible combinations of the three types of critical points in 2D weak lensing field, above a given threshold chosen here to be $\nu=0.3$ and with a specific Gaussian smoothing scale $R=15'$. We observed similar properties (shown by Figs.~\ref{fig:P_critical_points_wl} and \ref{fig:xi_critical_points_wl}) as for the clustering of 3D critical points measured from N-body simulations in previous works such as Ref.~\cite{Shim2021}. The angular separation where the correlation function reaches its maximum beyond the exclusion zone is largest for peaks, followed by that of voids and then saddle points. Saddle points are implied to have a weak exclusion effect with respect to peaks and voids, reflecting a shell-like structure in the distribution of saddle points around a given extremum. Meanwhile we also notice that our theoretical prediction is incapable of fully capturing the nonperturbative exclusion zone on small angular scales, as expected due to its very nonlinear nature. Another interesting feature is that the effect of BAOs is enhanced in 2D peak 2PCF (Fig.~\ref{fig:xi_pp_bao}), compared to the underlying weak lensing convergence power spectrum where the effect is suppressed due to the line-of-sight projection. We believe this enhancement is due to the derivative constraint (curvature) on the weak lensing convergence peak field as discussed in Sec.~\ref{sec:2d_weak_lensing} and can be used as a BAO probe for weak lensing data.

In order to validate the theoretical prediction, we chose the peak 2PCF as an example and compared it to the exact and yet computationally intensive MC integration result, which assumes a Gaussian distributed underlying density field. The two are almost identical on large angular separations down to approximately $100'$ (Fig.~\ref{fig:MC_result}). Interestingly, part of the exclusion zone on the outer edge can be described by the perturbative bias expansion. The non-Gaussian correction from the bispectrum contribution modifies the correlation function most significantly around the maximum region.

Eventually, the main purpose of this paper has been to provide the analytical framework for the clustering of critical points in 2D weak lensing field, which is definitely non-Gaussian on scales of interests. There are certain directions to further extend and apply this work. One is to serve as a benchmark test for N-body simulations that are used to measure weak lensing peaks or minima clustering, to prevent those statistics from being biased by simulation systematic effects. One can also combine fast and accurate analytical predictions on large angular scales with detailed measurement from simulations on small scales, which saves significant time and computational resources, and could allow us to achieve a hybrid summary statistic of critical points clustering. This hybrid summary statistic could then be exploited in inference tasks in the ongoing and next-generation weak lensing surveys to extract more information from the large-scale structure of the Universe. Additionally, we have observed features like BAOs and inflection points (observed in Ref.~\cite{Shim2021} from simulations) on large scales for weak lensing peak 2PCFs. These features could serve as independent standard rulers, independent of galaxy bias, to study the evolution of the large-scale structure and test different cosmological models, an area that is currently under-explored. We hope to address the possibility of the above applications in the near future.

\begin{acknowledgments}
The authors thank Takahiko Matsubara, Stella Seitz, Masahiro Takada and Zhenyuan Wang for fruitful discussions. The authors thank the Yukawa Institute for Theoretical Physics at Kyoto University. Discussions during the YITP workshop YITP-T-24-05 on ``Theory and Data Analysis Challenges for Cosmological Large-Scale Structure Observations" were useful to complete this work. This work has made use of the Infinity cluster hosted by the Institut d'Astrophysique de Paris. We warmly thank S.~Rouberol for running it smoothly.
\end{acknowledgments}

\bibliography{apssamp}

\appendix

\section{Generalization of the Gram-Charlier A series expansion with the Wiener-Hermite functionals}
\label{appendix:A}
In this paper, we adopt the following Fourier transform convention
\begin{equation}
    \label{eq:FT}
    \tilde{f}(\mathbf{k})=\int{\rm d}^2x e^{-i\mathbf{k}\cdot\mathbf{x}}f(\mathbf{x}) \ , f(\mathbf{x})=\int\frac{{\rm d}^2k}{(2\pi)^2}e^{i\mathbf{k}\cdot\mathbf{x}}\tilde{f}(\mathbf{k}) \ .
\end{equation}
To derive the probability density functional $\mathcal{P}(\tilde{f})$, we start with the partition function, or the moment generating functional of the density field
\begin{equation}
    \label{eq:partition_function}
    \mathcal{Z}(\mathbf{J})=\int\mathcal{D}\tilde{f}{\rm exp}\left[i\int\frac{{\rm d}^2k}{(2\pi)^2}\mathbf{J}(\mathbf{k})\tilde{f}(\mathbf{k})\right]\mathcal{P}(\tilde{f}) \ ,
\end{equation}
where $\mathcal{D}\tilde{f}$ is the same to that mentioned in Eq.~(\ref{eq:F_2nd_moment_Fourier}). On the other hand, the partition function can also be expressed in terms of a series expansion using the cumulant expansion theorem \citep{Ma1985}
\begin{align}
    \label{eq:cumulant_expansion}
    {\rm ln}\mathcal{Z}(\mathbf{J}) = &\sum_{n=1}^{\infty}\frac{i^n}{n!}\int\frac{{\rm d}^2k_1}{(2\pi)^2}\cdots\int\frac{{\rm d}^2k_n}{(2\pi)^2}\nonumber\\
    &\times\langle\tilde{f}(\mathbf{k}_1)\cdots\tilde{f}(\mathbf{k}_n)\rangle_{\rm c}\mathbf{J}(\mathbf{k}_1)\cdots\mathbf{J}(\mathbf{k}_n) \ ,
\end{align}
and if we take the exponential on both sides of the above equation, we would have
\begin{align}
    \label{eq:partition_function_cumulant_expansion}
    \mathcal{Z}(\mathbf{J}) &= {\rm exp} \left[ -\frac{1}{2} \int \frac{{\rm d}^2k}{(2\pi)^2} 
    P(k) \mathbf{J}(\mathbf{k}) \mathbf{J}(-\mathbf{k}) \right] \nonumber \\
    &\quad \times {\rm exp} \Bigg[ \sum_{n=3}^{\infty} \frac{i^n}{n!} 
    \int \frac{{\rm d}^2k_1}{(2\pi)^2} \cdots \int \frac{{\rm d}^2k_n}{(2\pi)^2} \nonumber \\
    &\quad \times \langle\tilde{f}(\mathbf{k}_1) \cdots \tilde{f}(\mathbf{k}_n)\rangle_{\rm c} 
    \mathbf{J}(\mathbf{k}_1) \cdots \mathbf{J}(\mathbf{k}_n) \Bigg] \ .
\end{align}
where we used the definition of the density field power spectrum similar to Eqs.~(\ref{eq:bispectrum}) and (\ref{eq:trispectrum})
\begin{equation}
    \label{eq:power_spectrum}
    \langle\tilde{f}(\mathbf{k}_1)\tilde{f}(\mathbf{k}_2)\rangle_{\rm c}=(2\pi)^2\delta_{\rm D}(\mathbf{k}_1+\mathbf{k}_2)P(k) \ .
\end{equation}

We invert the transformation in Eq.~(\ref{eq:partition_function}) and replace $\mathcal{Z}(\mathbf{J})$ with the expression in Eq.~(\ref{eq:partition_function_cumulant_expansion})
\begin{widetext}
\begin{eqnarray}
    \label{eq:PDF_f_1}
    \mathcal{P}(\tilde{f})&&=\int\hat{\mathcal{D}}\mathbf{J}\mathcal{Z}(\mathbf{J}){\rm exp}\left[-i\int\frac{{\rm d}^2k}{(2\pi)^2}\mathbf{J}(\mathbf{k})\tilde{f}(\mathbf{k})\right] \nonumber\\
    &&=\int\hat{\mathcal{D}}\mathbf{J}{\rm exp}\left[\sum_{n=3}^{\infty}\frac{i^n}{n!}\int\frac{{\rm d}^2k_1}{(2\pi)^2}\cdots\int\frac{{\rm d}^2k_n}{(2\pi)^2}\langle\tilde{f}(\mathbf{k}_1)\cdots\tilde{f}(\mathbf{k}_n)\rangle_{\rm c}\mathbf{J}(\mathbf{k}_1)\cdots\mathbf{J}(\mathbf{k}_n)\right]{\rm exp}\left[-\frac{1}{2}\int\frac{{\rm d}^2k}{(2\pi)^2}P(k)\mathbf{J}(\mathbf{k})\mathbf{J}(-\mathbf{k})\right] \nonumber\\
    && \times {\rm exp}\left[-i\int\frac{{\rm d}^2k}{(2\pi)^2}\mathbf{J}(\mathbf{k})\tilde{f}(\mathbf{k})\right]\nonumber\\
    &&={\rm exp}\left[\sum_{n=3}^{\infty}\frac{(-1)^n}{n!}\int{\rm d}^2k_1\cdots\int{\rm d}^2k_n\langle\tilde{f}(\mathbf{k}_1)\cdots\tilde{f}(\mathbf{k}_n)\rangle_{\rm c}\frac{\delta^n}{\delta\tilde{f}(\mathbf{k}_1)\cdots\delta\tilde{f}(\mathbf{k}_n)}\right]\mathcal{P}_{\rm G}(\tilde{f}) \ ,
\end{eqnarray}
\end{widetext}
where $\mathcal{P}_{\rm G}(\tilde{f})$ is the Gaussian probability density functional
\begin{eqnarray}
    \label{eq:gaussian_probability_functional}
    \mathcal{P}_{\rm G}&&(\tilde{f})\nonumber\\
    &&=\int\hat{\mathcal{D}}\mathbf{J}{\rm exp}\Bigg[-\frac{1}{2}\int\frac{{\rm d}^2k}{(2\pi)^2}P(k)\mathbf{J}(\mathbf{k})\mathbf{J}(-\mathbf{k})\nonumber\\
    && -i\int\frac{{\rm d}^2k}{(2\pi)^2}\mathbf{J}(\mathbf{k})\tilde{f}(\mathbf{k})\Bigg] \nonumber\\
    && \propto {\rm exp}\left[-\frac{1}{2}\int\frac{{\rm d}^2k}{(2\pi)^2}\frac{\tilde{f}(\mathbf{k})\tilde{f}(-\mathbf{k})}{P(k)}\right] \ ,
\end{eqnarray}
up to a normalization constant and in the last line of Eq.~(\ref{eq:PDF_f_1}), we used the following identity
\begin{eqnarray}
    \label{eq:pdf_identity}
    \int&&\hat{\mathcal{D}}\mathbf{J}i^n\mathbf{J}(\mathbf{k}_1)\cdots\mathbf{J}(\mathbf{k}_n)\mathcal{P}_{\rm G}(\mathbf{J})\nonumber\\
    &&\times{\rm exp}\left[-i\int\frac{{\rm d}^2k}{(2\pi)^2}\mathbf{J}(\mathbf{k})\tilde{f}(\mathbf{k})\right] \nonumber\\
    &&= (-1)^n\times(2\pi)^{2n}\frac{\delta^n}{\delta\tilde{f}(\mathbf{k}_1)\cdots\delta\tilde{f}(\mathbf{k}_n)}\mathcal{P}_{\rm G}(\tilde{f}) \ .
\end{eqnarray}
With Eq.~(\ref{eq:PDF_f_1}), we can apply the power series expansion to the exponential term. This will give the functional derivatives of the Gaussian probability density functional which can be expressed using Eq.~(\ref{eq:Wiener-Hermite_functionals}) as $\mathcal{H}_n(\mathbf{k}_1,\ldots,\mathbf{k}_n)\mathcal{P}_{\rm G}(\tilde{f})/(-1)^n$. Substituting this into the power series expansion, we can obtain the generalized Gram-Charlier A series expansion in Eq.~(\ref{eq:fourier_P_expansion}).

\section{Diagrammatic method to evaluate $\langle\mathcal{H}_n^{\star}\mathcal{H}_m^{\star}\mathcal{H}_l\rangle_{\rm G}$ factors}
\label{appendix:B}
Based on Appendix.~A of Ref.~\cite{Matsubara1995}, here we present the diagrammatic rules in Fourier space for the products of generalized Wiener-Hermite functionals $\langle\mathcal{H}_n^{\star}\mathcal{H}_m^{\star}\mathcal{H}_l\rangle_{\rm G}$. 
\begin{enumerate}
    \item Corresponding to each $\mathcal{H}_{n_i}^{\star}$ or $\mathcal{H}_{n_i}$, draw $n_i$ points labeled by $\mathbf{k}_i^{(1)},\ldots,\mathbf{k}_i^{(n_i)}$, representing each mode.
    \item Create $\sum_i n_i/2$ pairs out of all the points such that two points in the same $\mathcal{H}_{n_i}^{\star}$ or $\mathcal{H}_{n_i}$ are not paired. If $\sum_i n_i/2$ is an odd number, then $\langle\mathcal{H}_n^{\star}\mathcal{H}_m^{\star}\mathcal{H}_l\rangle_{\rm G}=0$.
    \item Associate factors $(2\pi)^2\delta_{\rm D}(\mathbf{k}_i^{(p)}+\mathbf{k}_j^{(q)})P(k_i^{(p)})$ ($p\in \{1,\ldots,n_i\},q\in \{1,\ldots,n_j\}$) for each pair if the two points are from $\mathcal{H}_{n_i}^{\star}$ and $\mathcal{H}_{n_j}^{\star}$ respectively. If the two points are from $\mathcal{H}_{n_i}^{\star}$ and $\mathcal{H}_{n_j}$ separately, the associated factors change to $\delta_{\rm D}(\mathbf{k}_i^{(p)}-\mathbf{k}_j^{(q)})$ ($p\in \{1,\ldots,n_i\},q\in \{1,\ldots,n_j\}$) for each pair. We then make products of these factors.
    \item Sum up these products from all possible pair configurations.
\end{enumerate}

With the above diagrammatic rules, it is very convenient to compute any $\langle\mathcal{H}_n^{\star}\mathcal{H}_m^{\star}\mathcal{H}_l\rangle_{\rm G}$ factors with $n+m+l$ equal to an even number. Below we show all non-zero results needed in deriving Eq.~(\ref{eq:F_2nd_moment_Fourier_nnlo}).
\begin{widetext}
\setlength{\jot}{3mm}
\begin{eqnarray}
\label{eq:H0H0H0}
    &&\langle\mathcal{H}^{\star}_0\mathcal{H}^{\star}_0\mathcal{H}_0\rangle_{\mathrm{G}} = 1 \ ,\\
\label{eq:H1H1H0}
    &&\langle\mathcal{H}^{\star}_1(\mathbf{k}_1)\mathcal{H}^{\star}_1(\mathbf{k}_1^{\prime})\mathcal{H}_0\rangle_{\mathrm{G}} = (2\pi)^2P(k_1)\delta_\mathrm{D}^2(\mathbf{k}_1 + \mathbf{k}_1^{\prime}) \ ,\\
\label{eq:H2H2H0}
    &&\langle\mathcal{H}^{\star}_2(\mathbf{k}_1, \mathbf{k}_2)\mathcal{H}^{\star}_2(\mathbf{k}_1^{\prime}, \mathbf{k}_2^{\prime})\mathcal{H}_0\rangle_{\mathrm{G}} =(2\pi)^4P(k_1)P(k_2)\delta_\mathrm{D}^2(\mathbf{k}_1 + \mathbf{k}_1^{\prime})\delta_\mathrm{D}^2(\mathbf{k}_2 + \mathbf{k}_2^{\prime}) + \mathrm{sym} \ ,\\
\label{eq:H0H3H3}
    &&\langle\mathcal{H}^{\star}_0\mathcal{H}^{\star}_3(\mathbf{k}_1, \mathbf{k}_2, \mathbf{k}_3)\mathcal{H}_3(\mathbf{k}_1^{\prime}, \mathbf{k}_2^{\prime}, \mathbf{k}_3^{\prime})\rangle_{\mathrm{G}} = \delta_\mathrm{D}^2(\mathbf{k}_1 - \mathbf{k}_1^{\prime})\delta_\mathrm{D}^2(\mathbf{k}_2 - \mathbf{k}_2^{\prime})\delta_\mathrm{D}^2(\mathbf{k}_3 - \mathbf{k}_3^{\prime}) + \mathrm{sym} \ ,\\
\label{eq:H1H2H3}
    &&\langle\mathcal{H}^{\star}_1(\mathbf{k}_1)\mathcal{H}^{\star}_2(\mathbf{k}_1^{\prime}, \mathbf{k}_2^{\prime})\mathcal{H}_3(\mathbf{k}_1'', \mathbf{k}_2'', \mathbf{k}_3'')\rangle_{\mathrm{G}} = \delta_\mathrm{D}^2(\mathbf{k}_1 - \mathbf{k}_1'')\delta_\mathrm{D}^2(\mathbf{k}_2^{\prime} - \mathbf{k}_2'')\delta_\mathrm{D}^2(\mathbf{k}_3^{\prime} - \mathbf{k}_3'') + \mathrm{sym} \ ,\\
    \label{eq:H3H3H0}
    &&\langle\mathcal{H}^{\star}_3(\mathbf{k}_1, \mathbf{k}_2, \mathbf{k}_3)\mathcal{H}^{\star}_3(\mathbf{k}_1^{\prime}, \mathbf{k}_2^{\prime}, \mathbf{k}_3^{\prime})\mathcal{H}_0\rangle_{\mathrm{G}} = (2\pi)^6P(k_1)P(k_2)P(k_3)\delta_\mathrm{D}^2(\mathbf{k}_1 + \mathbf{k}_1^{\prime})\delta_\mathrm{D}^2(\mathbf{k}_2 + \mathbf{k}_2^{\prime})\delta_\mathrm{D}^2(\mathbf{k}_3 + \mathbf{k}_3^{\prime}) + \mathrm{sym}\ ,\\
    \label{eq:H1H4H3}
    &&\langle\mathcal{H}^{\star}_1(\mathbf{k}_1)\mathcal{H}^{\star}_4(\mathbf{k}_1^{\prime},\ldots,\mathbf{k}_4^{\prime})\mathcal{H}_3(\mathbf{k}_1^{''},\mathbf{k}_2^{''},\mathbf{k}_3^{''})\rangle_{\mathrm{G}} \nonumber\\
    &&= (2\pi)^2\delta_\mathrm{D}^2(\mathbf{k}_1 + \mathbf{k}_1^{\prime})\delta_\mathrm{D}^2(\mathbf{k}_2^{\prime} - \mathbf{k}_1^{''})\delta_\mathrm{D}^2(\mathbf{k}_3^{\prime} - \mathbf{k}_2^{''})\delta_\mathrm{D}^2(\mathbf{k}_4^{\prime} - \mathbf{k}_3^{''})P(k_1)+\mathrm{sym} \ ,\\
    \label{eq:H2H3H3}
    &&\langle\mathcal{H}^{\star}_2(\mathbf{k}_1,\mathbf{k}_2)\mathcal{H}^{\star}_3(\mathbf{k}_1^{\prime},\mathbf{k}_2^{\prime},\mathbf{k}_3^{\prime})\mathcal{H}_3(\mathbf{k}_1^{''},\mathbf{k}_2^{''},\mathbf{k}_3^{''})\rangle_{\mathrm{G}} \nonumber\\
    &&= (2\pi)^2\delta_\mathrm{D}^2(\mathbf{k}_1 + \mathbf{k}_1^{\prime})\delta_\mathrm{D}^2(\mathbf{k}_2 - \mathbf{k}_1^{''})\delta_\mathrm{D}^2(\mathbf{k}_2^{\prime} - \mathbf{k}_2^{''})\delta_\mathrm{D}^2(\mathbf{k}_3^{\prime} - \mathbf{k}_3^{''})P(k_1) + \mathrm{sym} \ ,\\
    \label{eq:H0H4H4}
    &&\langle\mathcal{H}^{\star}_0\mathcal{H}_4^{\star}(\mathbf{k}_1,\ldots,\mathbf{k}_4)\mathcal{H}_4(\mathbf{k}_1^{\prime},\ldots,\mathbf{k}_4^{\prime})\rangle_{\mathrm{G}} =  \delta_\mathrm{D}^2(\mathbf{k}_1 - \mathbf{k}_1^{\prime})\delta_\mathrm{D}^2(\mathbf{k}_2 - \mathbf{k}_2^{\prime})\delta_\mathrm{D}^2(\mathbf{k}_3 - \mathbf{k}_3^{\prime})\delta_\mathrm{D}^2(\mathbf{k}_4 - \mathbf{k}_3^{\prime}) + \mathrm{sym} \ ,\\
    \label{eq:H1H3H4}
    &&\langle\mathcal{H}^{\star}_1(\mathbf{k}_1)\mathcal{H}_3^{\star}(\mathbf{k}_1^{\prime},\mathbf{k}_2^{\prime}, \mathbf{k}_3^{\prime})\mathcal{H}_4(\mathbf{k}_1^{''},\ldots,\mathbf{k}_4^{''})\rangle_{\mathrm{G}} =  \delta_\mathrm{D}^2(\mathbf{k}_1 - \mathbf{k}_1^{''})\delta_\mathrm{D}^2(\mathbf{k}_1^{\prime} - \mathbf{k}_2^{''})\delta_\mathrm{D}^2(\mathbf{k}_2^{\prime} - \mathbf{k}_3^{''})\delta_\mathrm{D}^2(\mathbf{k}_3^{\prime} - \mathbf{k}_4^{''}) + \mathrm{sym} \ ,\\
    \label{eq:H2H2H4}
    &&\langle\mathcal{H}^{\star}_2(\mathbf{k}_1, \mathbf{k}_2)\mathcal{H}_2^{\star}(\mathbf{k}_1^{\prime},\mathbf{k}_2^{\prime})\mathcal{H}_4(\mathbf{k}_1^{''},\ldots,\mathbf{k}_4^{''})\rangle_{\mathrm{G}} =  \delta_\mathrm{D}^2(\mathbf{k}_1 - \mathbf{k}_1^{''})\delta_\mathrm{D}^2(\mathbf{k}_2 - \mathbf{k}_2^{''})\delta_\mathrm{D}^2(\mathbf{k}_1^{\prime} - \mathbf{k}_3^{''})\delta_\mathrm{D}^2(\mathbf{k}_2^{\prime} - \mathbf{k}_4^{''}) + \mathrm{sym} \ ,
\end{eqnarray}
\end{widetext}
where all ``sym" expressions in the above equations denote all following additional terms that have the same form as the previous one but are composed of other pair configurations in the diagrammatic scheme. 

\section{Plots of $g_{ijklm}$ factors and $g_1$ function}
\label{appendix:C}
\begin{figure*}[t!]
\begin{minipage}{.33\linewidth}
\centering
{\label{main:g10000}\includegraphics[scale=.3]{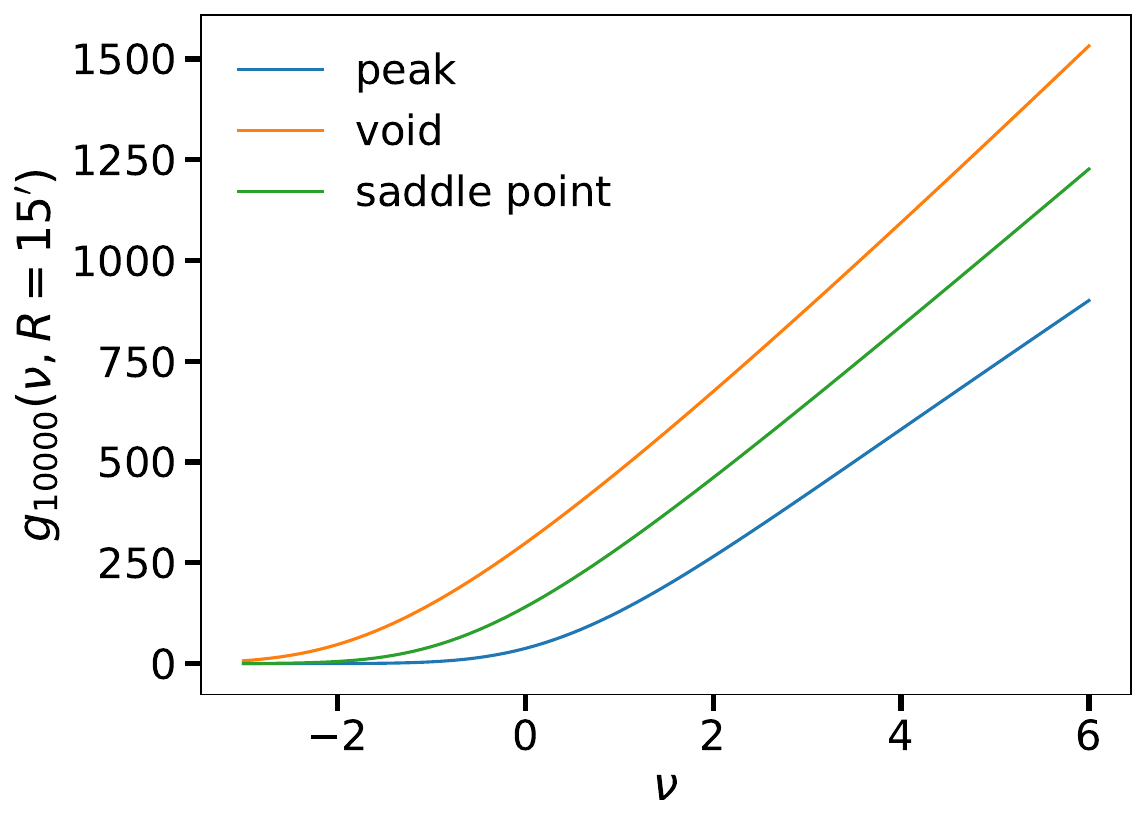}}
\end{minipage}%
\begin{minipage}{.33\linewidth}
\centering
{\label{main:g01000}\includegraphics[scale=.3]{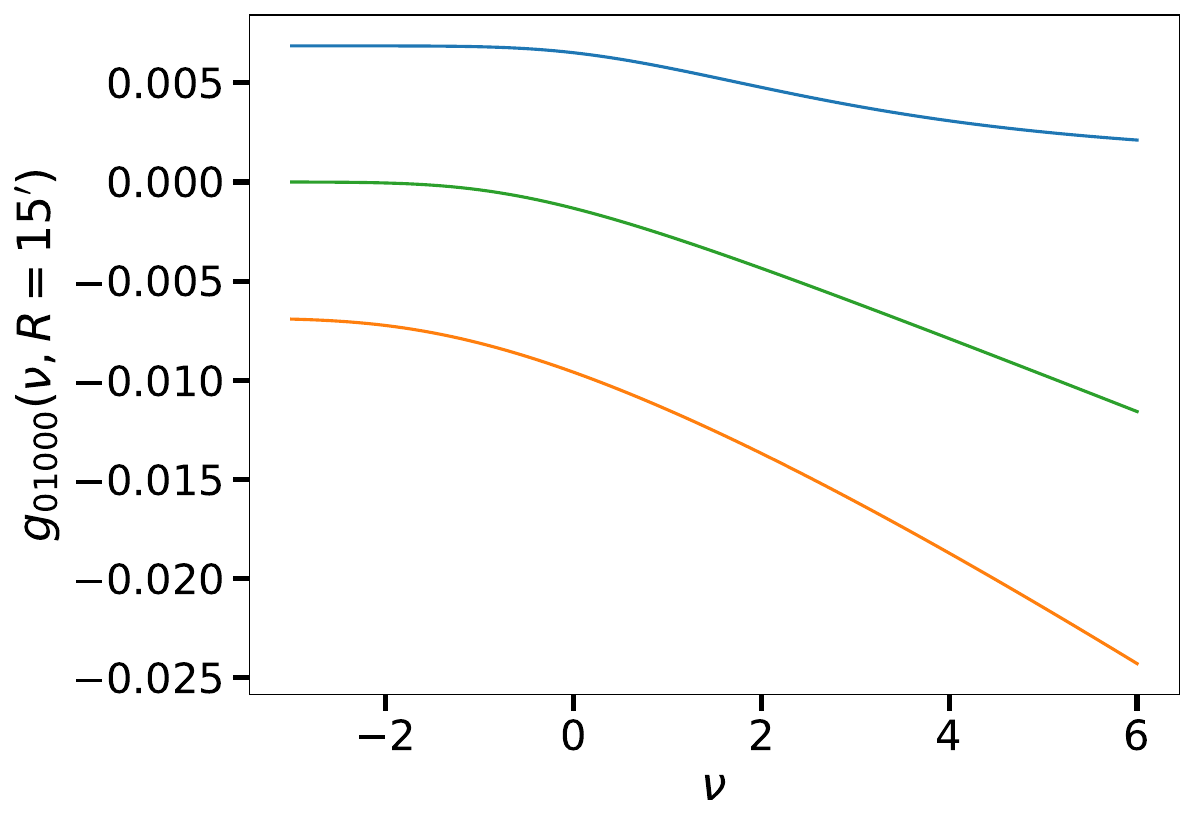}}
\end{minipage}%
\begin{minipage}{.33\linewidth}
\centering
{\label{main:g20000}\includegraphics[scale=.3]{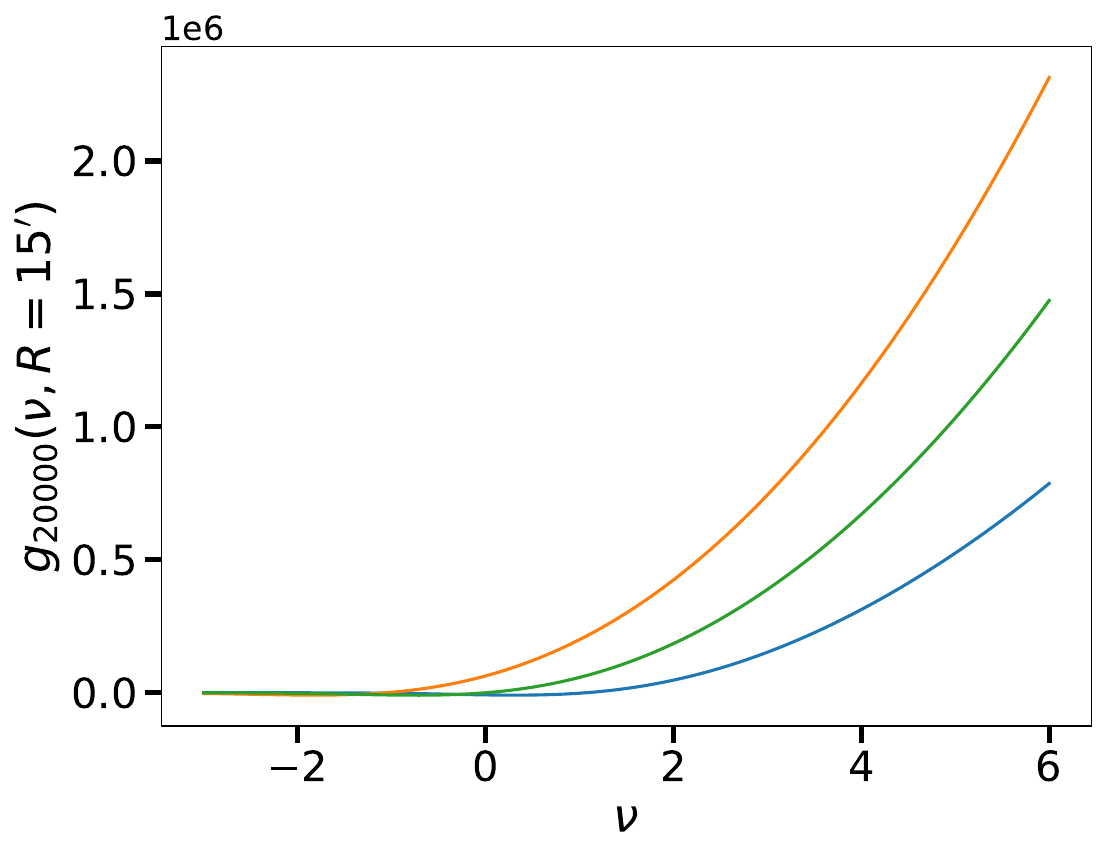}}
\end{minipage} \par\medskip
\begin{minipage}{.33\linewidth}
\centering
{\label{main:g11000}\includegraphics[scale=.3]{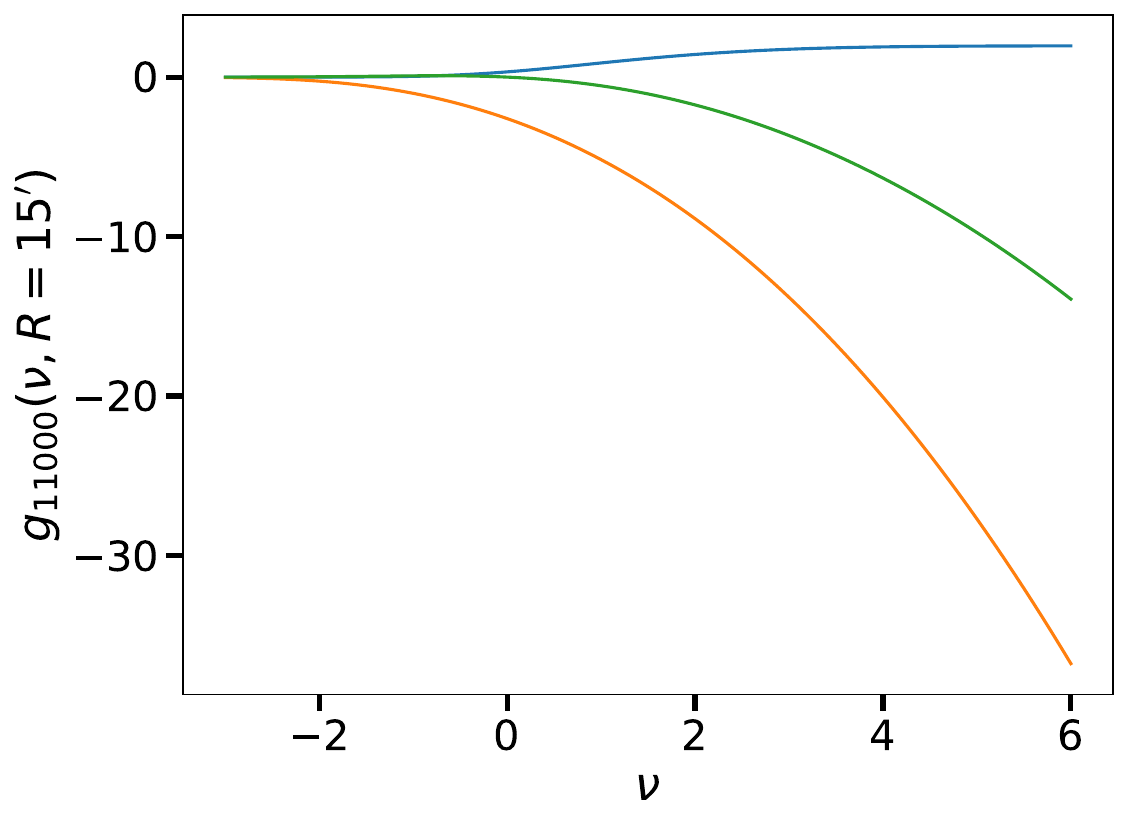}}
\end{minipage}%
\begin{minipage}{.33\linewidth}
\centering
{\label{main:g02000}\includegraphics[scale=.3]{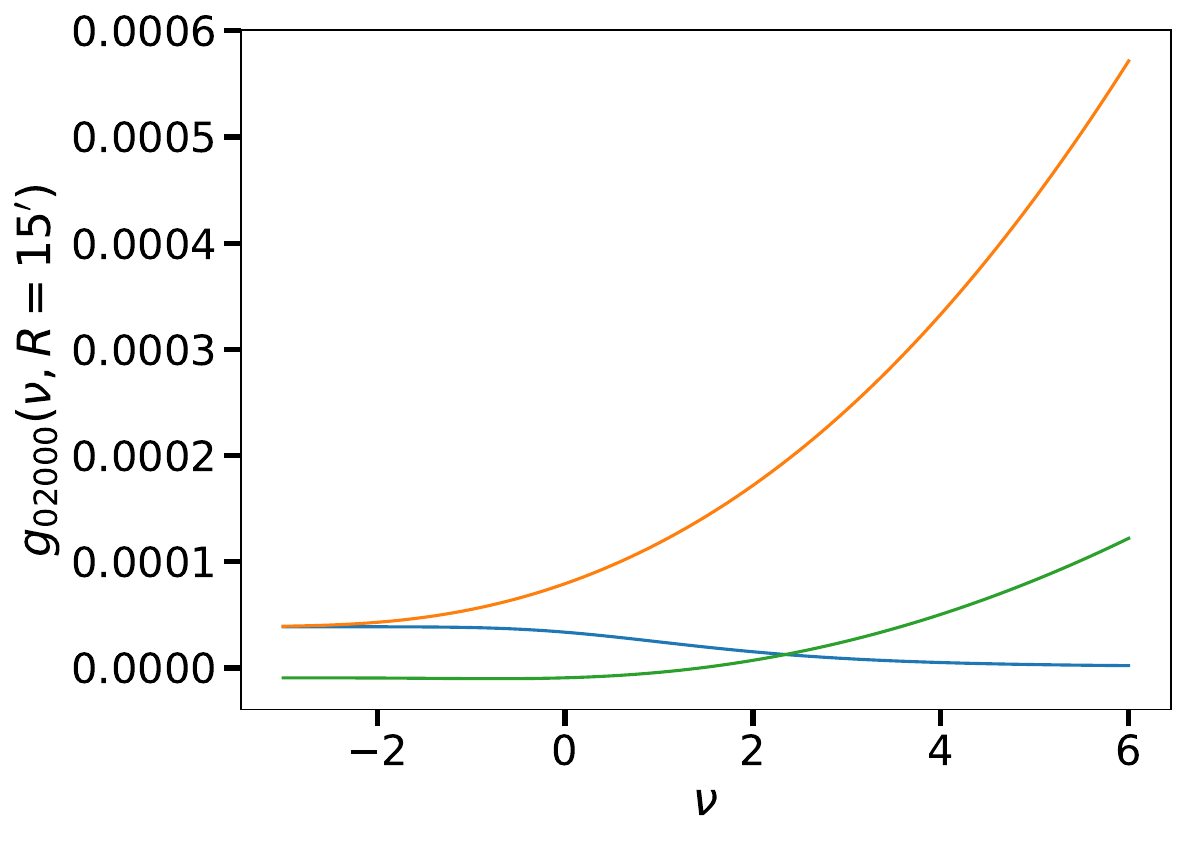}}
\end{minipage}%
\begin{minipage}{.33\linewidth}
\centering
{\label{main:g00100}\includegraphics[scale=.3]{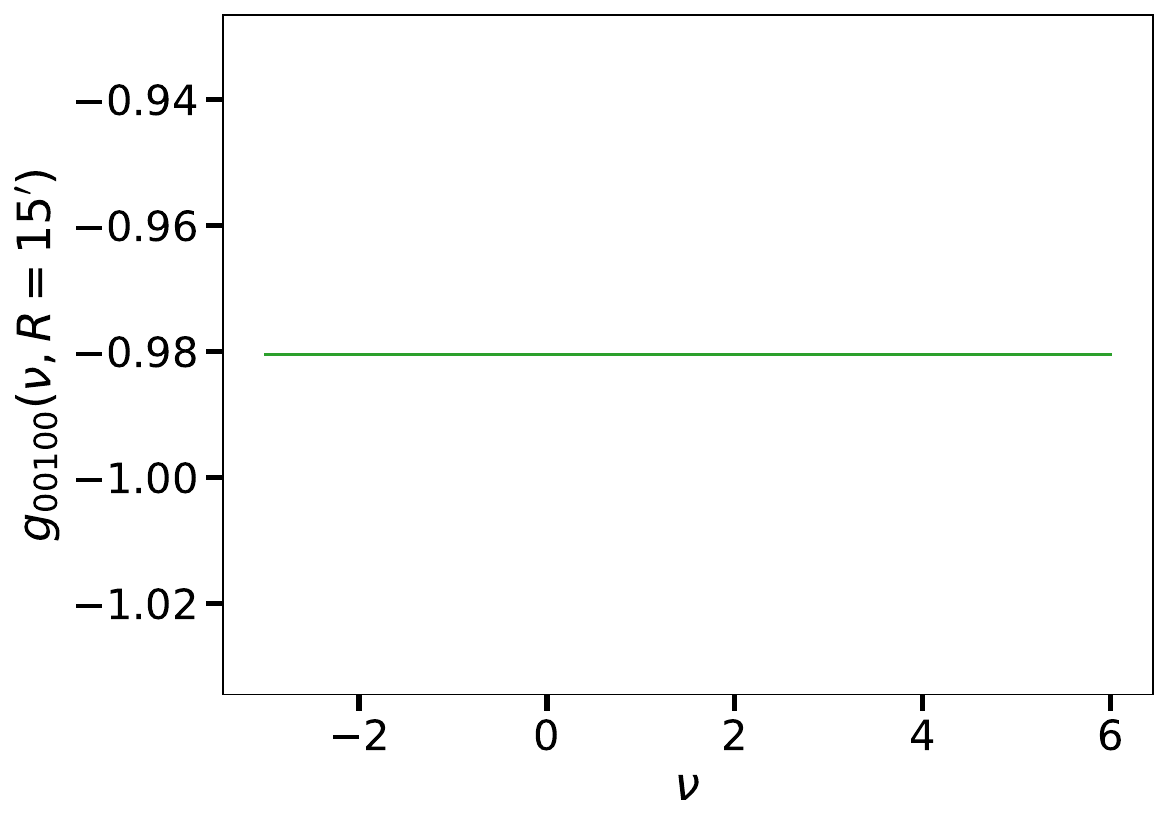}}
\end{minipage} \par\medskip
\centering
{\label{main:g00010}\includegraphics[scale=.3]{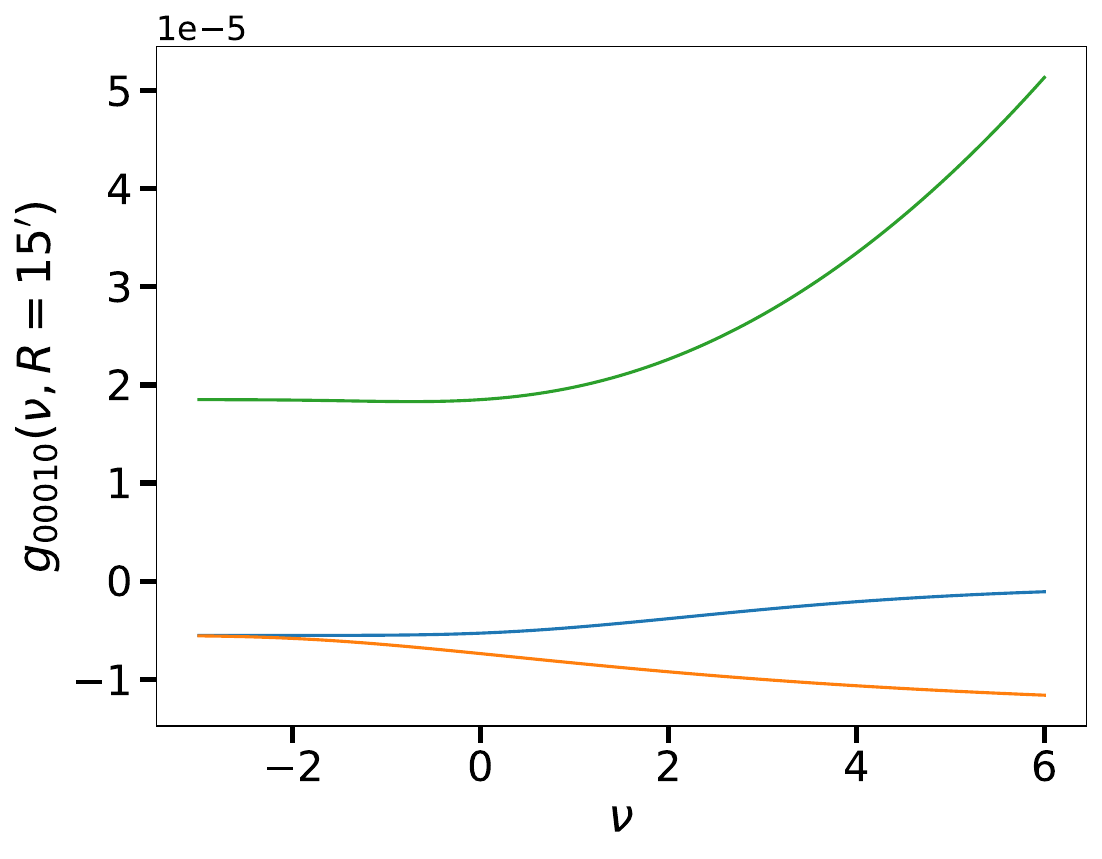}}
\caption{Plots of $g_{ijklm}$ factors as functions of the threshold $\nu$. The range of $\nu$ here is from $-3$ to $6$. The color curves in all the other sub-panels have the same representation as that denoted in the top left subplot. For $g_{00100}(\nu)$ specifically, peaks, voids and saddle points have the same constant function.}
\label{fig:7_g_ijklm}
\end{figure*}

Here we show the seven $g_{ijklm}(\nu)$ functions in Eqs.~(\ref{eq:g1}) and (\ref{eq:g2}) in Fig.~\ref{fig:7_g_ijklm} where we adopt the same cosmological parameters as those in Sec.~\ref{sec:2d_weak_lensing} and a smoothing scale of $15^{\prime}$.
\begin{figure*}[t!]
\begin{minipage}{.5\linewidth}
\centering
{\label{main:g1_k_critical}\includegraphics[scale=.45]{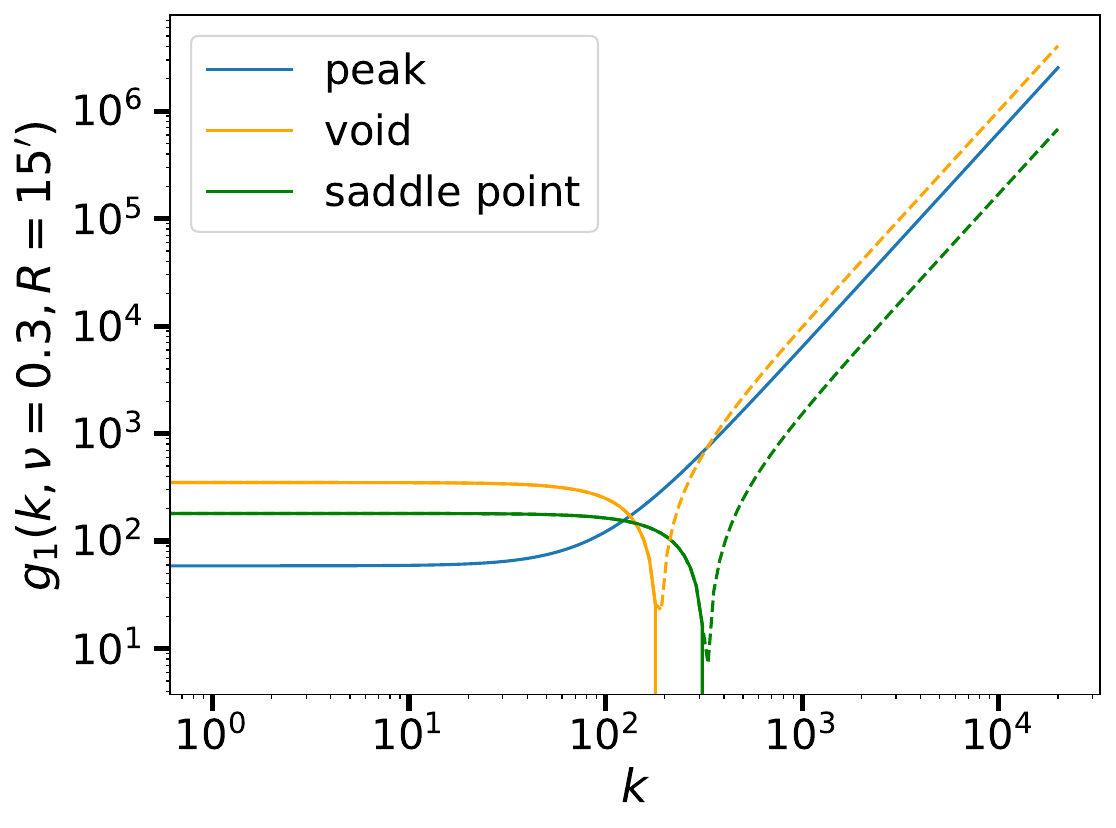}}
\end{minipage}%
\begin{minipage}{.5\linewidth}
\centering
{\label{main:g1_k_nu}\includegraphics[scale=.45]{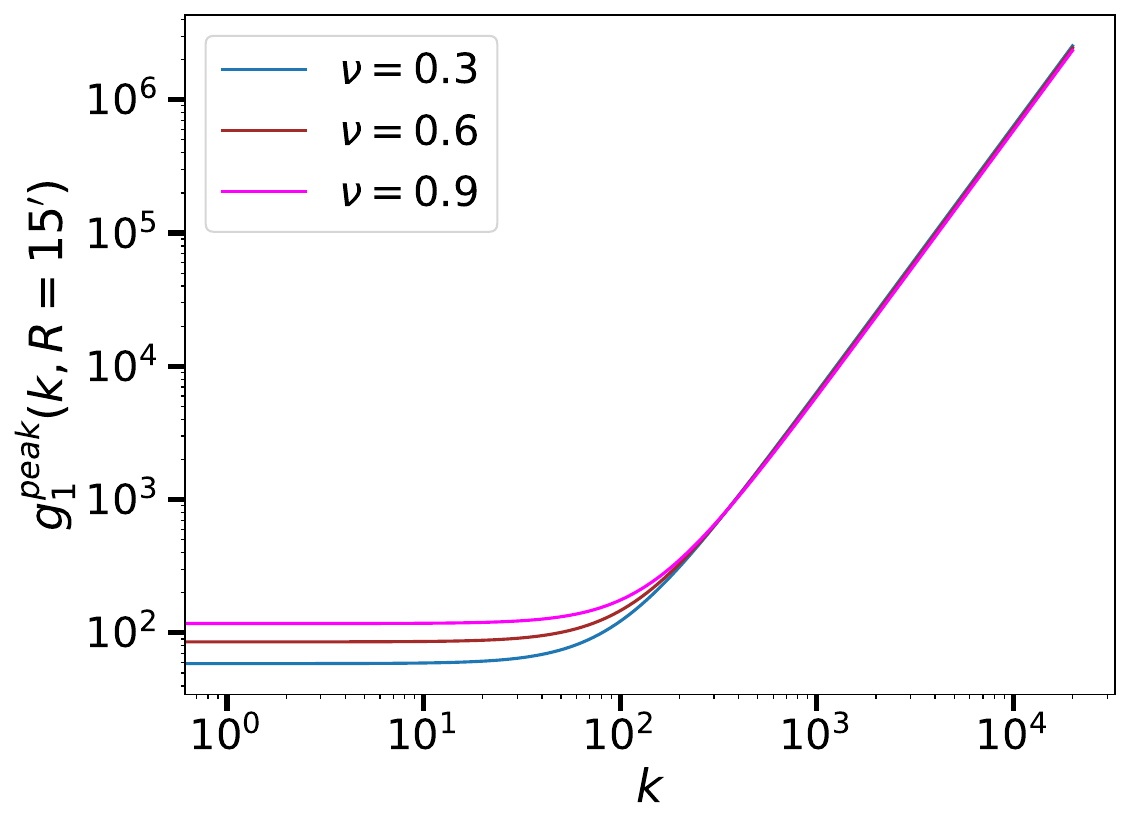}}
\end{minipage}
\caption{Plots of $g_1(\mathbf{k})$ in Eq.~(\ref{eq:g1}) under a Gaussian smoothing kernel with the smoothing scale $R=15^{\prime}$. The left panel shows the $g_1$ function of different critical point above the same threshold $\nu=0.3$. The dashed lines are the absolute value of the negative part of the function. The right panel shows the $g_1$ function of peak above different thresholds.}
\label{fig:g1_k}
\end{figure*}

By combing Fig.~\ref{fig:7_g_ijklm} and Eqs.~(\ref{eq:F_power_spectrum_nnlo}), (\ref{eq:g1}) and (\ref{eq:g2}), we can observe that $g_{10000}$ and $g_{20000}$ are the two dominant factors that determine the amplitude of the power spectrum for different critical points. In both subplots, the corresponding factor for voids is greater than that for saddle points, which in turn is greater than that for peaks within the range $0<\nu<6$. This numerically explains the amplitude relation among the power spectrum of peaks, voids and saddle points we computed in Sec.~\ref{sec:2d_weak_lensing}. 

As a schematic illustration, we show in the left panel of Fig.~\ref{fig:g1_k} the $g_1$ functions of all three types of critical points above the threshold $\nu=0.3$. And we do observe that on small to intermediate $k$ scales, voids have a larger function value than that of saddle points which in turn larger than that of peaks. Furthermore, in the right panel of Fig.~\ref{fig:g1_k}, we show how $g_1$ function of peaks, which has a quadratic form, varies along with different threshold $\nu$. For higher threshold, larger the leading order Gaussian response function $g_1$ value is, therefore a higher clustering amplitude in the 2PCF. This holds true for voids and saddle points as well.

\section{Angular integration of the peak power spectrum}
\label{appendix:Senpai_derivation}
We demonstrate here how to simplify the type of integrals that appear in Eq.~(\ref{eq:wl_power_spectrum_nlo}) to obtain the extrema power spectra of 2D fields. This appendix borrows from the derivation presented in Ref.~\cite{matsubara2020} and is only shown here for completeness.

We are interested in constrained integrals of the form
\begin{equation}
    \label{eq:A_def}
    A = \int_{\boldsymbol{k}_1+\boldsymbol{k}_2=\boldsymbol{k}}\left(\hat{\boldsymbol{k}}_1 \cdot \hat{\boldsymbol{k}}_2\right)^l X\left(k_1\right) Y\left(k_2\right) \ ,
\end{equation}
where $\hat{\boldsymbol{k}}_1 \cdot \hat{\boldsymbol{k}}_2$ is the cosine of the angle $\theta$ between ${\bm k}_1$ and ${\bm k}_2$. In 2D, this constraint can be explicitly written as
\begin{eqnarray}
    \label{eq:A_2D}
    A = &&\int {\rm d}^2 r e^{-i \boldsymbol{k} \cdot \boldsymbol{r}} \int \frac{{\rm d}^2 \boldsymbol{k}_1}{(2 \pi)^2} \frac{{\rm d}^2 \boldsymbol{k}_2}{(2 \pi)^2} e^{i\left(\boldsymbol{k}_1+\boldsymbol{k}_2\right) \cdot r}\nonumber\\
    && \times\left(\hat{\boldsymbol{k}}_1 \cdot \hat{\boldsymbol{k}}_2\right)^l X\left(k_1\right) Y\left(k_2\right) \ .
\end{eqnarray}
Rotational invariance of the system makes the result of the $\boldsymbol{k}_i$ integrals only dependent on the amplitude $r$ of $\boldsymbol{r}$ so that we can directly perform the angular integration replacing the exponentials by their angular averages given by Bessel functions of the first kind:
\begin{eqnarray}
    \label{eq:A_bessel}
    A = &&2\pi \int {\rm d} r J_0(kr) \int \frac{{\rm d}^2 \boldsymbol{k}_1}{(2 \pi)^2} \frac{{\rm d}^2 \boldsymbol{k}_2}{(2 \pi)^2} J_0(|\boldsymbol{k}_1+\boldsymbol{k}_2|r)\nonumber\\
    &&\times\left(\hat{\boldsymbol{k}}_1 \cdot \hat{\boldsymbol{k}}_2\right)^l X\left(k_1\right) Y\left(k_2\right) \ .
\end{eqnarray}

A theorem of Bessel function enables us to write
\begin{equation}
    \label{eq:bessel_theorem}
    J_0\left(\left|\boldsymbol{k}_1+\boldsymbol{k}_2\right| r\right)=\sum_{n=-\infty}^{\infty}(-1)^n J_n\left(k_1 r\right) J_n\left(k_2 r\right) e^{i n \theta} \ ,
\end{equation}
and the final trick consists in expressing $\hat{\boldsymbol{k}}_1 \cdot \hat{\boldsymbol{k}}_2 = {\rm cos}(\theta)$ as a sum of exponentials thanks to Euler's formula, and expand its $l^{\rm th}$ power thanks to the Binomial theorem. We get
\begin{equation}
    \label{eq:binomial_theorem}
    (\hat{\boldsymbol{k}}_1 \cdot \hat{\boldsymbol{k}}_2)^l = \frac{1}{2^l} e^{-il\theta} \sum_{m = 0}^{l} \binom{l}{m} e^{2im\theta} \ .
\end{equation}

Finally, combining Eqs.~(\ref{eq:bessel_theorem}) and (\ref{eq:binomial_theorem}) into Eq.~(\ref{eq:A_bessel}), and noticing that every integral in which $n \neq l - 2m$ is 0 since it gives the integral of $e^{i\theta}$ over the full circle, we get
\begin{eqnarray}
    A =&&2\pi \int {\rm d} r J_0(kr) \frac{1}{2^l} \sum_{m = 0}^l \binom{l}{m} \nonumber\\
    && \times (-1)^{l-2m} \int \frac{k {\rm d}k}{2 \pi} J_{l-2m}(kr)X(k)\nonumber\\
    &&\times\int \frac{k {\rm d}k}{2 \pi} J_{l-2m}(kr)Y(k) \ .
\end{eqnarray}

We have thus reduced the computation of $A$ to a product of two 1D integrals which are Fourier transforms (or more specifically Hankel) of $X$ and $Y$, and can thus be very easily implemented using traditional methods such as FFTs, and a final radial integration. This enables the efficient numerical evaluation of the extrema power spectra presented in this paper.

\section{Comparison of other critical point 2PCFs to MC integration}
\label{appendix:MC_comparison}
\begin{figure*}[ht!]
\begin{minipage}{.5\linewidth}
\centering
{\label{main:xi_vv_mc}\includegraphics[scale=.45]{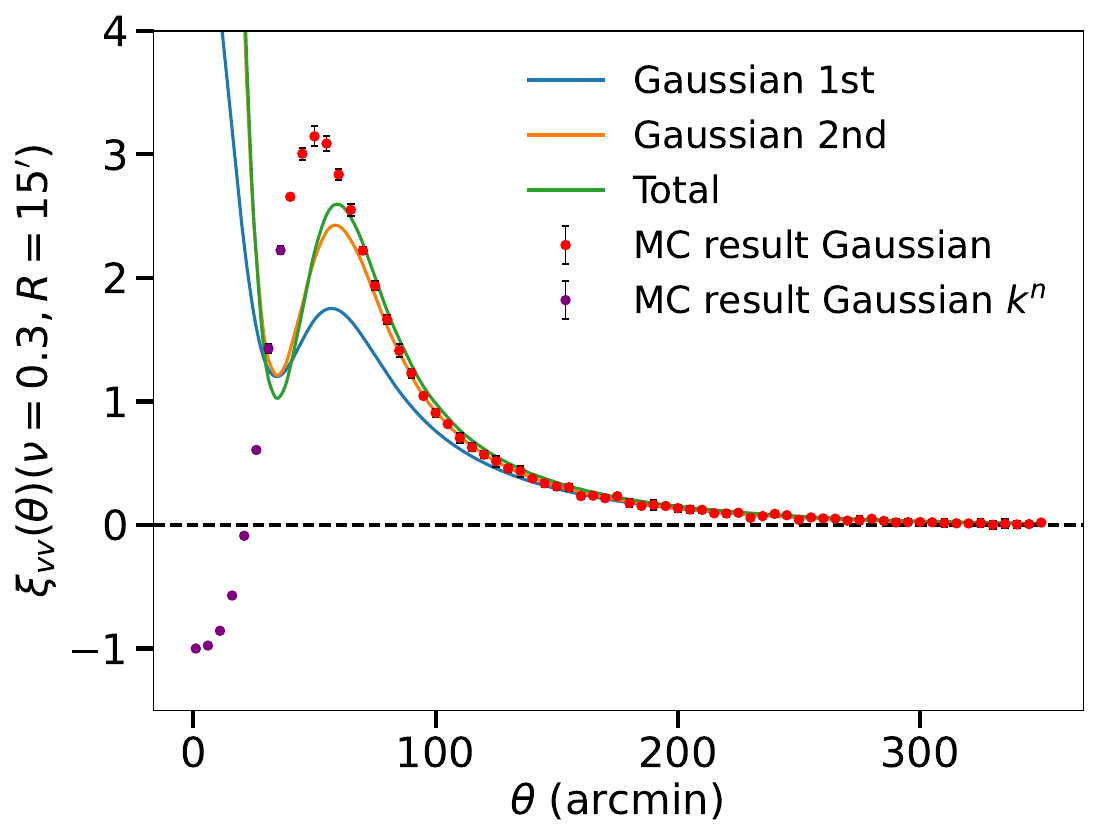}}
\end{minipage}%
\begin{minipage}{.5\linewidth}
\centering
{\label{main:xi_ss_mc}\includegraphics[scale=.45]{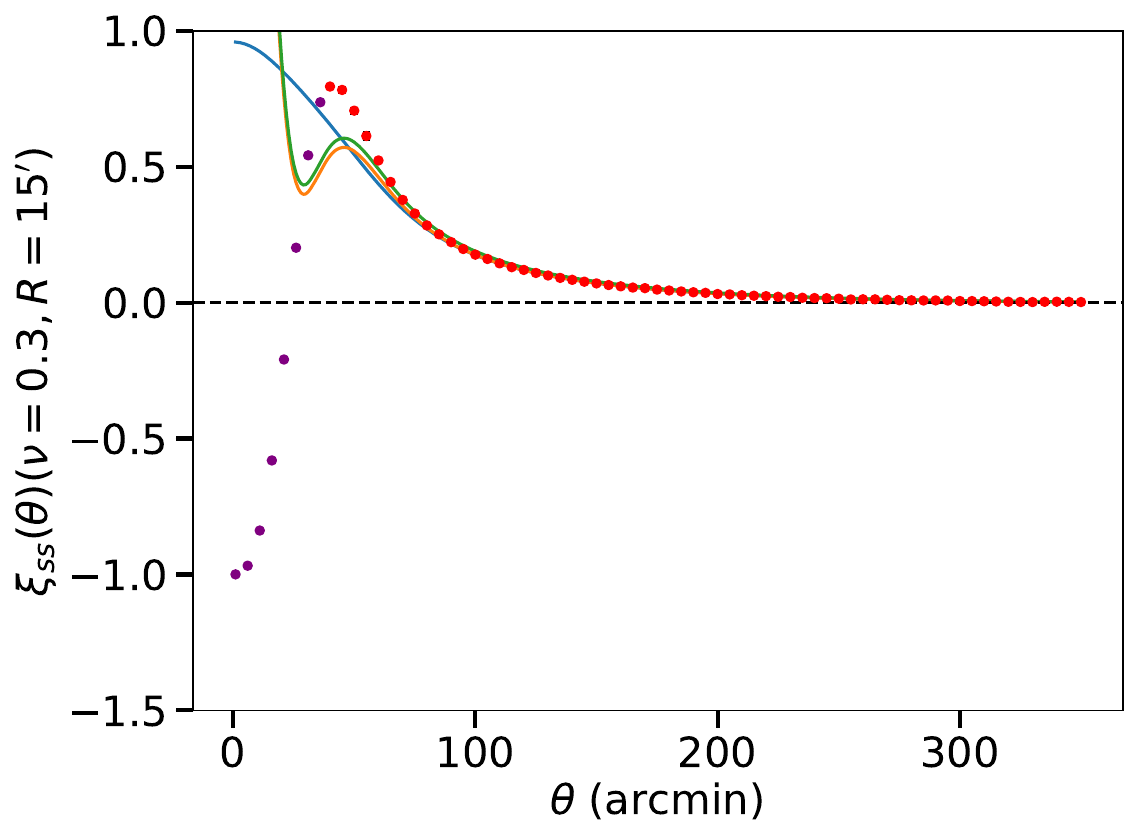}}
\end{minipage}
\caption{\textit{Left}: The auto 2PCFs of weak lensing voids (minima) above a threshold $\nu=0.3$ with a Gaussian smoothing scale of $15^{\prime}$. All colored curves and scattered dots with error bars have the same representation to those in Fig.~\ref{fig:MC_result}. \textit{Right}: The same auto 2PCFs as those in the left panel but for weak lensing saddle points.}
\label{fig:auto_2pcfs_void_saddle}
\end{figure*}

In the main text, we showed a comparison between our analytical predictions and the exact MC integration results with Gaussian assumption only for peak 2PCF. In this appendix, we show the same comparison but for all pairs of critical points. These include the auto 2PCFs of both voids and saddle points, as well as the cross 2PCFs among all three types of critical points. In Fig.~\ref{fig:auto_2pcfs_void_saddle}, we show the two auto 2PCFs. In both panels, the MC integration with the power law approximation of the weak lensing convergence power spectrum (purple dots) qualitatively exhibits the exclusion zone on small angular scales where our analytic theory is limited in its prediction as discussed in Sect~\ref{sec:MC_integration}. On large angular separations ($\theta > 100^{\prime}$), our analytic predictions from different orders of perturbative bias expansion converge and agree with the exact MC integration results under the Gaussian assumption. It is worth noticing that on the high amplitude part of both 2PCFs ($\theta\approx 60^{\prime}$ for voids and $45^{\prime}$ for saddle points), the 2nd-order Gaussian approximation has quite a discrepancy with respect to the MC integration result, this indicates that the 3rd-order Gaussian approximation term in NNLO would have a more significant role for clustering of voids and saddle points compared to what was shown for peaks. Similar to the peak clustering case, the bispectrum correction here changes the amplitude of the 2PCF but not the overall shape.

\begin{figure*}
\centering
{\label{main:2pcf_pv}\includegraphics[scale=.35]{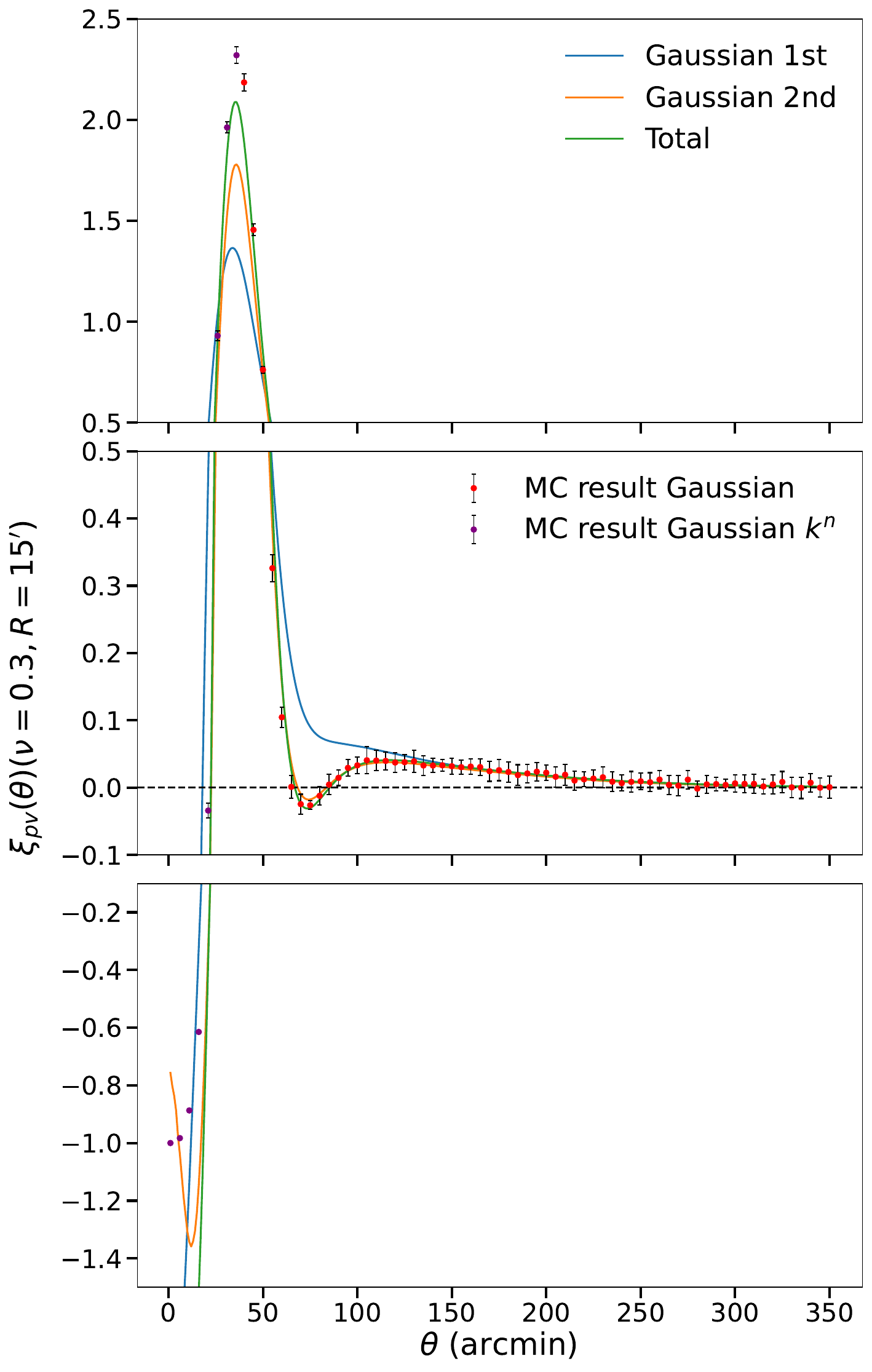}}\par\medskip
\begin{minipage}{.5\linewidth}
\centering
{\label{main:2pcf_ps}\includegraphics[scale=.35]{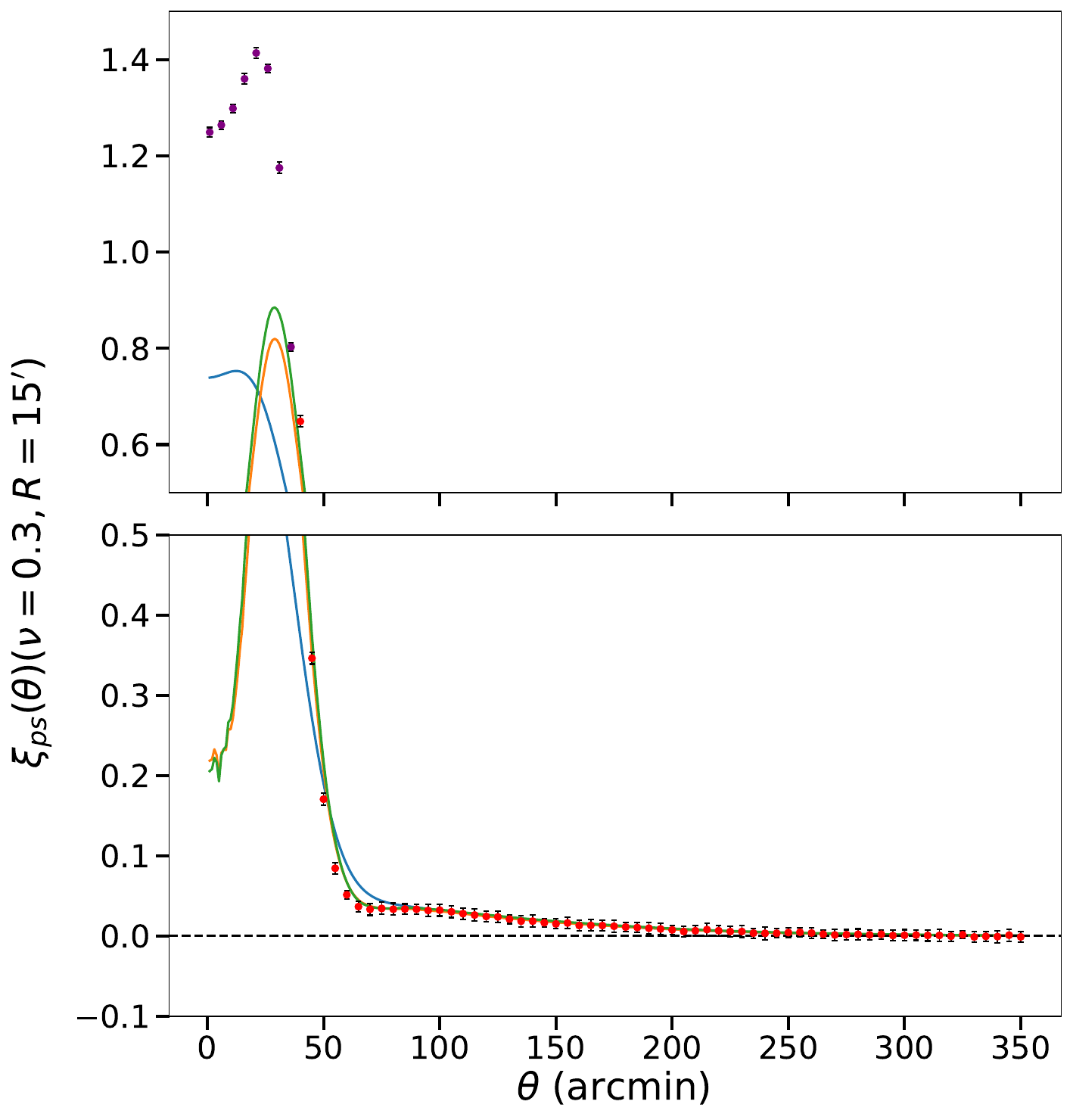}}
\end{minipage}%
\begin{minipage}{.5\linewidth}
\centering
{\label{main:2pcf_vs}\includegraphics[scale=.35]{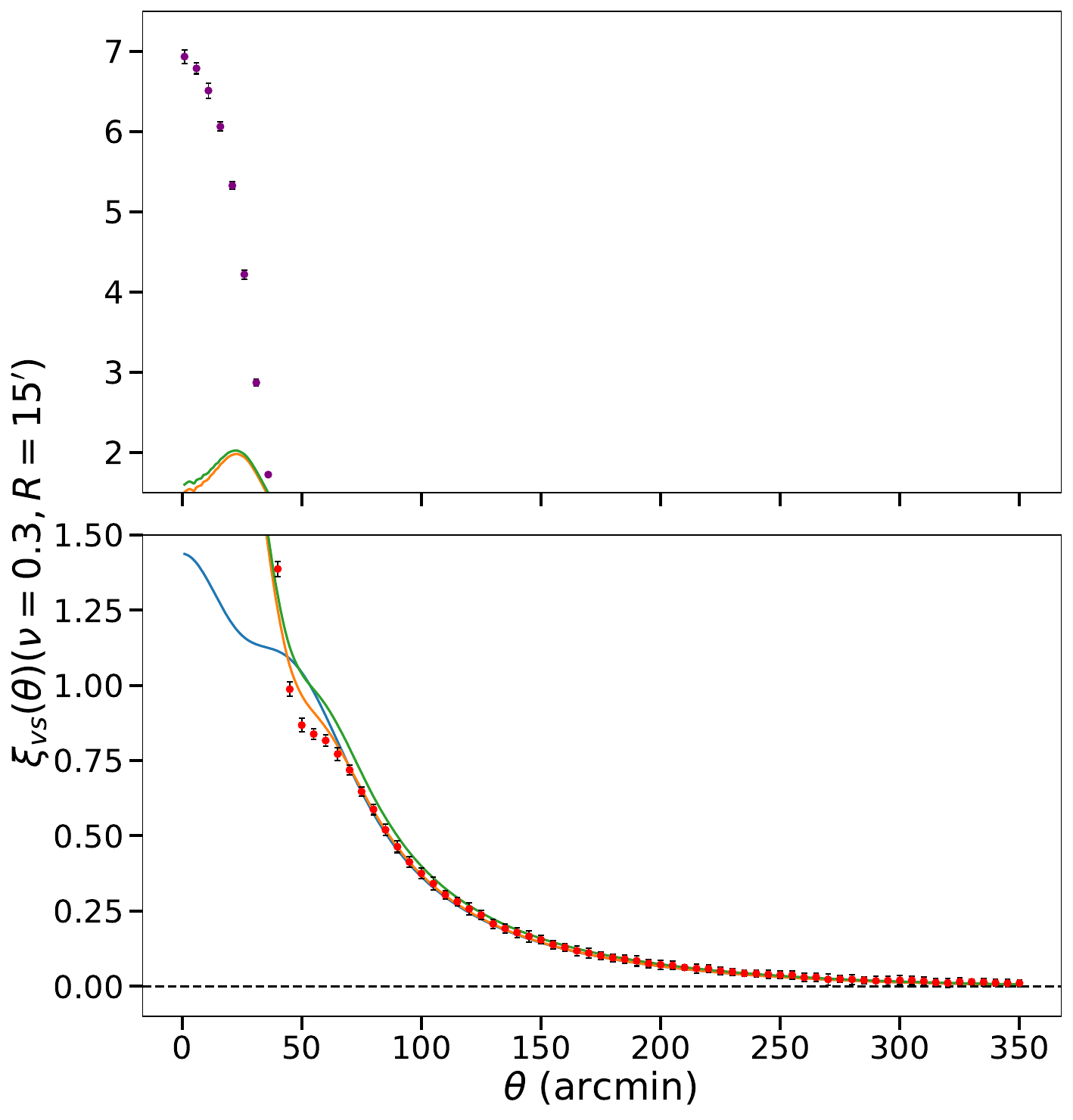}}
\end{minipage}
\caption{\textit{Top}: The cross 2PCF between peaks and voids above a threshold $\nu=0.3$ with a Gaussian smoothing scale of $15^{\prime}$. All colored curves and scattered dots with error bars have the same representation to those in Fig.~\ref{fig:MC_result}. The bottom left and right plots show the cross 2PCFs of peak-saddle point, and void-saddle point respectively. We divide each cross 2PCF plot into two or three panels, covering a range of linear scales with different intervals, allowing for a clearer examination of the small amplitude at larger separations.}
\label{fig:cross_2pcfs_MC}
\end{figure*}

In Fig.~\ref{fig:cross_2pcfs_MC} we show the comparison between all three cross 2PCFs among different types of critical points and their respective MC integration results. In the top panel, we observe that there is not only an exclusion zone on small angular scales between peaks and voids, but also a turnover feature with negative amplitude on $\theta\approx 70^{\prime}$, which is captured fairly accurately by our analytic predictions. This implies that there are two angular scales on which the clustering between a peak and a void above the same threshold is negatively correlated, different from what we have shown above. On the other hand, MC results in the bottom two plots confirm that there are no exclusion zones between saddle points and the other two types of critical points, although the amplitude of the cross 2PCFs on small angular scales is not well described by the analytic predictions.
 
\end{document}